\documentclass[onecolumn]{emulateapj}

\usepackage{lscape}
\usepackage{longtable}

\newcommand{\gf}{{\it gf}}
\newcommand{\vt}{v$_{\rm t}$}
\newcommand{\teff}{T$_{\rm eff}$}
\newcommand{\logg}{log~$g$}
\newcommand{\msun}{M$_{\odot}$}

\newcommand{\leps}{$\log{\epsilon}$}
\newcommand{\mfeh}{\rm{[m/H]}}

\newcommand{\mlogg}{\rm{log}\thinspace g}
\newcommand{\loge}{\log{\epsilon}}

\shorttitle{Abundances in Baade's Window}
\shortauthors{Fulbright et al.}

\begin{document}

\title{Abundances of Baade's Window Giants from Keck/HIRES Spectra:  II. The Alpha- and Light Odd Elements\footnote{Based on data obtained at the W. M. Keck Observatory, which is operated as a scientific partnership among the California Institute of Technology, the University of California, and NASA, and was made possible by the generous financial support of the W. M. Keck Foundation.}}

\author{Jon. P. Fulbright}
\affil{Observatories of the Carnegie Institution of Washington, 813 Santa Barbara St., Pasadena, CA 91101}
\affil{Department of Physics and Astronomy, Johns Hopkins University, 3400 North Charles Street, Baltimore, MD 21218}
\email{jfulb@pha.jhu.edu}

\author{Andrew McWilliam}
\affil{Observatories of the Carnegie Institution of Washington, 813 Santa Barbara St., Pasadena, CA 91101}
\email{andy@ociw.edu}

\author{R. Michael Rich}
\affil{Division of Astronomy, Department of Physics and Astronomy, UCLA, Los Angeles, CA  90095-1562}
\email{rmr@astro.ucla.edu}

\begin{abstract}
We report detailed chemical abundance analysis of 27 RGB stars towards the 
Galactic bulge in Baade's Window for elements produced by massive stars: O, 
Na, Mg, Al, Si, Ca and Ti.  All of these elements are overabundant in the 
bulge relative to the disk, indicating that the bulge is enhanced in Type~II 
supernova ejecta and most likely formed more rapidly than the disk.  

Our [Mg/Fe] ratios, confirmed by [Al/Mg], declines much more slowly with 
[Fe/H] than O, Si, Ca and Ti.  The [Mg/Fe] ratio stays above +0.25 dex up
to well above solar metallicity.  We attribute the rapid decline of [O/Fe] to a 
metallicity-dependent modulation of the oxygen yield from massive stars, 
perhaps connected to the Wolf-Reyet phenomenon.

The explosive nucleosynthesis alphas, Si, Ca and Ti, relative to Fe, possess 
identical trends with [Fe/H], consistent with their putative common origin.  
We note that different behaviors of hydrostatic and explosive alpha elements 
can be seen in the stellar abundances of stars in Local Group dwarf galaxies.  
We also attribute the decline of Si,Ca and Ti, relative to Mg, to 
metallicity-dependent yields for these explosive alpha elements from Type~II 
supernovae.  The production of explosive alphas in Type~Ia supernovae likely 
explains the absence of obvious difference between Mg and Si, Ca, Ti trends in 
the Galactic thin disk.

An alternative explanation for the [Mg/$<$SiCaTi$>$] increase with metallicity 
is an excess population of 30--35M$_{\odot}$ stars in the bulge that grew in 
importance with [Fe/H].  In this model the bulge formation timescale would 
have been significantly longer than our favored scenario of declining 
metallicity-dependent yields from Type~II supernovae.

The starkly smaller scatter of [$<$SiCaTi$>$/Fe] with [Fe/H] in the bulge, 
as compared to the halo, is consistent with expected efficient mixing for the 
bulge.  Since the metal-poor bulge [$<$SiCaTi$>$/Fe] ratios are higher than 
$\sim$80\% of the halo, the early bulge could not have formed from gas with 
the present-day halo composition.  If the bulge formed from halo gas, it 
occurred before $\sim$80\% of today's stellar halo; and the halo subsequently 
reduced its alpha/Fe ratios significantly.

The lack of overlap between the thick and thin disk composition with the bulge
does not support the idea that the bulge was built by a thickening of the disk 
driven by the bar.

The trend of [Al/Fe] is very sensitive to the chemical evolution environment: 
a comparison of the bulge, disk and Sagittarius dwarf spheroidal galaxy shows a
range of $\sim$0.7 dex in [Al/Fe] at a given [Fe/H]; presumably due to a range 
of Type~II/Type~Ia supernova ratios in these systems.
\end{abstract}



\section{Introduction}

We present the second in a series of papers on the abundances of a sample 
of Galactic bulge K giants.  While our long term aim is to undertake a 
large scale survey of the Galactic bulge, the first step is to determine 
the correct stellar abundance scale.  This is straightforward for most 
stellar populations, but has been a subject of debate for bulge and metal 
rich stars.  Paper I \citep{f06} reports 
our new iron abundance scale that supersedes \citep{mr94}.  
Here we employ our new iron abundance scale to derive the abundances of 
alpha elements in the Galactic bulge stars.  In the third paper, we turn 
to the iron peak, r-, and s-process elements.  In future papers we will 
use the methods developed here to extend our studies to larger samples 
using multiobject spectroscopy.  The initial aim is to settle the iron 
abundance scale and then to follow with well determined abundances for a 
large number of additional elements.  When issues concerning the abundance 
determination of these additional elements have been settled, it will be 
possible to investigate large samples with confidence.
 
The understanding that it is possible for stellar populations to be both 
old and metal rich was gained surprisingly early.  In referring to the 
bulge of M31, \citet{b63} states ``And the process of enrichment has 
taken very little time. ... So the CN giants that contribute most of the 
light in the nuclear region of the Nebula must also be called old stars; 
they are not young''.  The journey from insight to demonstration took 
nearly four decades, however.  Some significant steps along the way 
include the first integrated light spectroscopy of the bulge \citep{w78},
surveys of bulge M giants \citep{bb84} the first survey 
of bulge K giants \citep{r88}, the first detailed abundances in the 
Galactic bulge (MR94), and the first demonstration 
that the bulge is as old as metal rich globular clusters \citep{ort95}.
Recent studies that confirm the old, metal rich nature of the 
bulge population \citep{kr02,z03}, and most 
recently, the revised bulge abundance scale (Paper~I).

The events leading to the development of the new iron abundance scale are 
reviewed in Paper~I and in \citet{r06}. Our new abundance scale is 
based on the use of weak iron lines in Arcturus (HR3905); we can use the very 
well-constrained parameters for Arcturus to perform a differential analysis,
which effectively cancels out systematic problems with $gf$-values and
stellar atmospheres.  The adopted lines are weak enough to remain on the linear
portion of the curve of growth even for our most metal rich Galactic bulge 
giants.  The second advance is the use of multiple approaches to assess 
bulge membership and stellar parameters; this effort overcomes earlier 
problems with self-consistent abundance analysis that did not rely on 
photometry.

A significant thread in the abundance studies has been the discovery that 
alpha elements are enhanced in bulge giants.  This was demonstrated 
convincingly for the case of Mg and Ti (and to some extent Na and Al)
in MR94.  Preliminary studies of 
high resolution spectra obtained at Keck, based on the MR94 iron abundance 
scale and line list, confirmed the Mg enhancement and found all alpha 
elements enhanced \citep{rm00,mr04}.  It is 
widely accepted that at least Mg is enhanced in extragalactic spheroidal 
populations \citep{wfg92}.  The modern effort to revisit 
Whitford's (1978) spectroscopy of the integrated light of a Galactic bulge 
field compared to external galaxies \citep{puzia2002} supports moderate 
alpha enhancement in the bulge and shows that the bulge
falls on the lower end of the elliptical galaxy sequence.

The long standing operating paradigm is that alpha elements (e.g. O, Mg, 
Si, Ca) are produced in the supernovae of short lived massive stars while 
iron is predominantly produced in core collapse SNe that occur on a 
timescale that is 1-2 orders of magnitude longer \citep{t79,w89,t95,mcw97}. 
\citet{m90} combined the SN yields and sophisticated chemical evolution 
models.  \citet{m99} employ the SN yields of \citet{tnh96}
for Type Ia and \citet{ww95}
for Type II SNe. The yields from numerical models of SNe, along 
with corroborating abundance determinations in very metal poor stars, are 
the theoretical underpinning that support this paradigm.  The models of 
\citet{m90}, \citet{m99}, and 
\citet{m06} incorporate star formation predictions and the 
requirement that the bulge form rapidly ($<1$ Gyr).  The alpha enhancement 
has been known since MR94 and is the constraint (along with the 
observations of an old turnoff age) that underpin the empirical arguments 
for a very old Galactic bulge.

The growing evidence that the Galactic bulge formed early and rapidly 
finds support in numerous observations at high redshift.  The discovery of 
the Lyman break selected high redshift galaxy population by \citet{st96}
was the first direct identification of a population of galaxies at 
$z>3$ with high star formation rates and the potential (in terms of mass) 
to evolve into galaxies like the Milky Way.  Since that discovery there 
have been numerous other high redshift populations discovered that are 
equally plausible bulge progenitors; all share the property of having high 
rates of star formation and large stellar mass.  \citet{st96} and 
\citet{adel03} have also detected and studied metal enriched wind 
outflows from Lyman Break galaxies.  \citet{pett02} find winds and 
derive alpha enhancements in a gravitationally lensed Lyman break galaxy 
observed at high 
spectral resolution.  These offer additional evidence of high star 
formation and also indicate that winds should eventually be incorporated 
into the chemical evolution models.  However, the evolutionary path of 
these high redshift galaxies cannot unequivocally link them to a Milky Way 
like result in the present day Universe.

The union of these ideas--that there is overwhelming evidence for a rapid 
formation timescale for the bulge both from the chemical enrichment, 
turnoff age, and high redshift perspectives, is described in \citet{r06}.
If the bulge has undergone secular evolution \citep{kk04} 
then it has done so (it would seem) mostly in the dynamical sense, 
with its population having been built early and rapidly.  However, in the 
case of the bulge, as opposed to giant ellipticals, there are good reasons 
to suspect that the starburst formation scenario \citep{elm99} is 
not the complete picture. \citet{imm04} explore models in which 
the bulge forms from the accumulation of smaller units (as might be 
favored in the LCDM cosmology scenario). It may be possible to detect 
fossil evidence of these subclumps by correlations between abundance and 
kinematics, or differing element trends (e.g. [Mg/Fe] vs [Fe/H] at a 
different level).  A large scale survey program to study these trends will 
require significant samples of stars with abundances determined to very 
high precision.  Contamination by the thick disk and foreground 
disk could mimic the expected signal from a subclump.  Further, it should 
be noted that the bar, a rapid formation timescale, and high metallicity all 
point toward a formation model in which the mass is in place early; 
nonetheless, scenarios of inhomogeneous evolution should be explored with 
larger surveys.

Evidence for the presence of a bar structure in the bulge is now 
indisputable.  Observations by the {\sl COBE} satellite and subsequent 
modeling \citep{dw95} argue for the bar based on surface photometry.  
\citet{mao} find evidence for the bar from the stellar 
population and from stellar dynamics \citep{zhao94}.  A recent study 
\citep{soto06} finds that kinematics consistent 
with the bar set in at [Fe/H]$>-0.2$, with more metal poor stars being 
more consistent with an isotropic velocity ellipsoid.  Bars are also 
thought to be formed by secular evolution \citep{kk04}. 
There has been extensive theoretical discussion about the origin of the 
bar based on dynamical processes that are secular in nature \citep{c01}.
There is prima facie evidence for ongoing star formation in the 
nuclear region and modeling of the luminosity function in fields within 50~pc 
of the nucleus requires a constant star formation rate, with models 
dominated by early bursts of formation are ruled out by models of the
luminosity function \cite{figer04}.  A study of infrared 
abundances in old stars in the bulge by \citet{ram00} finds no 
abundance gradient from latitude 0$^{\rm o}$ to -$6^{\rm o}$, an observation 
expected if 
secular evolution is an important process in establishing the vertical 
thickness of the bar. At present, there is strong evidence from the 
turnoff age and chemical evolution that the bulge formed early and 
rapidly.  On the other hand, the bulge has been shown to have a clear bar 
structure and other properties (no vertical abundance gradient) more 
consistent with secular evolution.

There is at present a conflicted situation with regards to age and 
formation constraints from the turnoff and chemistry, and possible secular 
evolution from the presence of the bar structure. We believe that it is 
vital to not only measure abundances of alpha elements in a larger sample, 
but to also measure the abundances of more species, so as to provide 
better constraints for the chemical evolution models. We have the 
opportunity in the bulge to track the abundance trends of many elements in 
detail.  This is something that we will be unlikely to achieve for any 
other spheroidal population in the near future.  This approach can be used 
to ask more detailed questions, such as constraints on the initial mass 
function as well as setting limits on the timescale of the formation epoch 
to greater precision.  The presence or absence of vertical gradients in 
the alphas and [Fe/H] gives an additional important constraint on whether 
secular processes have played a role in the formation of the bulge.

We also undertake a detailed comparison between the bulge and thick 
disk/halo.  These well studied populations serve both as test samples to 
confirm the soundness of our abundance analysis and as a comparison sample 
against which the bulge abundance patterns can be compared.  This will 
enable us to relate our findings in the bulge to the formation of well 
studied populations in the Milky Way.

The paper is organized as follows.  In Section~2, we discuss the spectral
data and the creation of a new unblended line list for the elements under 
study.  We review the stellar parameters adopted for these stars and
their uncertainties in Section 3.  In Section~4, we describe the 
uncertainties of our 
abundance determinations, including a full quantitative analysis method.
In Section~5, we apply a simple correction for the varying alpha-enhancements
of our target stars.  In Section~6, we check our abundance results, 
showing that our adopted parameters return abundances in ionization equilibrium
and in agreement with previous literature studies.  Our primary bulge abundance
results are presented in Section~7, which includes the evidence for enhanced
alpha element ratios at all metallicities in the bulge.  Further discussion
of the abundances of Na and Al are in Section~8, including evidence that 
the Type II SNe that enriched the bulge included some metal-rich stars. 
In Section~9, we find that the metal-poor stars in the bulge 
show the same Na-O anticorrelation as seen in metal-poor globular clusters
and discuss what that may mean for the early formation of the bulge.

\section{Data and Line List}

The spectra used in this work are the same as those used in Paper~I.  
Briefly, most of the bulge star spectra were obtained with the Keck~I 
telescope using the HIRES echelle spectrograph \citep{vogt} at a resolution
of either 45,000 or 67,000.  Most of the disk giant spectra were obtained
with the echelle spectrograph on the Las Campanas Observatory 2.5-m du Pont
telescope.  The disk giant spectra were of higher signal-to-noise, but of
lower resolution (roughly 30,000).
We use the same continuum values as defined in that work, and we continue to
use the GETJOB program \citep{mcw95a} to measure equivalent width values
and their uncertainties.

An adjustment to our line-measurement methods was required around the Ca 
auto-ionization features at 6319, 6444 and 6362~\AA{} \citep{gr64,mm65}. 
These features are broad, several \AA ngstroms wide, and relatively 
shallow.  For the Mg~I lines near 6319\AA, within one of these features, 
we measured modified equivalent widths using the IRAF\footnote{IRAF is 
distributed by the National Optical Astronomy Observatory, which is operated
by the Association of Universities for Research in Astronomy, Inc., under
cooperative agreement with the National Science Foundation}
{\it splot} package, with the Ca auto-ionization line profile treated as 
continuum.  A full abundance analysis treatment of these lines would 
include the blend of the Ca auto-ionization feature with the Mg~I line 
profiles; this would require modification of the spectrum synthesis code 
MOOG, and a reliable Ca auto-ionization lifetime for the lower level.  
Since these options were not presently available to us, and because the 
Ca auto-ionization feature is quite shallow, we employ the first-order 
approximation for the analysis: namely, to treat the modified equivalent 
widths as normal EWs without the Ca autoionization line.  This 
approximation should work well when both the autoionization line and the 
target, Mg~I, line are unsaturated.  During the abundance analysis stage, 
we have found that lines measured this way yield abundances that agree 
with other, unaffected, lines of the same species.

In order to identify unblended lines useful for abundance analysis we 
employ the method used to find clean Fe lines in Paper~I:  spectrum 
synthesis was performed for all lines between 5000 and 8000~\AA\ using the 
Kurucz atomic line list\footnote{The most recent versions of the Kurucz 
line list can be found at http://kurucz.harvard.edu/.} plus the CN line 
list of MR94  for for Arcturus (HR5340), the 
Sun, and $\mu$ Leo (HR3905).  For every line we calculated the log(RW) value 
and compared it to the log(RW) value for the sum of all lines within 0.2~\AA{} 
of the line center.  Lines were then sorted in order of relative 
contamination to find the most unblended lines for each species.

Inspection of the spectral atlases of Arcturus \citep{hw00} and the Sun 
\citep{k93} and our Paper~I spectrum of $\mu$ Leo were performed to 
eliminate lines blended with unknown features, including telluric 
contamination. Additionally, we cut lines that are too strong ($> 200$~m\AA{} 
in the $\mu$ Leo synthesis, assuming solar ratios) or too weak 
($< 5$~m\AA{} in the Arcturus synthesis, again assuming solar ratios).  
In practice, we rarely used lines approaching these limits in the final 
analysis, but we wished to be inclusive at this point in order to account 
for possible errors in the Kurucz line list \gf-values, non-solar 
abundance ratios, etc.  For the purpose of identifying unblended lines we 
did not include hyperfine splitting for Na or Al lines.

For these initial analyses, we used a Kurucz model atmospheres with 
overshooting.  For each target line, we include the blending effects for 
any weak lines within 0.2~\AA.  The synthesis needed to remove these 
blending effects assumed solar element ratios with the exception of 
assuming [C/Fe] = $-0.2$ and [N/Fe] = +0.4.  These revised C and N 
abundances reflect the effects of dredge up on the red giant branch 
\citep{lr81} and will increase the strength of CN features 
blended with the features of interest. We attempted to find lines where 
the contaminants contributed less than 0.05 dex to the final line log(RW) 
in the synthesis of $\mu$ Leo. 

Several of the Mg~I lines are actually blends of three Mg~I lines at 
nearly identical (within 0.01~\AA) wavelengths.  The weaker lines usually
have \gf-values at least a factor of 10 weaker than the strongest line, but
we retain the weaker lines in the deblending routine.  

Other lines (such as the Na~I lines at 5682~\AA{} and 5688~\AA{}) have very
close blends, but we are forced to use these lines because other options
are not available (the 6154~\AA{} and 6160~\AA{} Na~I lines were not in
the wavelength coverage of every star).  In such cases, we use the ``blends''
package of MOOG \citep{moog} to deblend the full equivalent width of the 
whole blend.
The line parameters of the other features in the blend were taken from
the Kurucz line list.  For the 27 Baade's Window stars where at least one 
line of both doublets were used we found a mean difference of $-0.01 \pm 0.09$ 
(standard deviation) where the value given by the redder doublet was 
slightly larger. For the 6300~\AA{} [O~I] line we included the 
Ni~I blend \citep{ap01}, even though its effect is negligible in giants.

As in Paper~I, we will use Arcturus as the reference star for our 
differential analysis.  The advantage of a line-by-line
differential analysis is that systematic errors, due to various 
inadequacies of the model atmospheres, such as non-plane parallel 
geometry, 3D hydrodynamical variations, chromospheric effects, non-LTE, 
magnetic fields etc, may to a first approximation, cancel out when 
performed relative to similar stars; certainly the oscillator strengths of 
the atomic lines do not enter into the abundance determinations.  

To place our abundances on an absolute scale, we must
adopt \leps{} values for Arcturus for every element.  
Our method for determining these 
abundance values is similar to the method we used in Paper~I to determine the
\leps(Fe) abundance.
Errors in these determinations will create a systematic
zero-point offset for each element.  

We start with a line list as constructed above, but we change our 
selection criteria to only exclude lines that show any blending in both 
the Sun and Arcturus.  This adds a few lines that are blended or too 
strong in $\mu$ Leo, but we exclude many lines that have minor blends that 
were allowed into the main list.  The equivalent width (EW) values for 
these lines were measured in both the \citet{hw00} Arcturus atlas 
and the \citet{kfb84} solar atlas.

The measured line lists were then run through the MOOG stellar abundance
program using a model atmosphere appropriate for Arcturus:  a Fiorella
Castelli\footnote{http://wwwuser.oat.ts.astro.it/castelli/} 
alpha-enhanced AODFNEW atmosphere with \teff{} = 4290 K,
\logg{} = 1.55, [m/H] = $-$0.50, and \vt{} = 1.67 km/s. For the Sun we used a
Castelli solar-ratio ODFNEW atmosphere with
\teff{} = 5770K, \logg{} = 4.44, [m/H] = 0.00, and \vt{} = 0.93 km/s. 
A line-by-line differential analysis was conducted which yields the relative
abundance of Arcturus with respect to the Sun.  The Arcturus abundances
were then placed on an absolute scale by the use of \citet{lodd03} solar
abundance ratios, with the exception that we continue to use the 
solar \leps(Fe) value of 7.45 we derive in Paper~I.  The EW values of the lines
used in this analysis are given in Table~1 and the final adopted Arcturus
abundances are given in Table~2.  The final line list is given in Table~3.
Our EW measurements are given in Table~4.

\section{Stellar Parameters and Parameter Errors}

Paper~I gives our approach for deriving the stellar parameters for our
program stars.  In Paper~I, we mainly used Kurucz atmospheres with overshooting
enabled, but in this paper we will use the Castelli model atmospheres in 
order to create a simple correction for the variable [$\alpha$/Fe] ratios
found in our stars (see Section~5 for the description of the correction).

The parameter-finding methods using the alternate atmosphere models are 
exactly the same as performed in Paper~I.  The
choice of atmosphere grid has a small effect on the final adopted parameters
(see Table~9 of Paper~I).  We list the final adopted parameters for each
grid for each star in Table~5.  

In this paper, we will derive uncertainties for our abundance measurements
using a method based on the work of \citet{mcw95b}.  One component of the
error analysis is deriving the uncertainties in the stellar parameters.
The listed ($\sigma$T) value is the standard
deviation of the three (or two for a few stars mentioned in Section~7 of 
Paper~I) \teff{} values derived by the three different \teff-setting methods.  
The $\sigma$[m/H] value comes from the standard deviation of the Fe~I lines 
used to measure the [Fe/H] abundance.  It would be more correct to use the
derived value of $\sigma$[Fe~I/H] for $\sigma$[m/H] (and therefore make
this an iterative process), but we found from experience that using the
less exact standard deviation value makes minimal difference.  The standard
deviation of the Fe~I lines ends up being slightly larger than the weighted 
error for these lines.  The $\sigma$\vt{} value comes from
Equation 2 from Paper~I.

The value of $\sigma$\logg{} required a Monte Carlo simulation.  The derived
\logg{} values come from Equation~1 in Paper~I, where \logg{} is a function
of stellar mass, \teff, and M$_{\rm bol}$.  The error in M$_{\rm bol}$ depends
on photometry, reddening, distance and bolometric correction uncertainties.  
We use the bolometric corrections of \citet{a99}, which give BC(V) as a 
function of \teff{} and [m/H].  While it would be possible to use normal
analytical propagation of error techniques to calculate $\sigma$\logg,
we decided to use a Monte Carlo simulation to assure there were no 
hidden dependencies.

To simplify the analysis, we adopted standard values for the uncertainties
of several of the inputs.  For the bulge sample, we assume:  
$\sigma$M$ = 0.1$~\msun, $\sigma$V$ = 0.05$~mag, and 
$\sigma$A$_{\rm V} = 0.10$~mag.  We assume a distance error of about 500~pc,
which converts to a distance modulus uncertainty of 0.13~mag.  We include
the stated fitting uncertainty of 0.024~mag to the BC(V) function of
\citet{a99}.  For the disk giant sample, we set the distance and A$_{\rm V}$ 
errors to zero.  Therefore, the main variation in the value of $\sigma$\logg{}
between stars is due the the differences in \teff{}, $\sigma$T, [m/H] and 
$\sigma$[m/H].  

The values of $\sigma$T for individual stars was calculated to be the 
standard deviation of the values derived by the independent methods
from Paper~I.  However, we use the standard deviation of the distribution
of (\teff$_{,i} - $\teff$_{\rm ,Final}$) for all the stars in order to calculate
what the ensemble $\sigma$T value should be.  For the Kurucz atmospheres,
this value is 49~K, while it is 38~K and 41~K for the ODFNEW and AODFNEW
atmospheres, respectively.  For stars where the measured $\sigma$T  
is less than this value, we used the ensemble value.  If the $\sigma$T 
is larger than the ensemble value, we kept the larger value to
account for potential cases where exceptional mistakes in the input values
may be increasing the \teff{} error.  The value in Table~5 is the value
used in the error analysis.

\section{Abundance Error Analysis}

The error analysis of the abundance determinations in this series of papers
is based on the work of \citet{mcw95b}.  In our work, however, we will
include the terms for the uncertainty on [m/H].  This increases
the number of terms in Equations~A5 and~A16 of \citet{mcw95b}:  

\begin{eqnarray}
\lefteqn{\sigma^{2}_{\loge} = \sigma^{2}_{\rm EW} + 
\left({\partial \loge\over\partial {\rm T}}\right)^2 \sigma^2_{\rm T} 
+ 
\left({\partial \loge\over\partial \mlogg}\right)^2 \sigma^2_{\rm G} 
+
\left({\partial \loge\over\partial \mfeh}\right)^2 \sigma^2_{\rm M}  
+ 
\left({\partial \loge\over\partial \xi}\right)^2 \sigma^2_{\rm V} 
} \nonumber \\
 & & \mbox{} +  
\left({\partial \loge\over\partial {\rm T}}\right)
\left({\partial \loge\over\partial \mlogg}\right)
\left(\sigma_{\rm TG} + \sigma_{\rm GT}\right)
+
\left({\partial \loge\over\partial {\rm T}}\right)
\left({\partial \loge\over\partial \mfeh}\right)
\left(\sigma_{\rm TM} + \sigma_{\rm MT}\right)
\nonumber \\
 & & \mbox{} +
\left({\partial \loge\over\partial {\rm T}}\right)
\left({\partial \loge\over\partial \xi}\right)
\left(\sigma_{\rm TV} + \sigma_{\rm VT}\right)
+  
\left({\partial \loge\over\partial \mlogg}\right)
\left({\partial \loge\over\partial {\mfeh}}\right)
\left(\sigma_{\rm GM} + \sigma_{\rm MG}\right)
\nonumber \\
 & & \mbox{} +  
\left({\partial \loge\over\partial \mlogg}\right)
\left({\partial \loge\over\partial {\xi}}\right)
\left(\sigma_{\rm GV} + \sigma_{\rm VG}\right)
+  
\left({\partial \loge\over\partial \mfeh}\right)
\left({\partial \loge\over\partial {\xi}}\right)
\left(\sigma_{\rm MV} + \sigma_{\rm VM}\right)
\end{eqnarray}
For the sake of brevity, in the subscripts we use the T for \teff, G for
\logg, M for the atmospheric [m/H] value, and V for the microturbulence 
value v$_t$.

We treat
each line independently (including the Fe~I and Fe~II lines from Paper~I)
and then combine the results using weighted means to obtain the final
\leps(X) and [X/Fe] ratios.  Our analysis
requires two major objectives:  calculating the value of the partial
derivatives of the abundance with respect to the various parameters and 
the value of the covariance factors.

The calculation of the partial derivative values initially seems
straight-forward--just change each parameter individually and calculate
the effect on the abundance derived from each line.  However, some of our
stellar parameters lie at the limits of our stellar atmosphere grid, so some
changes (such as decreasing the \logg{} value or increasing the [m/H] 
value) may move the test atmosphere off the grid for some stars.
Therefore, we only change the parameters in a direction of increasing
stability (increasing \teff, \logg, \vt, and W$_{\lambda}$; decreasing 
[m/H]), but use the method of second differences to increase the 
accuracy of the value of the derivative at the initial point.  The step
size we use for each difference is 100~K for \teff, 0.3~dex for \logg{}
and [m/H], 0.3~km/s for \vt, and 5~m\AA{} for W$_{\lambda}$.

We use Monte Carlo experiments to determine the covariance terms.  The
covariant terms are defined by:
\begin{equation}
\sigma_{\rm AB}=\frac{1}{N}\sum_{i=1}^N \left(A_i - \overline{A}\right)
\left(B_{\it i} - \overline{B}\right).
\end{equation}
where A and B are different parameters (\teff, \logg, etc.).  For example,
$\sigma_{\rm GM}$ would be determined by using $\sigma$\logg{} to
randomly pick a new value of \logg{} and then re-run the analysis to determine
what effect this change had on the final [m/H] value.  
In this study, A is the 
independent parameter and $\sigma_{\rm AB}$ describes the dependence of
parameter B on uncertainties in parameter A.  

Note that in comparison to previous formulation, we do not assume that
the covariance coefficients commute.
The non-commutative property of the covariance coefficients can be demonstrated
by an example:  Photometric \teff{} values are to first order independent 
of the adopted microturbulence value.  Therefore in this case, 
$\sigma_{\rm TV} = 0$, but the reverse is not true:  we know the weakest Fe~I
lines often come from high-excitation lines and the strongest lines from 
low excitation lines.  That means a change in the \teff{} value can
affect the slope of the abundance versus line strength plot used to set the
\vt{} values.  

Rather than running
the analysis many times to build up the Monte Carlo sample, we found that
we could speed up the process by determining a functional form of the
change in one parameter as determined by a change in the other parameter.
For example, for the case of $\sigma_{\rm GM}$, we would run the
analysis to determine the new [m/H] for a series of steps in \logg{} around
the final value.  In the case of \logg{}, we used steps of 0.10~dex to 
$\pm 0.50$~dex around the adopted \logg{} value.  We used similar step
sizes and ranges (going out many to many times the measured uncertainties of
each parameter) for \teff, [m/H] and \vt.  

After all the steps were completed, we fit a quadratic function to the
points (excluding cases where the steps may have moved off the atmospheric
grid) and then used that fit as a look-up function in the Monte Carlo.  
We could then run a very large number of test cases in a reasonable
amount of time (the calculation of all the covariance terms took about
30~minutes per star on a modern workstation).  

\begin{figure}[h]
\plotone{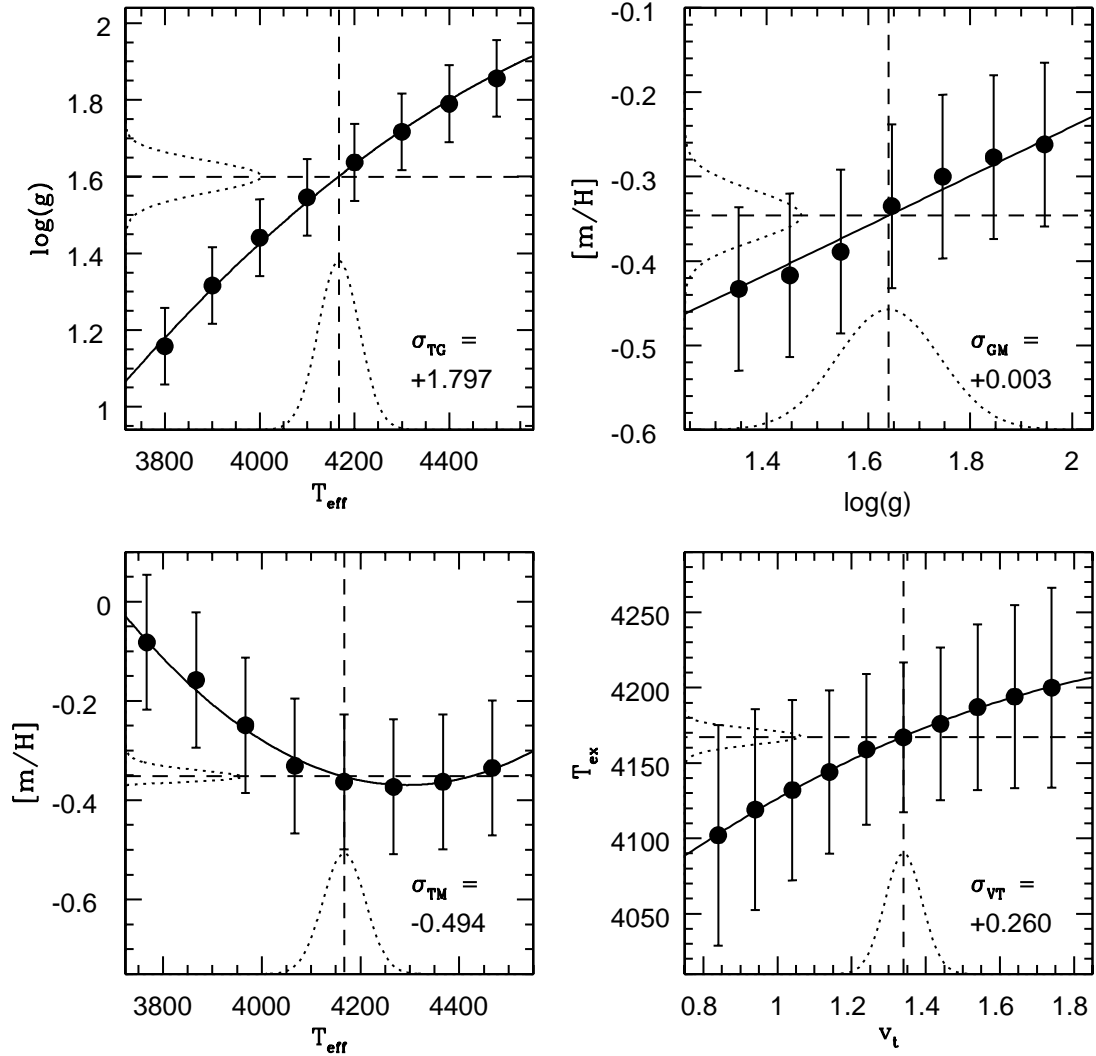}
\caption{Four example plots that help describe our method for determining the
covariance coefficients between the stellar parameters.  In each panel, one 
parameter was stepped through several values and the response by the other 
parameter was measured after reanalyzing the star.  In this case, the
star tested was I-194.  In each panel, the solid points are 
test cases, and the error bars are the derived error in the dependent
parameter.  The solid line is a quadratic fit to the points.  The dotted
line along the x-axis is a Gaussian with a full-width
half-maximum consistent with the parameter uncertainty.  We then ran 
a Monte-Carlo 
simulation using the probability distribution function, the quadratic fit and 
Equation~2 to calculate the covariance coefficient.}\label{fig-jff1-covar}
\end{figure}

Graphical examples of this fitting methods for some of the covariant terms 
are given in Figure~\ref{fig-jff1-covar}.  In each panel, the independent 
parameter is along
the x-axis.  The solid points give the results for some of the steps (the 
full range is not given in order to show detail), and the line is the
quadratic fit.  The error bars on each point show the error estimate for
the final value of the dependent parameter--for some parameters like \vt,
this uncertainty is often much larger than the variation in the parameter.  
The Gaussian displayed in the panel is based on the uncertainty of the
independent parameter used in the calculation and is an indication of 
the probability distribution used in the final Monte Carlo simulation.

For a few cases, we find that our error estimates are larger than the 
star-to-star scatter seen within a population.  This suggests that we have
over-estimated the errors.  For most lines, the dominant term in Equation~1
is the $\sigma^{2}_{\rm EW}$ term.  This is a consequence of having 
well-constrained stellar parameters that is a result of having a large
number of weak Fe lines available.  We believe that our estimates of our
equivalent width errors may be highly correlated due to our 
continuum-determination method.  The EW errors contain a component due to
the photon statistics within the line itself and a component due to the
measurement error in the continuum.  The latter component is linked between
lines--if the continuum is set too high for one line, it is likely too high
for many other lines as well.  We cannot easily correct for this covariance.
If systematic problems with the continuum placement exist, then the 
star-to-star scatter may be smaller than the values calculated here because 
most stars will be affected in a similar way.  An alternative explanation
is that we have have overestimated the stellar parameter errors used in
the calculation, meaning the
results from Equation~1 may overestimate the abundance errors.

The final abundances and error estimates for our sample for both grids
are given in Table~6.  The procedure we use to obtain Our ``Final'' 
abundances are described in the next section.

\section{Atmospheric Alpha-enhancement Corrections}

The true compositions of most of the stars in our sample are not matched by the
assumed compositions of either the ODFNEW or AODFNEW grids.  The ODFNEW grid
assume solar abundance ratios, while the AODFNEW grid assumes that all of the
alpha elements (O, Ne, Mg, Si, S, Ar, Ca and Ti) are enhanced over the solar
[$\alpha$/Fe] ratios by $+0.4$ dex.  Only a few 
stars of the disk sample with about solar composition are matched well by 
ODFNEW atmospheres, while only the most metal-poor stars in both the disk and 
Baade's Window sample have all of the alpha elements enhanced by about the
same amount as the AODFNEW grid. The choice of
which grid is applied does affect the final abundances greatly, as can be seen
by examining Table~6, as well as Table~7 of Paper~I.

Our correction method used a linear interpolation between the results of the 
two grids:

\begin{equation}
{\rm [X/Fe]}_{\rm Final} = \frac{({\rm [X/Fe]}_{+0.4} - {\rm
[X/Fe]}_{0.0})}{+0.4}{\rm [Mg/Fe]}_{\rm Final} + {\rm [X/Fe]}_{0.0}
\end{equation}

Where [X/Fe]$_{0.0}$ and [X/Fe]$_{+0.4}$ are the results for the ODFNEW and
AODFNEW for a given ratio (the same method was used on the individual \leps{} 
and [Fe/H] values as well) and [Mg/Fe] is final ratio for that star. 
The final corrected abundance values are given in Table~6. 

This equation assumes that [Mg/Fe] is an appropriate surrogate for the 
[$\alpha$/Fe] ratio, and is better than using some mean of several 
alpha-element ratios (for example, [(Mg + Si + Ca + Ti)/Fe]).  Our reasoning is 
that the continuous opacity of K-giants is dominated by H$^-$. Mg is the 
largest electron donor in the atmospheres of solar composition K giants, 
followed by Fe, then Si, Al, Ca, and Na; in the deeper, hotter, atmosphere 
layers the contribution from Si exceeds Fe.  In total these alpha-element 
and alpha-like elements dominate over the electrons contributed from Fe, 
particularly for alpha-enhanced atmospheres.  Given the relative electron 
contribution it would make some sense to use an average of [Mg/Fe] and 
[Si/Fe] abundance ratios for the hotter giants; but we found that the 
abundance corrections do not change very much if this index was used 
instead of [Mg/Fe].  Similarly, using [(Mg + Si + Ca + Ti)/Fe] as the index
makes only a small difference.  The largest abundance corrections are for 
Fe~II, Si~I and Ti~I (up to $\sim 0.1$ dex in both cases).

One problem with this method is that inaccurate [Mg/Fe] ratios can cause
problems.  We find subsolar [Mg/Fe] ratios for some of the metal-rich disk
giants.  These values do not agree with what is observed in disk dwarfs
(see Section~6.2).  We believe that this is due to specific difficulties 
presented by the du Pont spectra.  While the signal-to-noise ratio for these
stars is very high, the resolution is relatively low, which increases problems
with line contamination and continuum determination.  In addition, the 
the Mg~I lines near 6319\AA{} lie on or near a set of bad columns on the 
du Pont echelle CCD, making most of these lines unusable.  The remaining
lines available in the metal-rich stars are either the strong 5711\AA{} line
or weaker lines in the far red.  The CCD for the du Pont echelle suffers from
strong fringing in the far red.  Therefore, to counter this unfortunate set
of difficulties, we adopt the ODFNEW abundance values for those stars for
which [Mg/Fe] $< 0$ as the final abundances.

In this paper we find that the bulge [Mg/Fe] ratio stays high at all 
[Fe/H] values, but the other [$\alpha$/Fe] ratios drop with increasing 
metallicity.  This means that [Mg/Fe] is not a good indicator of the mean 
molecular weight of the atmosphere, which is heavily influenced by the 
[O/Fe] ratio.  However, short of creating custom atmospheres from ODF 
functions specifically designed for the unique mix of elements found in 
bulge stars, all the atmospheres we could calculate would not be correct 
for the stars.  Our choice is a reasonable compromise, and we are 
fortunate that the magnitude of the corrections is small.

\section{The Element Ratios and Trends with [Fe/H]}

\subsection{Ionization Equilibrium of Ti and Fe}

\begin{figure} 
\plotone{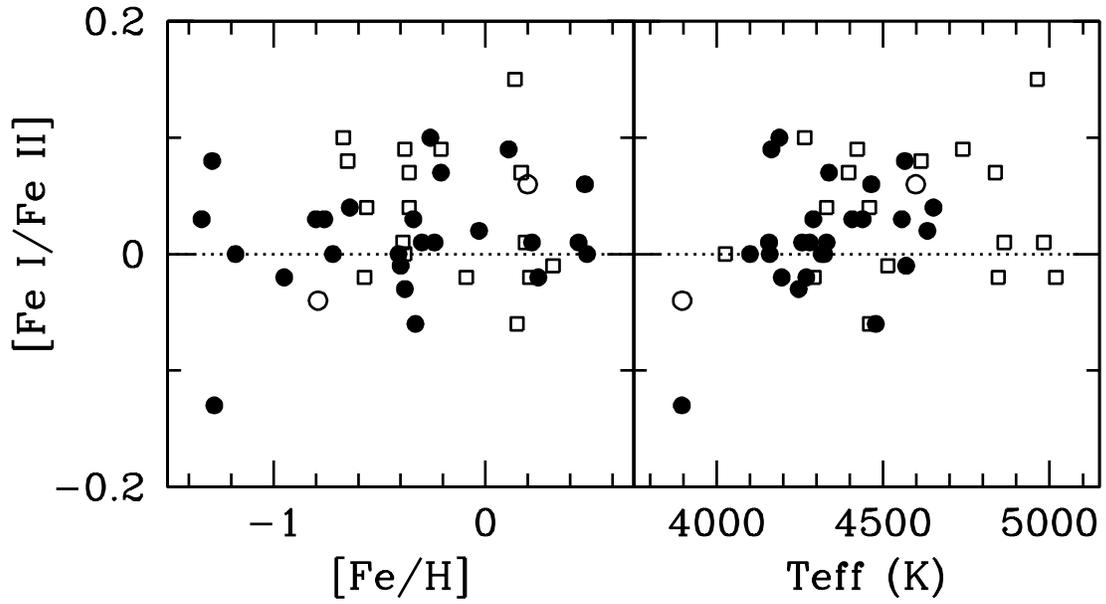} 
\caption{{\bf Left:} A plot of [Fe~I/Fe~II] versus [Fe/H] for our bulge
giants (filled circles), Baade's Window disk giants (open circles), and solar 
neighborhood disk stars (open squares).  \quad {\bf Right:} The [Fe~I/Fe~II] 
ratio versus T$_{\rm eff}$.
The mean [Fe~I/Fe~II] ratio of $\sim$$+$0.01 $\pm$0.05 dex shows no trend
with metallicity or temperature, and indicates that any non-LTE overionization 
of iron must be
constant across 1000K in temperature and 2 dex in metallicity. }
\label{fig-fe1fe2} 
\end{figure}

In Figure~\ref{fig-fe1fe2} we show the [Fe~I/Fe~II] abundance ratios in 
our bulge and disk stars; the figures indicate that ionization equilibrium 
for iron is properly determined, with the mean ratio at 0.01 $\pm$0.05 dex 
for our bulge stars, and 0.04 $\pm$0.05 dex for our disk star sample.

For our sample of bulge and disk stars there is no clear evidence of 
systematic trends in the [Fe~I/Fe~II] ratio with [Fe/H] or T$_{\rm eff}$.  
This indicates that any non-LTE overionization of iron, if it exists, must 
be constant across the temperature and metallicity range covered by our 
sample, and equal to the non-LTE overionization present in Arcturus.  
Qualitatively, in red giant stars one expects non-LTE overionization to 
increase with the transparency of the atmosphere and the amount of UV 
flux; thus increasing overionization is expected for decreasing 
metallicity and increasing temperature.

The absence of trends in our ionization plots leads to the conclusion that 
iron overionization changes by less than 0.05 dex over a range from $-$1.5 
to $+$0.5 dex in metallicity, and 4000 to 5000K in temperature. This upper 
limit on the range is similar to the predicted total iron overionization 
for Arcturus, as computed by \citet{s85}.

\begin{figure} 
\plotone{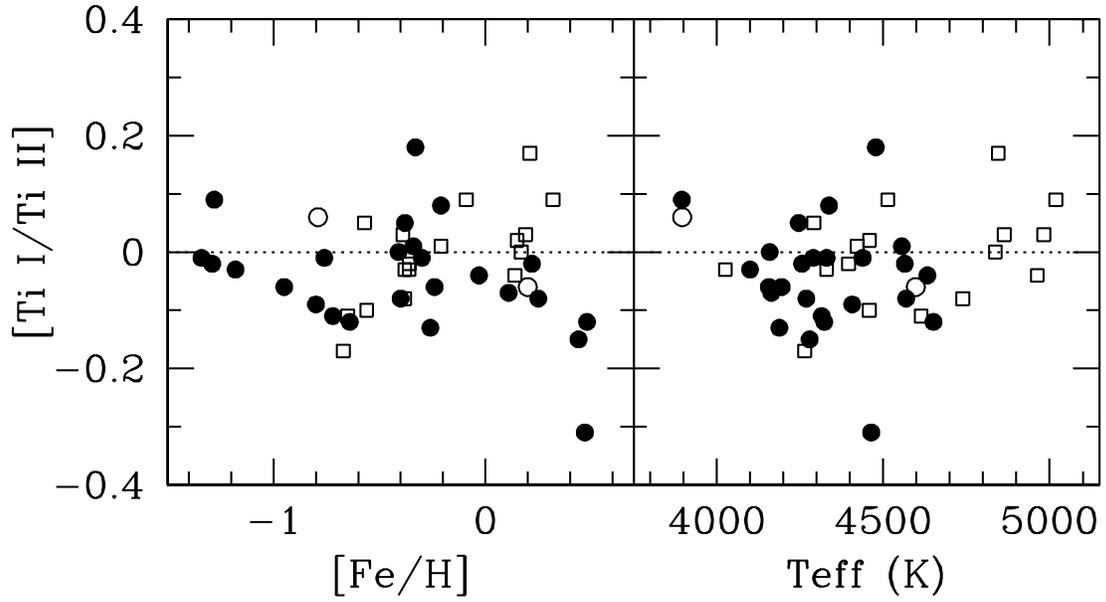} 
\caption{{\bf Left:} A plot of [Ti~I/Ti~II] versus [Fe/H] for our bulge
giants (filled circles), Baade's Window disk giants (open circles), and solar 
neighborhood disk stars (open squares).  \quad {\bf Right:} The [Ti~I/Ti~II] 
ratio versus T$_{\rm eff}$. 
Ionization equilibrium is evident for the sample, with no evidence of
systematic trends or shifts more than 0.05 dex, over 1000K in temperature 
and 2 dex in metallicity.}
\label{fig-ti1ti2} 
\end{figure}

Figure \ref{fig-ti1ti2} shows the [Ti~I/Ti~II] abundance ratios in our 
bulge and disk stars.  For our bulge star sample we, again, find no 
convincing trend in the neutral to ion ratio over a 2 dex range in 
metallicity and 1000K in temperature; this despite the lower 
ionization potential of Ti relative to Fe, which favors
overionization.  There may be a small positive slope in 
the [Ti~I/Ti~II] versus [Fe/H] plot for our disk giants, perhaps resulting
from an increase in non-LTE overionization with decreasing metallicity.  
Nonetheless, theis apparent trend is sensitive to one or two points; thus,
our evidence for a change in Ti ionization in the disk stars is marginal.

Numerous previous abundance studies \citep{pro00,lb85}
found Ti~II more abundant than Ti~I by $\sim$0.1--0.2 dex; this 
might be due to systematic problems with the $gf$ scales between ionized 
and neutral lines, or, possibly, due to non-LTE over-ionization.  Although 
non-LTE effects have long been a concern in red giant stars \citep{rul80},
our differential technique allows us to measure the same Ti 
abundance values in red giants from both neutral and ionized species, at 
least because the non-LTE overionization effect is approximately the same 
in all our bulge red giant stars.

\subsection{Literature Samples and Our Disk Sample}

\begin{figure}[h]
\plotone{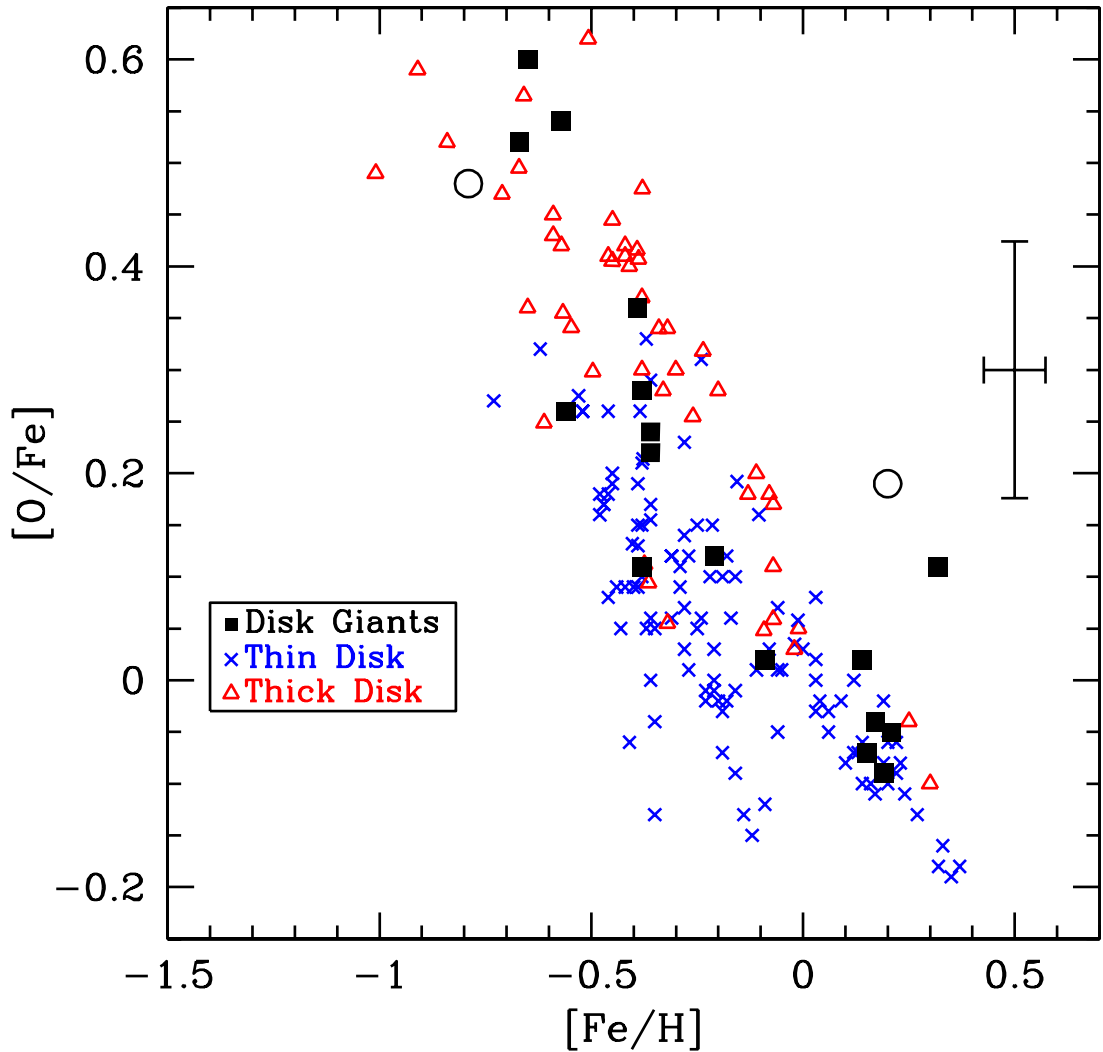}
\caption{A comparison of our [O/Fe] abundances for our local
disk giant sample (filled squares) against previous observations of nearby 
stars.  Most of the stars from the literature samples are FGK-type dwarfs  
(blue crosses: Reddy et al. 2003; Bensby et al. 2005; Brewer \& Carney 2006), 
and thick disk stars (red triangles: Prochaska et al. 2000; Bensby et al. 2005; 
Brewer \& Carney 2006; black triangles: Fulbright 2000).  Overall, our disk
oxygen abundances are in good agreement with that of other samples.  The one
exception is $\mu$~Leo, the most metal-rich disk giant in our sample.  As
discussed in the text, other researchers have found the abundances for 
several light elements in this star have been shown to be enhanced with 
respect to the rest of the disk.  The two open circles
are the two non-bulge giants in the Baade's Window sample, included for
completeness.  The more metal-rich 
star of the two, IV-025, is about halfway between the Sun and the bulge 
and is probably a disk population star. 
The other star Baade's Window star, II-122 lies about 7~kpc beyond the bulge
and about 1 kpc from the plane of the disk.  It is likely either a thick disk
or halo giant (see Section 7.1 of Paper~1).} \label{fig-ofe1dg}
\end{figure}

\begin{figure}[h]
\plotone{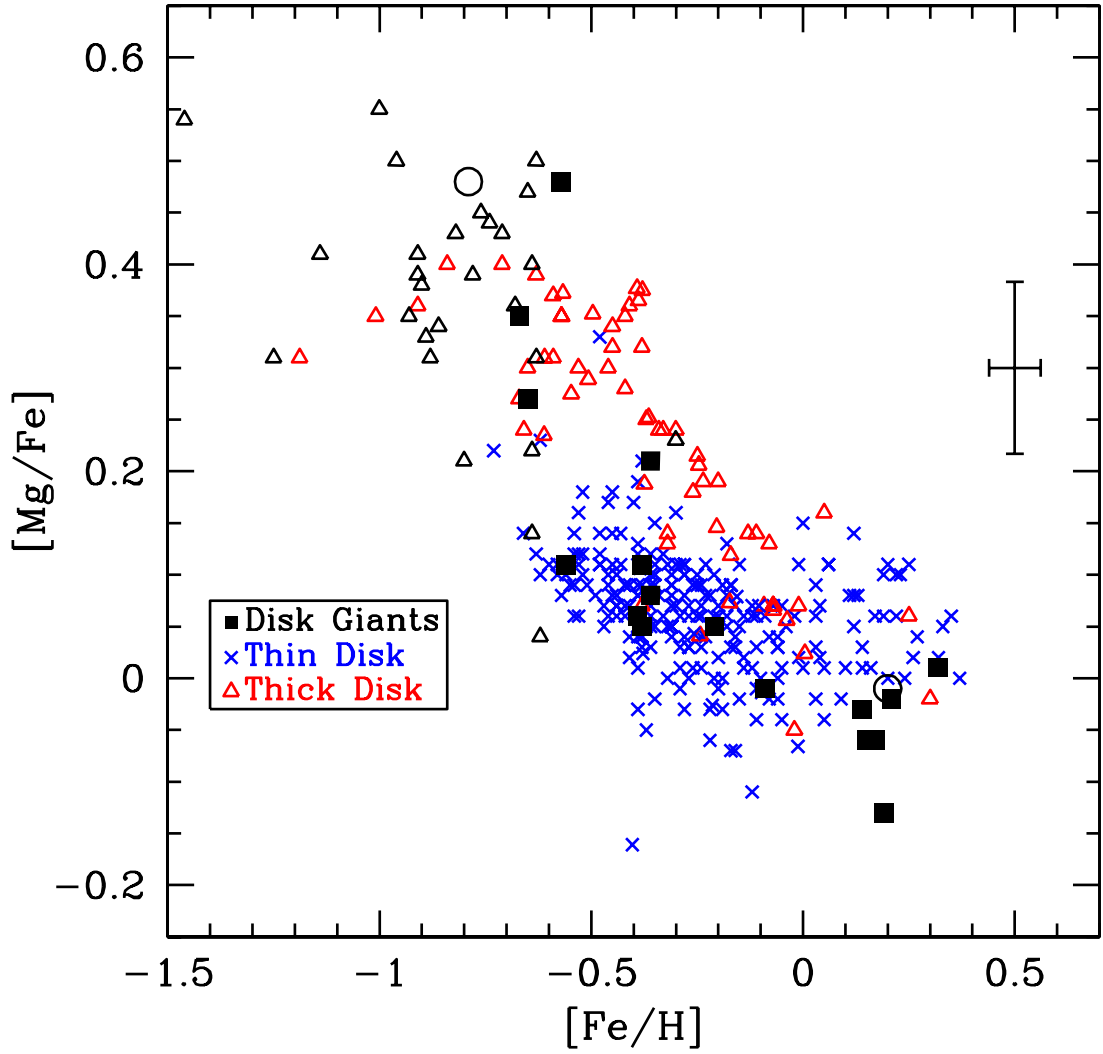}
\caption{Similar to Figure~\ref{fig-ofe1dg}, but for [Mg/Fe].  A few of the most
metal-rich disk giants show lower [Mg/Fe] ratios as compared to the 
comparison sample.  It is unknown whether this is indicative of the true
nature of the stars or the result of some flaw in our analysis.  One     
of the metal-rich disk giants with low [Mg/Fe] is HR4450, which is most 
likely a binary.}
\label{fig-mgfe1dg}
\end{figure}

\begin{figure}[h]
\plotone{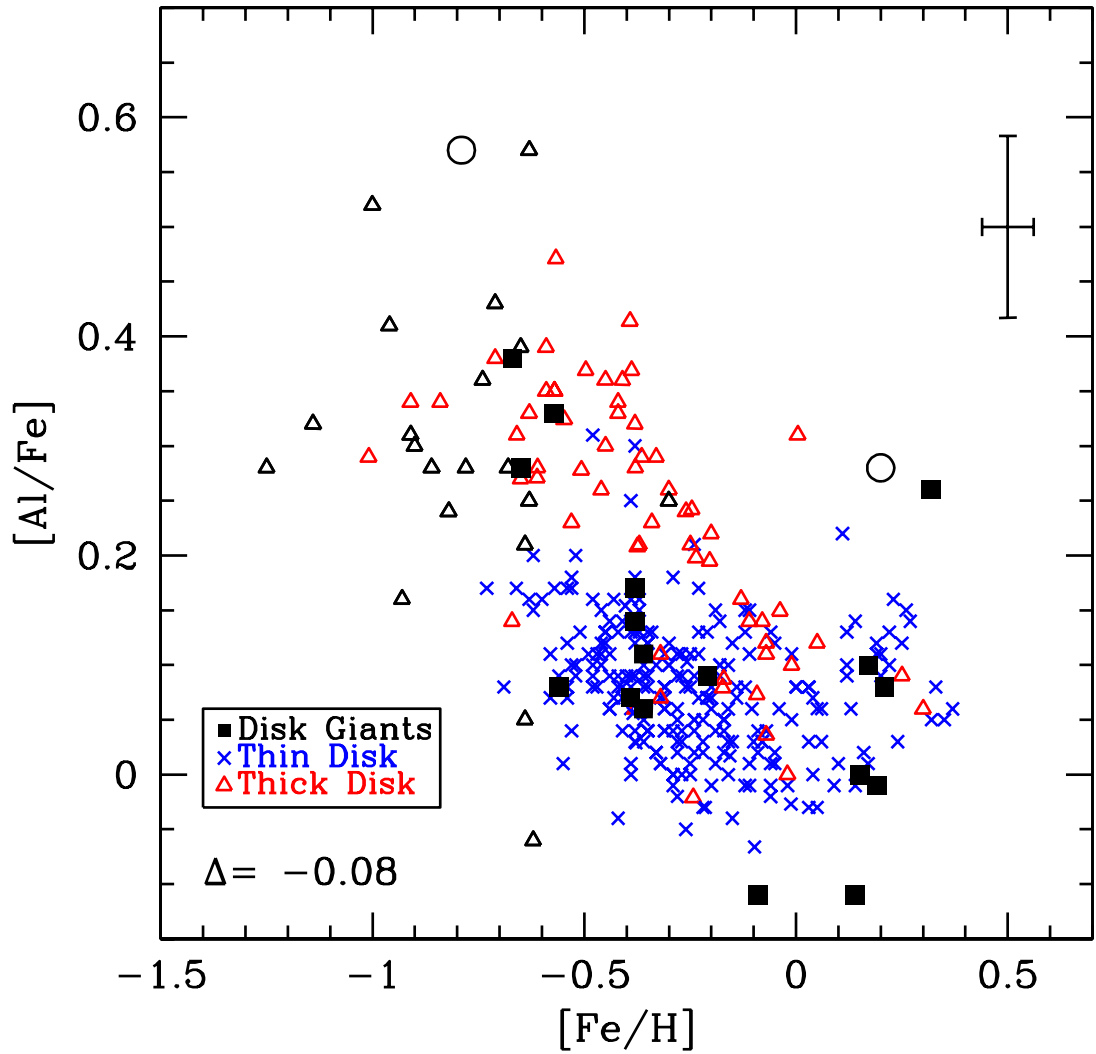}
\caption{Similar to Figure~\ref{fig-ofe1dg}, but for [Al/Fe].  Our data points
have been shifted down by 0.08 dex to adjust for possible zero-point offsets.
The shift brings our metal-poor disk giants into better agreement, but it
does worsen the fit for two metal-rich stars. }
\label{fig-alfe1dg}
\end{figure}

\begin{figure}[h]
\plotone{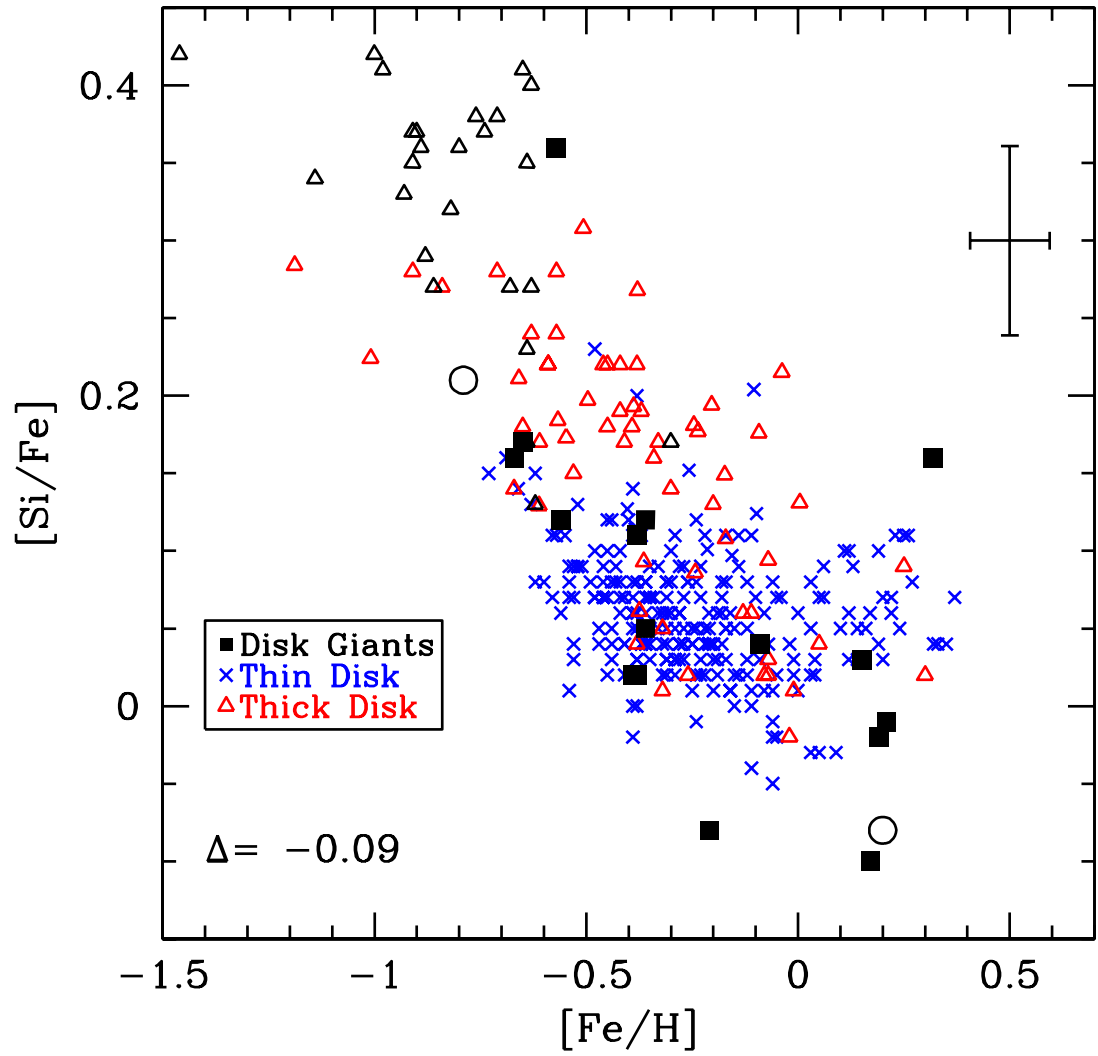}
\caption{Similar to Figure~\ref{fig-ofe1dg}, but for [Si/Fe].  The metal-rich
disk giants show a wide range of values.  Like [Al/Fe], the fit to the
metal-poor disk giants indicated the the need for a $-0.09$~dex shift in
our zero-point to better match the literature data.}
\label{fig-sife1dg}
\end{figure}

\begin{figure}[h]
\plotone{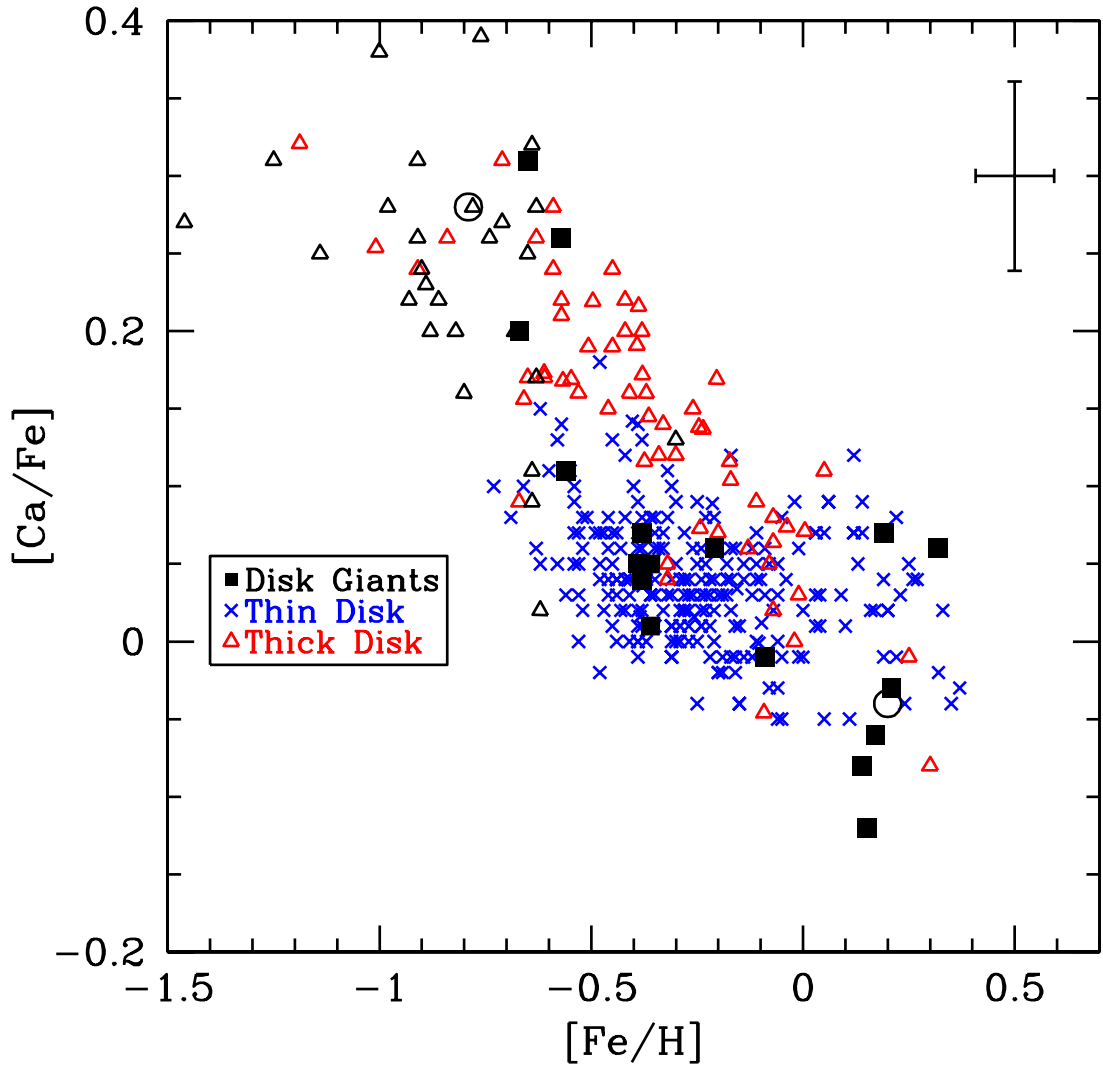}
\caption{Similar to Figure~\ref{fig-ofe1dg}, but for [Ca/Fe].  A few of the metal-rich
disk giants lie lower than similar-metallicity dwarf stars, although one of the
abnormally-low disk giants is the likely binary HR4450.  The metal-poor dwarfs
and giants are in good agreement. }
\label{fig-cafe1dg}
\end{figure}

\begin{figure}[h] 
\plotone{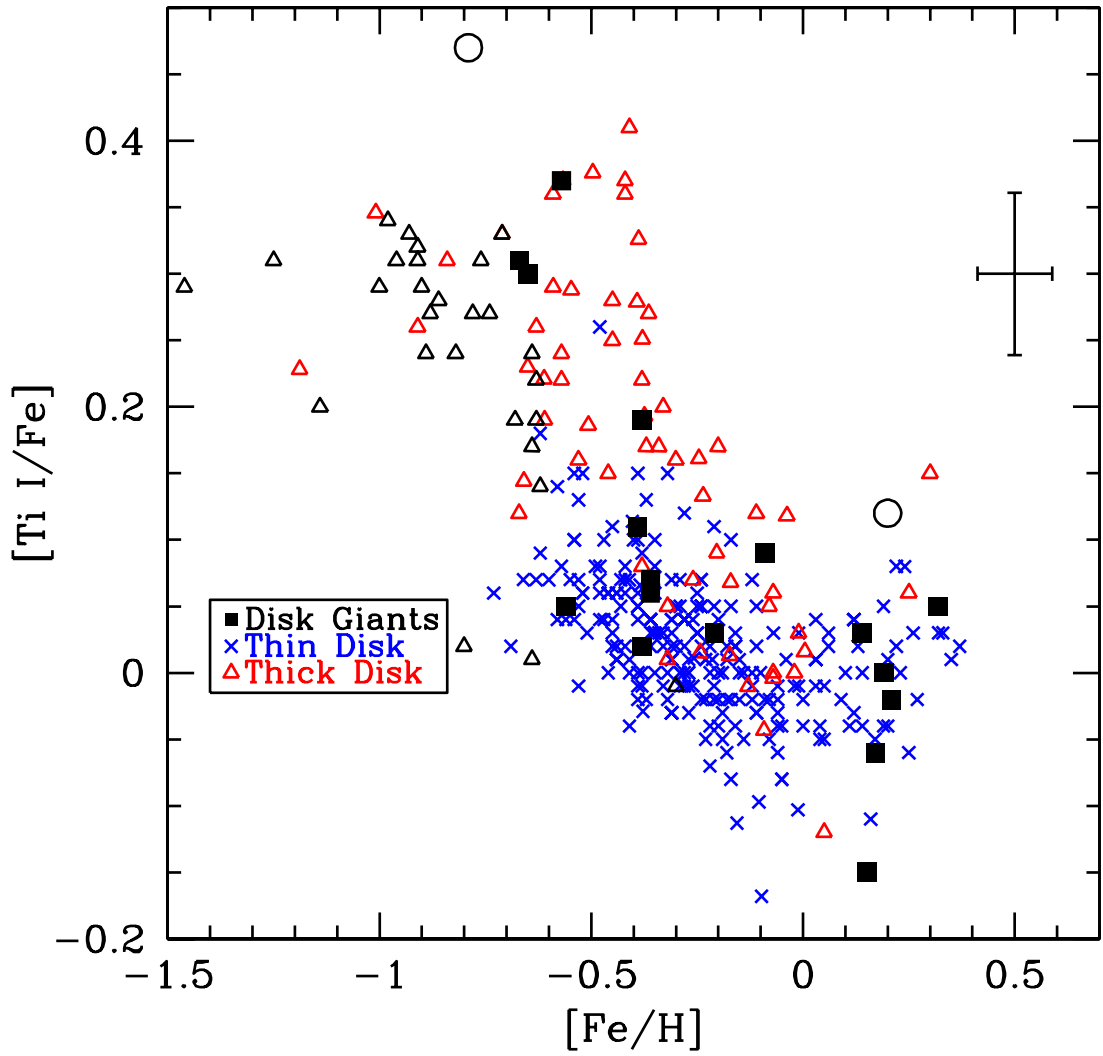}
\caption{Similar to Figure~\ref{fig-ofe1dg}, but for [Ti/Fe].  Only the Ti~I results
are used for the disk giants.  Again, the agreement between the dwarfs and
giants is good for most of the stars.}
\label{fig-tife1dg}
\end{figure}

Figures~\ref{fig-ofe1dg} through \ref{fig-tife1dg} 
display the [X/Fe] versus [Fe/H] ratios for 
the elements studied here for the local disk stars in our sample, compared 
to abundance results from various surveys of the thick and thin disk 
populations \citep{f00,r03,b05,bc06,pro00}.
The literature data have been 
included directly from the source papers, so systematic errors are likely 
to exist between the samples.  

The purpose of Figures~\ref{fig-ofe1dg} through \ref{fig-tife1dg} 
are to show that the abundance ratios we 
derive for the disk sample are very similar to what has been observed in 
earlier works which mainly analyzed dwarf and subgiant stars.  The good 
agreement between our disk giant results and the previous results is 
strong evidence that our analysis methods are sound and any differences, 
greater than $\sim$0.10 dex, between our bulge and disk samples are real 
and are not an artifact of our analysis methods.

Inspection of the composition of our disk giant sample, compared to the 
literature abundances in Figures~\ref{fig-ofe1dg} through \ref{fig-tife1dg},
suggests that we 
have both thin and thick disk giants present. To check this we computed 
the spatial velocities of our disk star sample, based on the SIMBAD 
database, and found that only HR2035 and HR5340 are certain kinematic 
thick disk members, following the definition used by \citet{b03};
while HR4382 and HR1184 lie in the kinematic gray 
zone between thin and thick disks.  Three of our four disk giants below 
[Fe/H]=$-$0.5 are thick disk, or possible thick disk members; curiously, the 
one star in that group that is not kinematically thick disk, HR2113, strongly 
resembles the thick disk chemical composition.  The presence of thick disk 
stars in our disk sample below [Fe/H]$\sim$$-$0.5 is not surprising, due 
to the paucity of thin disk stars at this metallicity, and the peak of the 
thick disk metallicity function near [Fe/H]$\sim$$-$0.6 \citep{rd06}.

Inspection of Figures~\ref{fig-ofe1dg} through \ref{fig-tife1dg} 
reveals no offset between our disk star 
abundances and the literature results for Na, Mg, and Ca.  For Al and Si 
the comparison is complex: our thin disk stars with [Fe/H]$\sim$$-$0.4 and 
thick disk stars with [Fe/H]$\sim$$-$0.6 appear high, by $\sim$0.08 and 
0.09~dex respectively.  We have shifted the data points for these element
ratios in our plots (but not Table~6) to help facilitate comparisons between
our results and the literature.
Contrary to this apparent shift most of our disk 
stars near solar metallicity compare well with the literature; although 
$\mu$~Leo and the two Baade's Window disk stars in our sample also appear 
high, with approximately the same shift as the more metal-poor stars.  We 
may understand this apparent contradiction if our abundances for the most 
metal-rich disk stars are affected by the lower resolving power of the 
DuPont echelle spectrograph (R$\sim$30,000).  This might be expected from 
line blanketing in metal--rich stars that would reduce the apparent 
continuum level in lower resolving power spectra.  Since our spectra of 
$\mu$~Leo and the two Baade's Window disk giants have significantly higher 
resolving power (R$\sim$60,000 and 45,000 respectively) they do not show a 
reduction in abundance ratios due to resolving power.  Our bulge star 
results should not be affected by this problem because most of our 
bulge star spectra have resolving power of 45,000, except for a handful of 
the metal-rich stars with resolving power 60,000. The lessons here are 
that analysis of standard disk stars provides a powerful way to identify 
systematic errors; but standard and program stars should be observed using 
the same equipment.

Although it is not clear which analysis is to blame for the apparent 
shifts, it is notable that a few of our [X/Fe] trends are slightly higher 
than the results from several other studies.  One possibility is that each 
study used the same set of $gf$ values, which might suffer from zero-point 
error.  It is difficult to understand how our line by line differential 
analysis could be in error, and we have no suggestion for how this might 
have occurred.

One difference is that the literature samples for the most part use atmosphere 
grids with solar abundance ratios.  We have attempted to correct for this, 
and for most of the metal-poor stars the adopted atmospheres should be 
alpha-enhanced.  Our analysis
has found that using alpha-enhanced models increases the [X/Fe] ratios for
most of the elements here in giant stars.  A similar analysis for prototypical
dwarf stars using the lines for several of the literature sample finds that
these abundances determinations are less sensitive to the alpha-enhancement
adopted.  

We believe that the conclusions we reach in this paper are not significantly
affected by these two shifts.  For example, we will see that for most stars
the element ratios are enhanced in the bulge in comparison to the disk.
For some of these same ratios the metal-rich disk giants are lower than
what is seen in disk dwarfs.  If this offset is due to some fundamental 
problem with the analysis method and not due to difficulties related to the 
lower resolution data then we would need to shift all our data higher,
increasing the differences between the disk and bulge.

The metal-rich disk giant $\mu$ Leo stands out as being enhanced in O, Na,
Al, and other elements.  This has been noted in earlier analyses
of this star \citep{gs90,c96,sr00}.  For example, \citet{gs90} find 
[Na/Fe]~$=~+0.56$ and [Al/Fe]~$=~+0.40$ and \citet{sr00} find 
[Na/Fe]~$=~+0.38$ compared to our values of $+0.43$ and $+0.34$.  Our [Si/Fe]
results of $+0.25$ is higher than the \citet{gs90} result of $+0.12$, but
for the rest of the elements our analysis is consistent with the earlier works.

\begin{figure} 
\plotone{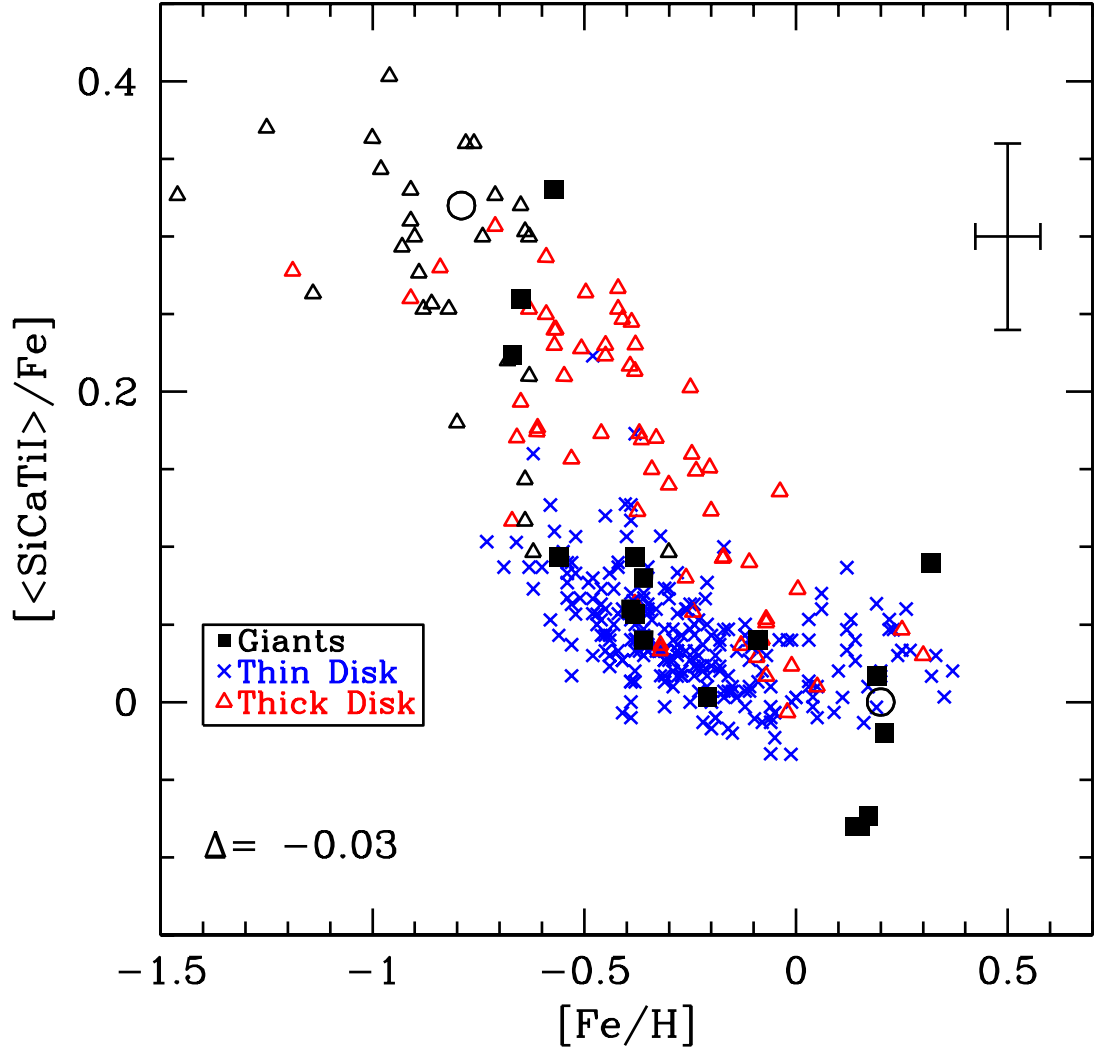} 
\caption{[$<$SiCaTiI$>$/Fe] for our sample of disk giants (filled 
squares), compared to the thin disk stars (blue crosses: Reddy et al. 
2003; Bensby et al. 2005; Brewer \& Carney 2006), and thick disk stars 
(red triangles: Prochaska et al. 2000; Bensby et al. 2005; Brewer \& 
Carney 2006; black triangles: Fulbright 2000).  Due to the $-0.09$~dex 
shift in [Si/Fe], our data points in this plot have been shifted down by
0.03~dex.}
\label{fig-casitife1dg} 
\end{figure}

In Figure \ref{fig-casitife1dg} we compare the average of the [Si/Fe], 
[Ca/Fe] and [Ti~I/Fe] ratios (henceforth [$<$SiCaTiI$>$/Fe]) for our disk 
giant sample with recent literature values from analysis of thin and thick 
disk stars. It appears that our [$<$SiCaTiI$>$/Fe] ratios are 
systematically higher than the disk trends by $\sim$0.03~dex, consistent 
with the shift of [Si/Fe] noted earlier.  In Figure 
\ref{fig-casitife1dg} we again compare our disk giants with the 
literature values, but with a 0.03~dex downward shift applied to our 
[$<$SiCaTiI$>$/Fe] results.  In this plot most of our disk giants overlap 
with the appropriate thin or thick disk trends in the literature.  We note 
that three of our most metal-rich stars in Figure \ref{fig-casitife1dg} lie 
below the literature thin disk trend; which we believe results from the 
lower dispersion of the DuPont spectra, as discussed earlier.

\begin{figure} 
\plotone{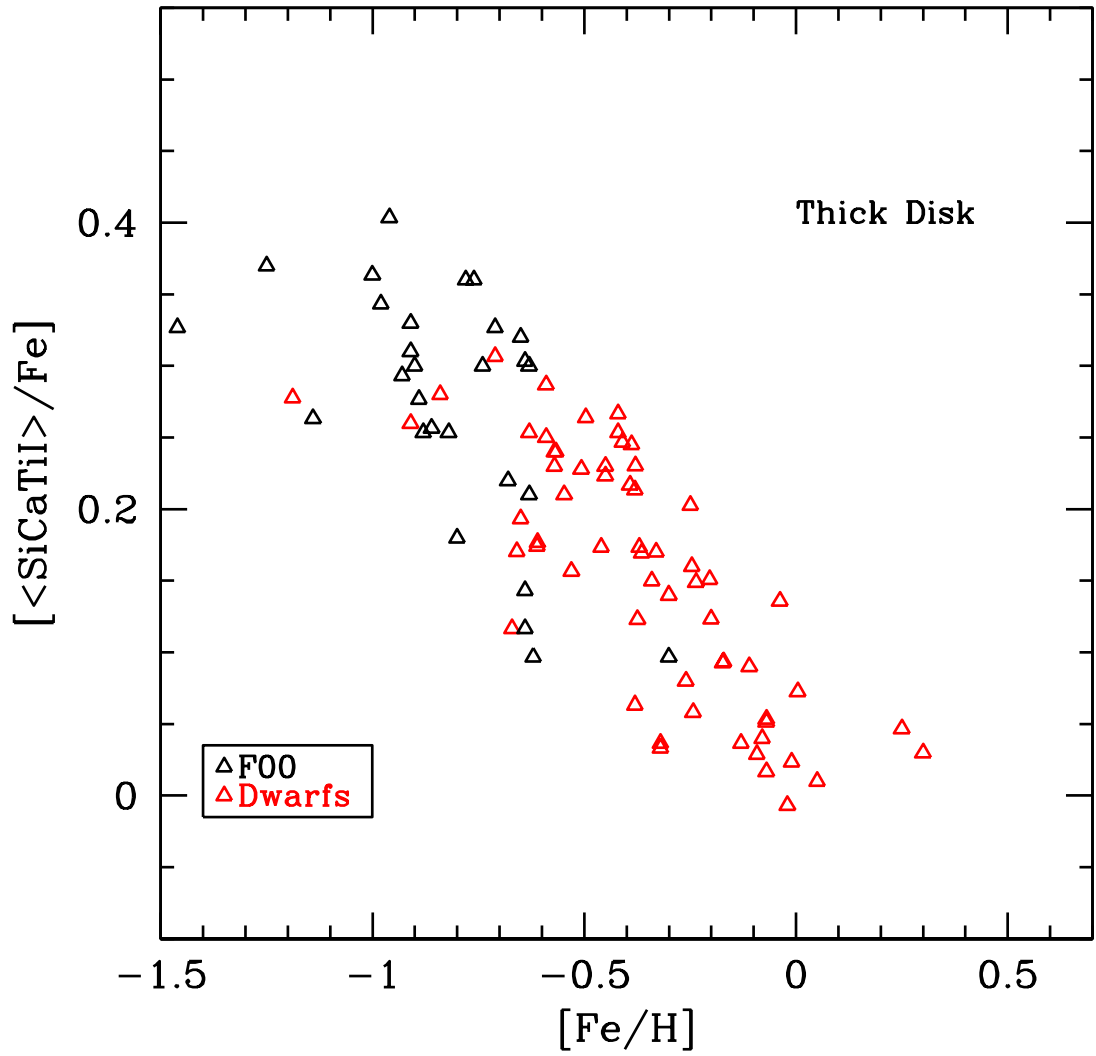} 
\caption{A comparison of the [$<$SiCaTiI$>$/Fe] for thick disk dwarf stars 
(red triangles: Prochaska et al. 2000; Bensby et al. 2005; Brewer \& 
Carney 2006) and thick disk giants from (black triangles: Fulbright 2000).  
The dwarf and giant data fit nicely together, without any detectable 
shifts.}\label{fig-jfcasiti1-thick} 
\end{figure}

In Figure \ref{fig-jfcasiti1-thick} we compare the [$<$SiCaTiI$>$/Fe] 
results from \citet{f00}, for red giant stars identified as thick 
disk members by \citet{venn04}, with results for thick disk dwarfs 
from \citet{pro00}, \citet{b05}, and \citet{bc06}.
The figure shows that the thick disk dwarf results segue nicely 
to the thick disk giant abundances of \citet{f00}, with no 
discernible offset between the giant and dwarf results.

\section{The Bulge Alpha-Element Abundance Trends}

In Figures~\ref{fig-jff5-ofebulgedisk} through 
\ref{fig-jff11-tifebulgedisk}, we present the [X/Fe] versus [Fe/H] ratios 
in all our sample stars, for O, Na, Mg, Al, Si, Ca, and Ti.  For each of 
these elements, which are thought to be produced mainly by massive stars 
that end as Type~II supernovae, the bulge abundance trend differs from 
that of the local disk, with the bulge showing higher ratios for all but 
the most metal-poor stars, where the bulge matches the Galactic halo 
composition.

For all the so-called ``alpha'' elements (O, Mg, Si, Ca, and Ti) 
our bulge abundances show a drop of the [X/Fe] ratio with increasing 
[Fe/H], but even at solar metallicity the mean trend for the bulge
population has a value of $\sim 0.2$ greater than the thin disk,
and with only a few exceptions, is greater than the disk at all
[Fe/H].  

The general abundance trends for Mg, Si and Ca are similar to the findings 
of MR94: very high [Mg/Fe] for most bulge stars, declining only slightly 
with [Fe/H], and more steeply declining [Si/Fe] and [Ca/Fe] with 
increasing [Fe/H]; although, the current [Si/Fe] and [Ca/Fe] are slightly 
higher than MR94.  The present work shows a declining [Ti/Fe] ratio with 
[Fe/H], but enhanced over the disk trend; whereas MR94 found [Ti/Fe] 
enhanced for most stars.  We believe that the MR94 [Ti/Fe] results are in 
error, and most likely resulted from poor model atmosphere temperatures.  
Since the present study is, in every way, superior to MR94 we prefer the 
current results.  For [O/Fe] the MR94 results were so uncertain as to be 
of little use: they were based on a single very badly blended line, in a 
heavily blanketed region with spectra of low resolving power.

The classical explanation for the drop of the [$\alpha$/Fe] ratios with 
increasing [Fe/H] in disk stars has been the effect of Type~Ia supernova 
\citep{t79}.  Type~Ia supernova are believed to create large amounts of 
the Fe-group elements and lesser amounts of particular alpha elements, thus 
``diluting'' the [$\alpha$/Fe] ratios of stars formed later.

Therefore, the alpha enhancements in our bulge stars, relative to the 
disk, is consistent with a higher Type~II/Type~Ia ratio in the Galactic bulge.  
While this may result from a more rapid formation timescale for the bulge 
than the disks, it could also be due to a lower binary fraction in the 
bulge, or a bulge IMF skewed to higher mass stars. We note that analysis 
of the Mg abundance results from MR94 by \citet{m99} and \citet{fws03} 
found that chemical enrichment of the bulge took about 500 Myr.  Since the 
current results for Mg are similar to MR94, we assume that an analysis by 
\citet{m99} of the Mg abundances from this work would yield a similar 
enrichment timescale.  This is plenty of time for some Type~Ia ejecta to 
be included into the star-forming material.

We should also state that other groups have recently published studies of bulge
giants.  \citet{ro05} published results for Baade's Window M-giants based 
on near-IR Keck/NIRSPEC data.  The stars studied covered only a narrow range
of metallicity ($-0.5 <$[Fe/H]$< +0.0$), but in that interval their derived
abundance ratios are in good agreement with ours.  \citet{cs06} used Pheonix
near-IR data for Baade's Window giants (including a number of stars in common 
with this paper), and, again, the general trends are in agreement.  In addition,
there have been a number of studies of globular clusters in the bulge
in the optical \citep{car01,bar06} and near-IR \citep{mel03,orig05}.  A 
review of these results, including the implications to the
formation of the bulge, will be included in \citet{fulb06c}.

\subsection{Hydrostatic vs. Explosive Alpha Element Abundances}

\begin{figure}
\plotone{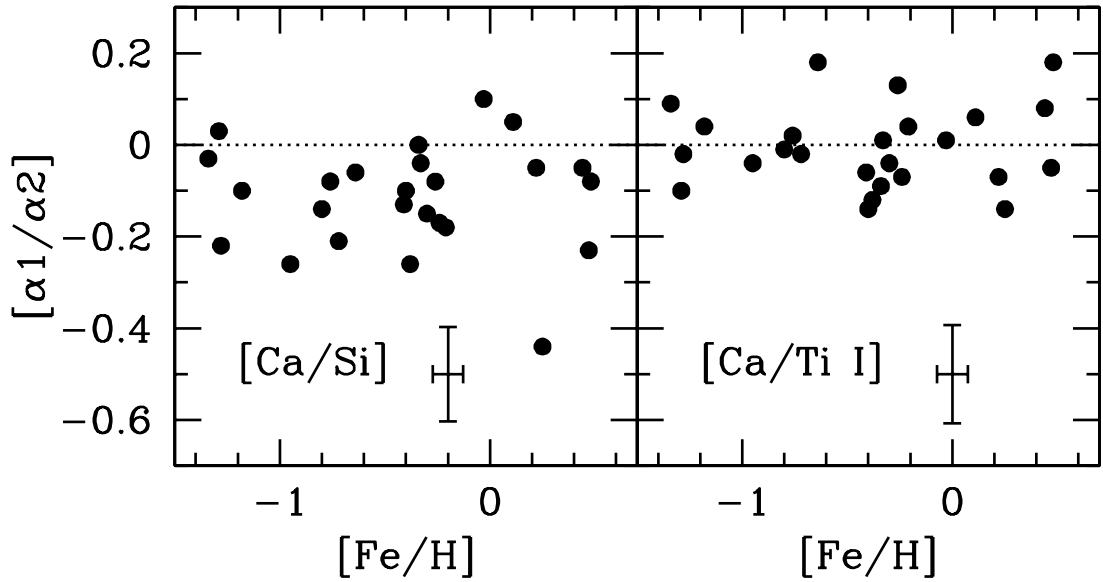}
\caption{A plot of [Ca/Si] and [Ca/Ti~I] for our sample of bulge giants 
shows no obvious trend with [Fe/H], indicating that Ti and Si both track 
Ca.  We have not applied the $-0.09$ zero-point shift to the Si abundances
to the left panel.  If applied, the [Ca/Si] ratio in the bulge becomes
roughly solar at all metallicities. } 
\label{fig-casicati1}
\end{figure}

\begin{figure}
\plotone{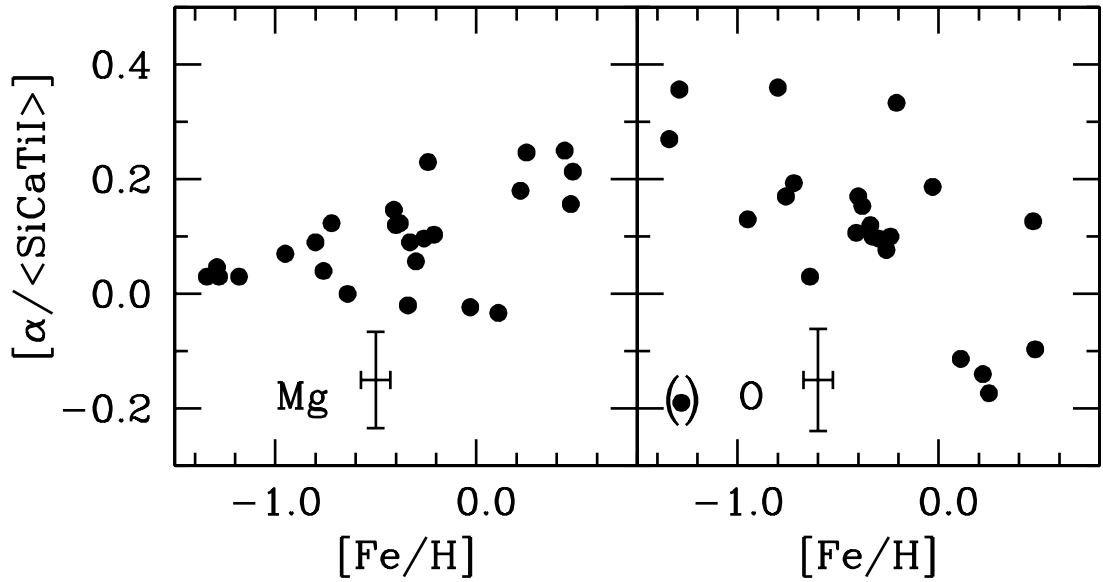}
\caption{[O/$<$SiCaTiI$>$] and [Mg/$<$SiCaTiI$>$] for our sample of bulge 
giants (filled circles).  The data are consistent with a linear decline 
in oxygen, relative to Si, Ca and Ti, over the entire [Fe/H] range; 
although points above solar [Fe/H] are distinctly lower than the trend at 
lower metallicity.  The two stars whose compositions reflect 
proton-burning products, with O, Na and Al abundances that do not 
reflect the primordial composition, have been omitted for clarity.  An 
approximately linear increase in magnesium, relative to Si, Ca and Ti, is 
evident over the full range of [Fe/H].}\label{fig-mgocasiti1}
\end{figure}

We investigate the relative abundance trends within the alpha-element 
group, by plotting the ratios of pairs of alpha elements, with the goal to 
determine whether all alpha elements have the same slope of [X/Fe] with 
[Fe/H].  Examples of the plots are presented in 
Figures~\ref{fig-casicati1} and \ref{fig-mgocasiti1}. If two alpha 
elements possess the same trend with metallicity, the ratio will be flat 
with [Fe/H].  Figure \ref{fig-casicati1} suggests that Ca, Si and Ti~I 
abundances track each other: the points could be fit with lines of zero 
slope and small zero-point shifts, so the abundances of these three 
elements may be averaged to reduce measurement scatter.  Note that 
the mean [Ca/Si] value of $\sim -0.1$~dex in the left panel of 
Figure~\ref{fig-casicati1} is roughly consistent with our possible 
zero-point shift for [Si/Fe] evident from the 
disk stars.  It is, perhaps, not surprising that these three 
alpha elements (Si, Ca and Ti) show similar trends with [Fe/H], since they 
are all thought to be produced in the explosive nucleosynthesis phase of 
Type~II supernovae (e.g. Woosley \& Weaver 1995, henceforth WW95).  The 
similar behavior of Si, Ca and Ti trends, and the putative common origin 
for these three elements provides justification for averaging these 
elements in our abundance plots.

\begin{figure}
\plotone{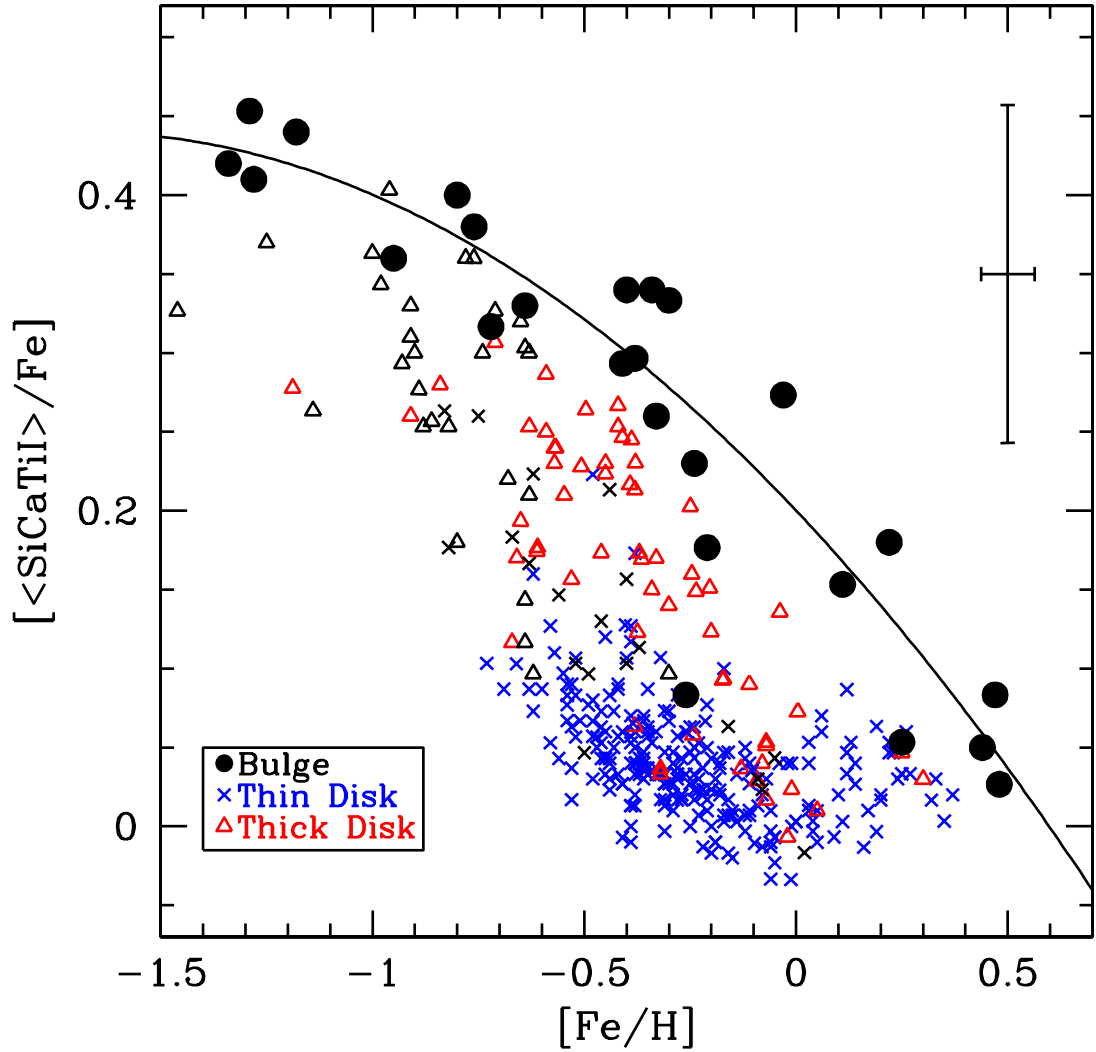}
\caption{[$<$SiCaTiI$>$/Fe] for our sample of bulge giants (filled circles)
shifted by $-0.03$~dex (filled circles), 
compared to the thin disk stars (blue 
crosses: Reddy et al. 2003; Bensby et al. 2005; Brewer \& Carney 2006), 
and thick disk stars (red triangles: Prochaska et al. 2000; Bensby et al. 
2005; Brewer \& Carney 2006; black triangles: Fulbright 2000).  The bulge 
alpha/Fe ratios are enhanced over the thin disk by $\sim$0.2 dex. The 
bulge is enhanced relative to the metal-rich thick disk,  by varying amounts, 
with the minimum difference of $\sim$0.1 dex near [Fe/H]=$-$1 dex.  The solid 
line shows a weighted cubic polynomial fit to the bulge data, assuming no error 
on the [Fe/H] values; the rms scatter of the bulge points about the curve is 
0.053 dex.}
\label{fig-jfcasiti1_bulge04s}
\end{figure}

Figure \ref{fig-jfcasiti1_bulge04s} compares the average 
[$<$SiCaTiI$>$/Fe] ratio in our Galactic bulge stars (shifted by $-0.03$ 
dex) with thin and thick disk stars.  The solid line indicates a quadratic 
fit to the bulge points. A remarkably small $rms$ scatter of the bulge 
points, at $\sim$0.053 dex, is evident; this is reduced to 0.039 dex if 
the lowest bulge point is removed. This value for the measured dispersion 
is much smaller than our prediction; but our analysis provides absolute 
uncertainties, whereas the measured dispersion ignores correlated errors; 
thus, the two are not necessarily inconsistent.

As noted previously, the bulge [$\alpha$/Fe] ratios lie about 0.2 dex above 
the thin disk trend, and this is the case for [$<$SiCaTiI$>$/Fe]. Figure 
\ref{fig-jfcasiti1_bulge04s} also shows a separation between the 
[$<$SiCaTiI$>$/Fe] in the thick-disk and bulge at [Fe/H]$\sim$$-$0.5, with 
the bulge stars more enhanced than the thick disk stars by $\sim$0.1 dex. 
However, the bulge and thick disk [$<$SiCaTiI$>$/Fe] trends merge near 
[Fe/H]=$-$1. Towards higher metallicity the thick disk alpha enhancement 
declines steeply to the thin disk level by solar metallicity, as noted by 
\citet{bc06}.  In the bulge, however, the [$<$SiCaTiI$>$/Fe] 
ratio declines more slowly, only approaching the combined thin and thick 
disk value near [Fe/H]$\sim$$+$0.5.

\begin{figure}
\plotone{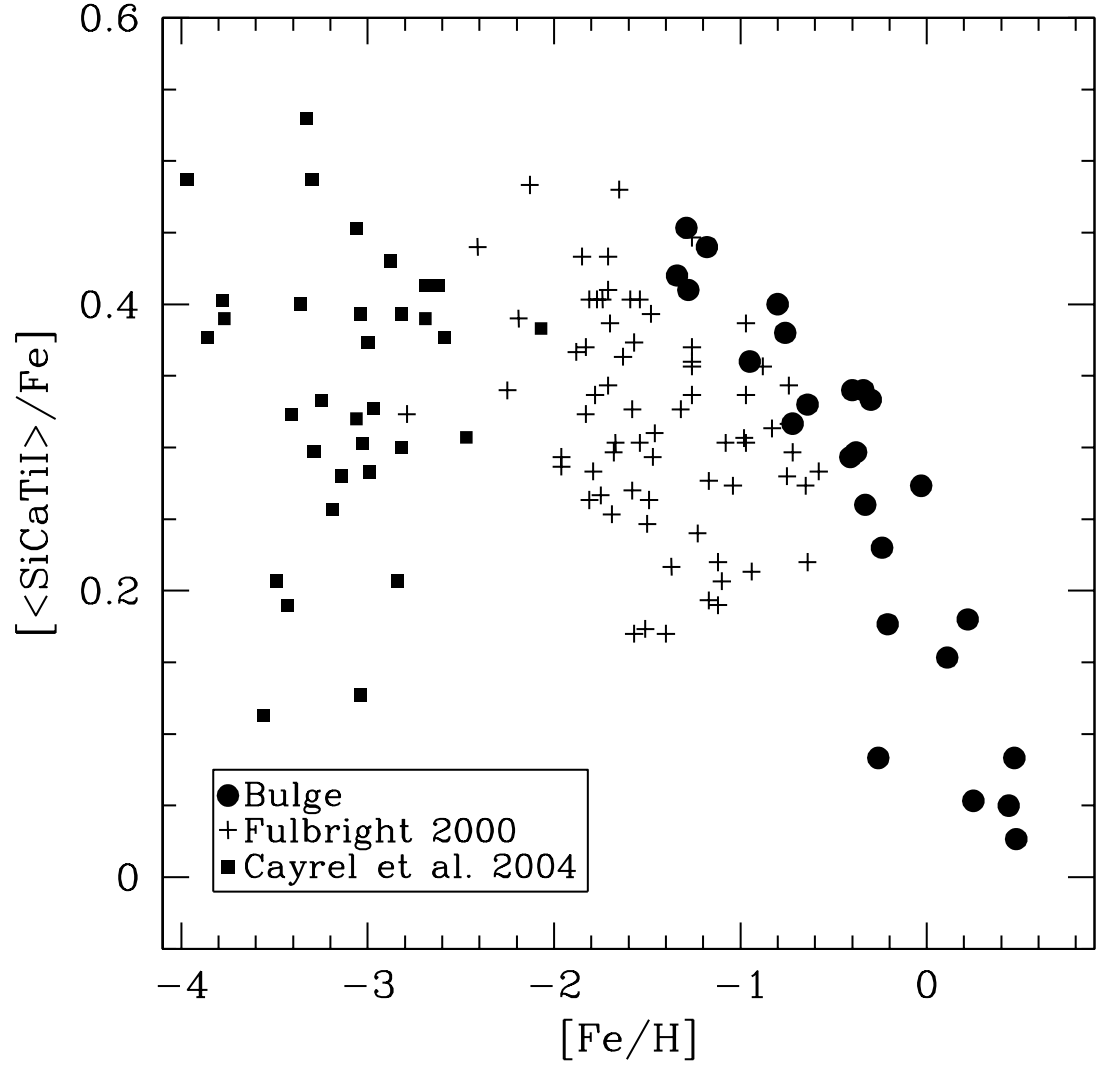}
\caption{[$<$SiCaTiI$>$/Fe] for our sample of bulge giants (filled 
circles), shifted by $-0.03$~dex, compared to the Galactic halo results of 
Fulbright (2000, plus 
signs) using population identifications by Venn et al (2004), and halo stars 
from Cayrel et al. 
(2004, filled squares).  Note the tighter trend seen in the bulge 
composition; the most metal-poor bulge stars have [$<$SiCaTiI$>$/Fe] about 
equal to the highest seen in the halo.}\label{fig-casiti1h}
\end{figure}

In Figure~\ref{fig-casiti1h} we compare the bulge [$<$SiCaTiI$>$/Fe] 
with halo abundance results from \citet{cayrel04} and Fulbright 
(2000, henceforth F00). For the F00 halo membership we use the population 
assignments of \citet{venn04}. Figure~\ref{fig-casiti1h} displays 
two clear differences between the bulge and halo samples: that the bulge 
forms a much tighter trend with [Fe/H] than the halo, and that the maximum 
[$<$SiCaTiI$>$/Fe] for our bulge sample lies at the upper envelope 
(approximately the top 20\%) of the distribution of this ratio in the 
halo.  For our bulge stars more metal poor than [Fe/H]=$-$1.0 dex the mean 
[$<$SiCaTiI$>$/Fe] value is $\sim$0.13 dex higher than for similar 
metallicity halo stars.

In Figure~\ref{fig-jfcasiti1-thick} we noted the good agreement between 
F00 thick disk giant abundances and literature values from dwarf stars; 
this, and the similarity of the mean F00 [$<$SiCaTiI$>$/Fe] value with the 
\citet{cayrel04}extreme metal-poor halo stars, suggests that there 
are no serious zero-point problems with the F00 halo abundances.  Thus, we 
assume that in Figure~\ref{fig-casiti1h} the 0.13 dex difference 
between the metal-poor bulge stars and halo giants is real.

The reduced scatter seen in the bulge stars indicates that the bulge 
composition evolved much more homogeneously than that of the halo.  The high 
[$<$SiCaTiI$>$/Fe] in the bulge suggests either that Type~II SNe in the 
bulge had more massive progenitor masses than in the halo (equal to the 
highest mass function in the halo), or that the halo experienced more 
nucleosynthesis contributions from Type~Ia SNe than the bulge.

The range of alpha/Fe ratios seen in the halo certainly indicate that it 
experienced a very inhomogeneous enrichment history.  More extreme 
evidence of alpha/Fe dispersion in the halo is already well established 
\citep{ns97,bwz97,f02}.

\citet{wg92} proposed that the bulge formed from Galactic 
spheroid (halo) gas, based on the similarity of the specific angular 
momentum of these two systems, and because the low mean metallicity 
indicates that 90\% of the spheroid gas was lost.  Our abundances provide
an interesting test of this bulge formation idea: since the [$<$SiCaTiI$>$/Fe] 
ratios in 
our most metal-poor bulge stars is higher than in the halo, and because 
our most metal-poor bulge stars, at [Fe/H]$\sim$$-$1.3 dex, are close to 
the mean metallicity for the halo, our metal-poor bulge stars could not 
have been made from halo gas with the average metallicity and composition 
seen today.  However, our metal-poor bulge stars could have been produced 
from halo gas providing that the halo composition at that time was similar 
to the top $\sim$20\% of the halo [$<$SiCaTiI$>$/Fe] ratios seen today; 
i.e. providing that the mean halo composition changed with time.  Thus, 
our results suggest that if the bulge formed out of halo gas, then the 
onset of bulge formation occurred before $\sim$80\% of the 
chemical enrichment of today's stellar halo had occurred.

In contrast to the explosive nucleosynthesis elements Si, Ca and Ti, it is 
clear that O and Mg behave in differently.  Figure~\ref{fig-mgocasiti1} 
indicates a decline in [O/$<$SiCaTiI$>$], by about 0.5 dex, over the 2 dex 
[Fe/H] range of our bulge sample. The O deficiency, relative to Si, Ca and 
Ti is greatest at $\sim$$-$0.2 dex for the most metal-rich bulge stars 
([Fe/H]$\sim$$+$0.5). On the other hand Figure~\ref{fig-mgocasiti1} shows 
the inverse behavior for Mg: a steady {\em increase} in [Mg/$<$SiCaTiI$>$] 
over the entire metallicity range, reaching approximately $+$0.2 dex at 
[Fe/H]$\sim$$+$0.5 dex.  At solar metallicity [Mg/Fe] is enhanced by 
$\sim$$+$0.3 dex.

It is interesting that in abundance plots for Local Group dwarf galaxies 
by \citet{venn04} the available hydrostatic alpha element, Mg, shows a 
much steeper slope with [Fe/H] than the explosive alpha elements, Ca and 
Ti.  This supports our idea that these two families of alpha elements 
should be treated separately.

Since Mg and O are both thought to be produced during the hydrostatic 
nuclear burning phase in Type~II SN progenitors, the strikingly different 
appearance of the trends for these two elements is somewhat surprising.  
The observed, opposite, trends of O and Mg, with [Fe/H], may be the result 
of a metallicity dependency of the yields of these two elements.  
Metal-dependent winds from the supernova progenitor may present a path to 
reduce the oxygen yield.  In this regard it is of interest that the 
frequency of the Wolf-Rayet phenomenon, thought to be due to stripping of 
the outer envelopes of massive stars through stellar winds, increases 
dramatically above [Fe/H]$\sim$$-$1 \citep{maeder80,maeder91}.  Detailed 
nucleosynthesis calculations of the yields from massive stars 
\citep{maeder92,mm02} show that near solar metallicity there is a 
reduction of the yield of oxygen and an increase in the carbon yield, due 
to mass-loss driven by metallicity-dependent stellar winds.  These effects 
are prominent only for stars with mass greater than M$\sim$30~M$_{\odot}$ 
and metallicity greater than Z$\sim$0.004 (roughly [Fe/H]$\ge$$-$0.7).  
If this is the case, then magnesium abundances are preferred, over oxygen, 
as an indicator of Type~II SN products.  The \citet{maeder92} and \citet{mm02}
predictions also indicate enhanced carbon yields from 
Wolf-Reyet stars, so bulge carbon abundances will provide a test of our 
proposed explanation for the steep decline in [O/Fe]. Another potential 
test comes from the F/O ratio, which \citep{ma00} predict is significantly 
higher from Wolf-Reyet stars; some support for this idea comes from 
\citet{ren04} and \citet{zl05}.  However, \citet{pal05} are more 
pessimistic about the contribution of WR stars to the evolution of F 
abundances.

The simple addition of Type~Ia SN material cannot solely explain the drop 
in the [alpha/Fe] ratios seen in Baade's Window, because the [Mg/Fe] ratio 
is nearly flat with [Fe/H], whereas the [O/Fe], [Si/Fe], [Ca/Fe] and 
[Ti/Fe] show steep declines with metallicity.  The [Mg/Fe] ratio does show 
a slight decline, of about 0.2, dex over the full [Fe/H] range; but even 
the most metal-rich bulge stars, near [Fe/H]=$+$0.5, have 
[Mg/Fe]$\sim$$+$0.25--0.30 dex.

It is unlikely that our high [Mg/Fe] ratios are due to systematic 
measurement errors: our disk giant sample (see 
Figure~\ref{fig-mgfe1dg}) indicates that, if anything, we may have 
slightly underestimated the [Mg/Fe] ratios for the most metal-rich stars.  
As discussed in the Introduction, MR94 and other studies have found
high [Mg/Fe] ratios in the Milky Way bulge, and extragalactic studies
of spheriodal systems often find high [Mg/Fe] ratios.  Therefore it is
not too suprising that we confirm the high [Mg/Fe] ratio found in the
bugle my MR94.

\begin{figure}
\plotone{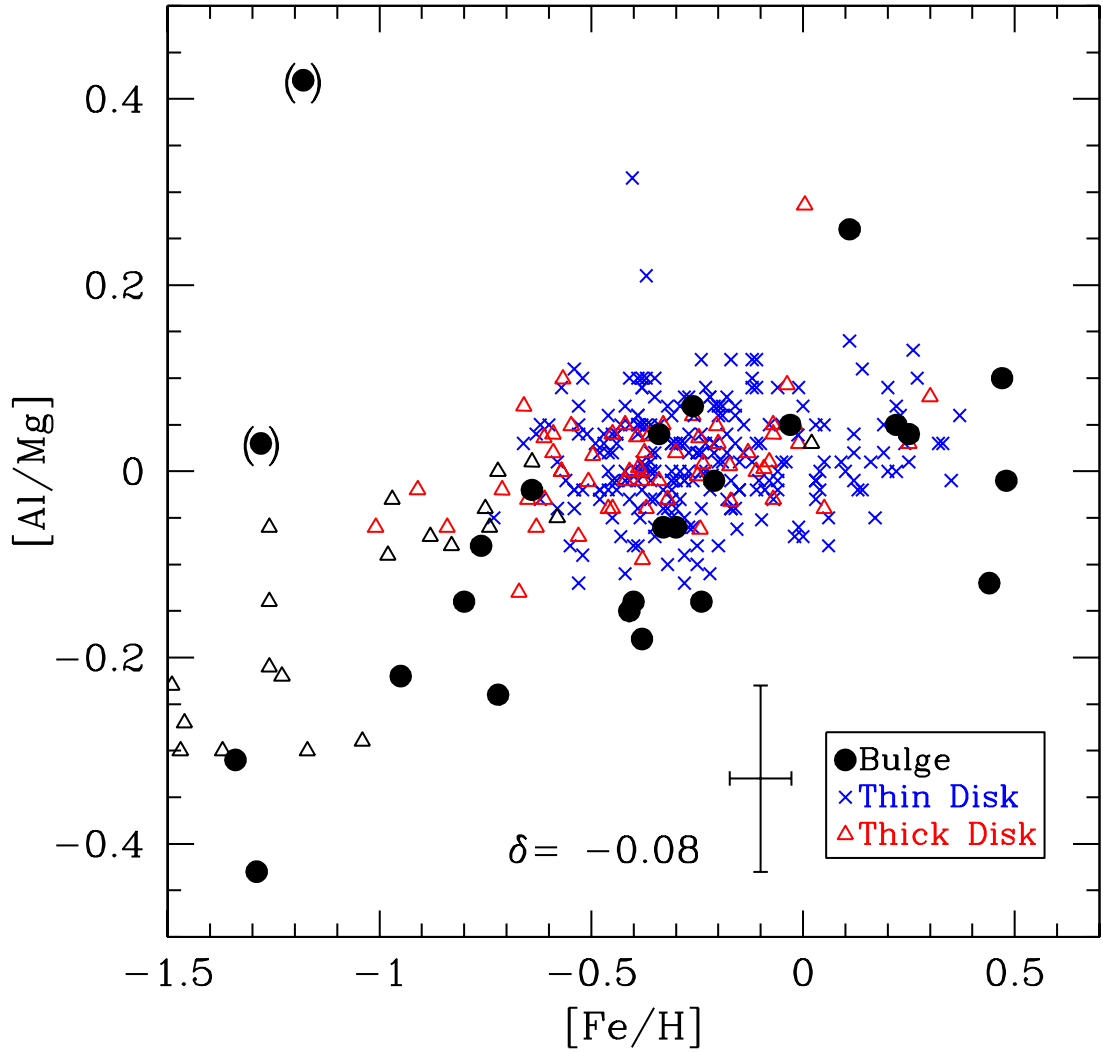}
\caption{A comparison showing [Al/Mg] versus [Fe/H] trends in the Galactic 
bulge (filled circles), thin disk (blue crosses), and thick disk (red 
triangles).  Our data points have been shifted by $-0.08$~dex to account
for possible zero-point offsets in the Al results.
The trend for the three populations appears identical; thus, 
our bulge Al abundances corroborate our unusual Mg values.  Removing the 
shift from our Al abundances would not affect this conclusion.  Points 
slightly above, and separated from, the mean trend may be due to enhanced 
contribution from high mass Type~II SNe. The two points in parentheses 
represent stars we have identified as being affected by envelope proton 
burning, with Al abundances that do not reflect their primordial 
compositions.}\label{fig-almgfe}
\end{figure}

A confirmation of our high Mg abundances comes from the plot of [Al/Mg] 
with [Fe/H]: Figure~\ref{fig-almgfe} shows that the [Al/Mg] ratio in 
the bulge compared to the thick and thin disks share considerable overlap.  
If we assume that our Mg abundances are spuriously large by 0.25 
dex, then the bulge [Al/Mg] would have no overlap at all with the trend 
seen in the Galactic disk.  The agreement between the observed [Al/Mg] 
ratio in the bulge with the disk indicates that if Al is used as a proxy 
for the Mg abundance we confirm our high Mg abundances in the bulge.

Our observed differences in the trends within the alpha-element family 
poses a problem for the standard explanation of the decline in [$\alpha$/Fe] 
with metallicity \citep{t79}. If iron-rich ejecta from Type~Ia 
supernovae in the bulge diluted the O/Fe, Si/Fe, Ca/Fe and Ti/Fe ratios 
with increasing [Fe/H], then the Type~Ia SNe would also have to create 
large amounts of Mg and Al in order to keep the Mg/Fe and Al/Fe ratios 
high; furthermore, the Type~Ia SNe would have to produce these two 
elements in the amount required to maintain the [Al/Mg] ratio trend 
(heretofore thought to be due almost entirely to Type~II SN).  In such a 
scenario it would be necessary to explain why the Type~Ia SN in the bulge 
produce Mg and Al, but Type~Ia SN in the disk do not. Nucleosynthesis 
calculations for Type~Ia SN \citep{iwa99} predict only trace 
amounts of Mg or Al production, about 1/200 of that required.  Given these 
considerations we dismiss the possibility that an unusual Type~Ia SNe 
produced the observed bulge Mg and Al abundances; we believe that the 
reason must lie with Type~II SNe.

While a deficiency of O, relative to Mg, might be understood as the result 
of stellar winds in massive stars, it is unclear that stellar winds could 
explain the enhancement of Mg over Si, Ca and Ti.  We favor the idea that 
the bulge composition reflects the metallicity-dependent yield ratios of 
Type~II SNe than the disk.  In order to understand why the disk does not 
also show Mg significantly enhanced over Si, Ca and Ti we require the 
production of Si, Ca, and Ti by Type~Ia SNe; thus, the disk values of Si, 
Ca and Ti would be $\sim$0.2 dex lower without the contribution of Type~Ia 
SN.  The idea that Type~Ia produce Si, Ca and Ti is supported by the 
composition of dwarf spheroidal galaxies from \citet{sm02}, \citet{ms05} 
and the work of \citet{s03}
and Geisler et al. (2004), as reviewed by \citet{venn04}.  We 
do not speculate on the underlying cause for the metallicity-dependent 
yield differences in Type~II SNe, between Mg, Al and the Si, Ca and Ti 
group.  Unfortunately, the WW95 yield trends with metallicity, for Mg, Si 
and Ca in Type~II SNe, are not consistent with our enhanced Mg abundances, 
relative to Si and Ca; indeed, the predictions favor a decrease in the 
Mg/Si and Mg/Ca ratios from [Fe/H]$\sim$$-$1 to 0; thus, opposite to the 
observations.  It is significant that the WW95 predictions did not include 
metallicity-dependent winds; we speculate that inclusion of these winds 
may show that very massive Type~II SN progenitors contribute 
nucleosynthesis products, such as Mg, at high metallicity only.  This 
possibility seems reasonable, given that the WW95 study identified 
problems with material ejected from the most massive SN falling back onto 
the remnant. In this scenario for understanding the unusual Mg abundances 
in the bulge, the bulge formed very quickly, with very little contribution 
from Type~Ia SNe or intermediate and low mass stars.

We note that the Type~II SN nucleosynthesis predictions of WW95 indicate 
large Mg and O yields, with little or no production of Si, Ca and Ti, from 
massive Type~II SNe (near 30--35M$_{\odot}$).  If we seek an understanding 
of our Mg/$<$SiCaTiI$>$ abundance ratios using the WW95 predictions then we 
would conclude that the 30--35M$_{\odot}$ stars became more important with 
increasing [Fe/H], as proposed by MR94. This suggests the existence of a 
population of massive stars, in addition to the normal IMF, that increases 
in importance with [Fe/H]; this might occur by a steepening of the IMF 
with metallicity.  In this scenario it would be possible for the bulge to 
have formed over a timescale, long enough for Type~Ia SNe to reduce the 
[alpha/Fe] ratio for Si, Ca and Ti; but with an extra population of 
high-mass stars that increased in importance with higher [Fe/H].  If 
the bulge composition is shown to possess chemical signatures from stars 
with long main sequence lifetimes, such as low-mass AGB stars, then this 
model would be favored over metallicity-dependent Si, Ca and Ti yields 
from Type~II SNe described above; at the same time it is still necessary 
to evoke metallicity-dependent SN yields to understand the bulge oxygen 
abundance trend.  One difficulty with this model, particularly for longer 
timescale bulge formation, is the question of why it did not also occur in 
the Galactic disk.  Another weak point is that the scenario requires the 
increasing production of Fe from Type~Ia SNe to be matched by an 
increasing yield of Mg and Al, in order to maintain the gentle downward 
slope in [Mg/Fe] with [Fe/H].

\section{Abundances of Aluminum and Sodium}

\begin{figure} 
\plotone{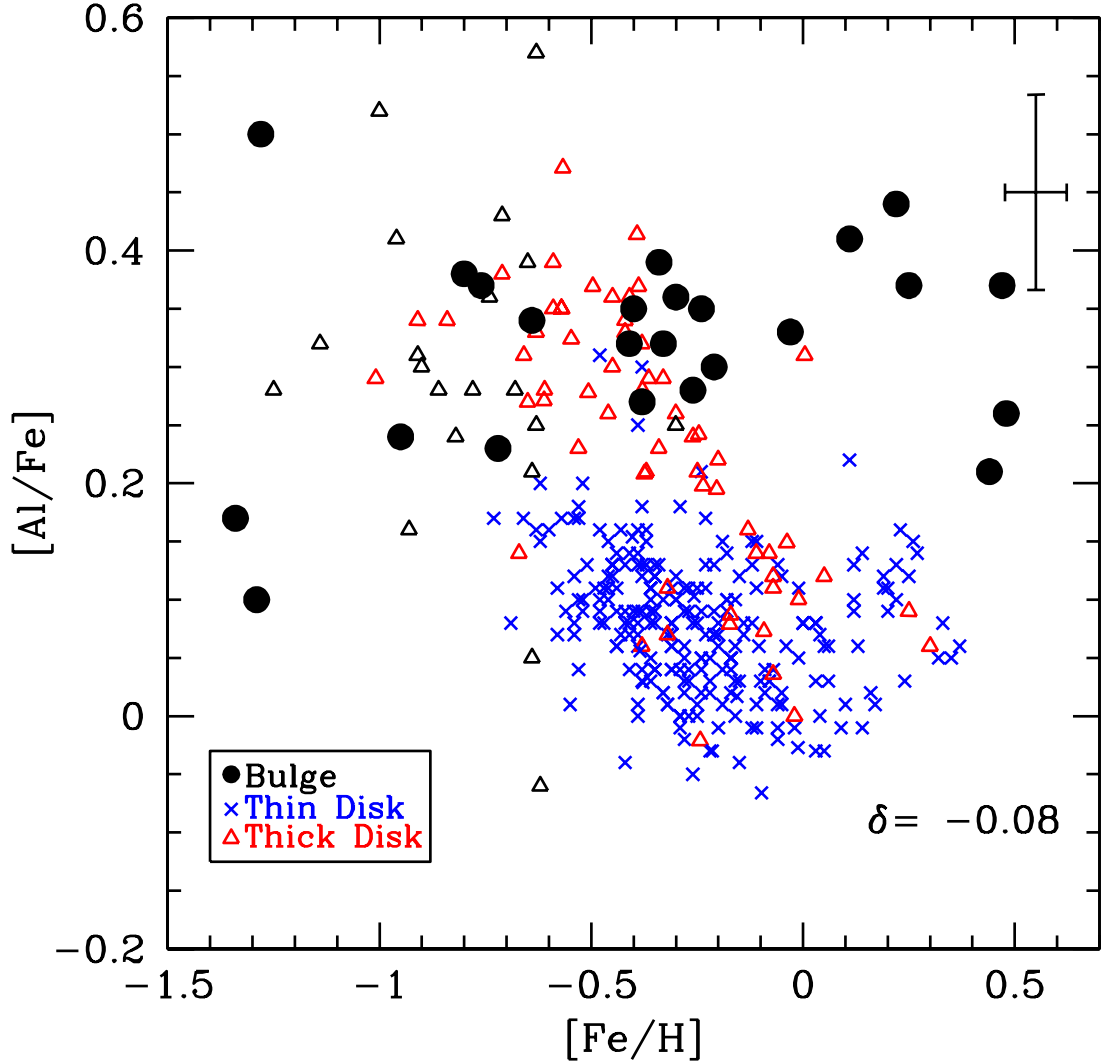} 
\caption{A comparison of [Al/Fe] versus [Fe/H] in the Galactic bulge 
(filled circles), with the Galactic thin and thick disks.  Like 
Figure~\ref{fig-alfe1dg}, our data points have been shifted by $-0.08$~dex.
The two stars 
affected by proton burning have been omitted.  The bulge [Al/Fe] ratio 
continues to rise with [Fe/H] to at least [Fe/H]$\sim$$+$0.45~dex; whereas the 
thin and thick disks show declining [Al/Fe] with [Fe/H].  Notice the 
overlap between the thick disk and bulge around [Fe/H]$\sim$$-$0.6.  
There is a increase in [Al/Fe] for the bulge from $\sim$$+$0.15 dex near
[Fe/H]$\sim$$-$1.3 to $\sim$$+0.40$ in the most metal-rich stars; this 
most likely reflects the increasing yield of Al with metallicity from
Type~II supernovae.  The thin disk comparisons are with the data of 
Reddy et al. (2003, R03),
Bensby et al. (2005, B05) and Brewer \& Carney (2006, BC06), whilst the
thick disk data points are taken from Fulbright (2000), Prochaska et al. 
(2000, P00), B05 and BC06.}
\label{fig-alfenosgr} 
\end{figure}

As noted earlier both [Al/Fe] and [Na/Fe] are generally enhanced in the 
bulge stars (see Figures \ref{fig-jff6-nafebulgedisk} and 
\ref{fig-alfenosgr}):  [Al/Fe] by $\sim$0.3 dex, and [Na/Fe] by $\sim$0.2 
dex, and both show increasing ratios with increasing [Fe/H].  This work
confirms, and vastly improves upon, the resutls of MR94, who found Al
enhancements at all metallicities and Na enhanced for only the most metal-rich
bulge stars.  One might 
ask: why, if they are Na and Al produced by Type~II SNe, don't they also 
decline with metallicity like O, Mg, Si, Ca and Ti?

While Al and Na are thought to be produced mainly by Type~II SNe, their 
trends with metallicity in the disk and halo are different from the alpha 
elements: where the alpha/Fe ratios are enhanced in the halo, Al/Fe is 
deficient, but steadily increases with [Fe/H].  In the Galactic disk, 
[Al/Fe] gently declines as [Fe/H] increases, presumably due to the 
addition of Fe from Type~Ia SNe.  While Na/Fe is predicted to follow the 
general form exhibited by Al/Fe, it defies such expectations in the halo, 
remaining approximately constant; this suggests that there was probably a 
primordial source of Na involved in the evolution of the halo.  In the 
disk there is a very gentle downward slope of [Na/Fe] with increasing 
[Fe/H].

Predictions of Al and Na in Type~II SN were discussed by
\citet{a71}, who explained the yield of odd-Z elements as a function
of the neutron excess ($\eta$ = (n - p)/(n + p)).  In general, the
larger the neutron excess, the higher the relative yield of these 
odd-Z elements.

The gentle decline in [Al/Fe] and [Na/Fe] in the Galactic disk can be 
understood as the combination of a decrease in the ratios due to the 
addition of iron from Type~Ia SNe, combined with enhanced Al and Na yields 
from increasing neutron excess, at higher [Fe/H], due to Type~II SNe.  In 
the bulge, however, there was less iron from Type~Ia SNe than in the disk, 
thus permitting the effect of the increase in Al and Na yields with 
increasing [Fe/H] to be seen as higher [Al/Fe] and [Na/Fe] ratios (Figures 
\ref{fig-alfenosgr} and \ref{fig-jff6-nafebulgedisk}), with increasing
[Fe/H].  The curious thing is that Al is significantly more 
enhanced than Na in the bulge, which suggests that the ratios were 
affected by more than the absence of iron from Type~Ia SNe; this might be 
explained if there was an additional source of Na in the disk.  Alternatively,
if significant amounts of Na is produced during H- and He-burning stages
(as suggested by WW95), then the mechanism that causes the decrease in the
O abundances (oxygen is mainly created during He-burning) may also be affecting
the Na abundances.  Sodium will not show the same decrease in production as O
because Na is also produced during C-burning.  

\begin{figure}
\plotone{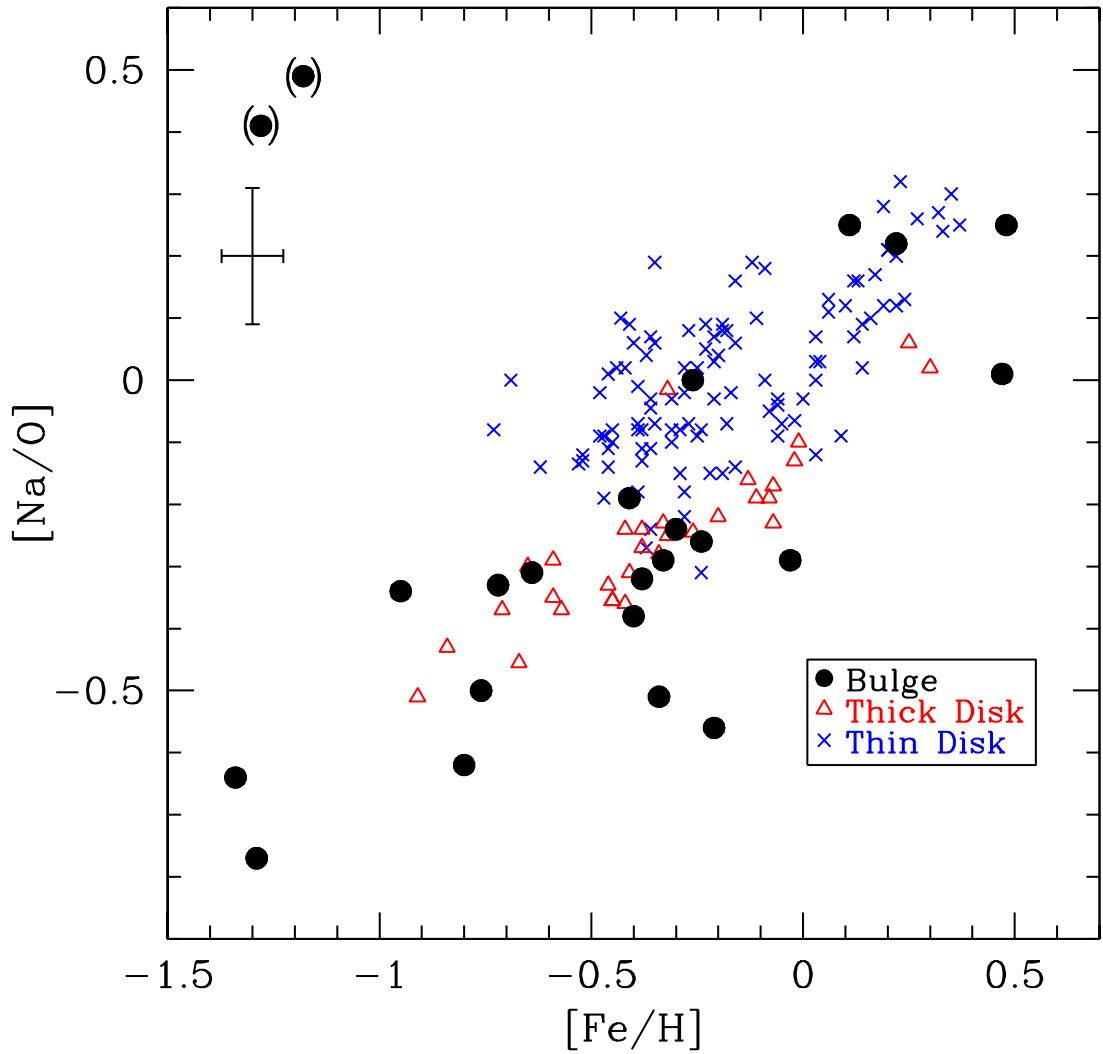}
\caption{A comparison showing [Na/O] versus [Fe/H] trends in the Galactic 
bulge (filled circles), thin disk (blue crosses), and thick disk (red 
triangles). The thick disk results show scatter significantly smaller than 
the bulge and thin disk; perhaps suggesting either larger measurement 
uncertainties in the bulge and thin disk, or more than one source of Na. 
The two points in parentheses represent stars we have identified as being 
affected by envelope proton burning, with Na and O abundances that do not 
reflect their primordial compositions.}\label{fig-naofe_10s}
\end{figure}

The odd/even trend in the bulge can be discerned from 
Figures~\ref{fig-almgfe} and \ref{fig-naofe_10s}. As mentioned 
earlier, the [Al/Mg] trend with [Fe/H] in the bulge is similar in form, 
and has considerable overlap with, the disk relation; the small scatter is 
likely due, entirely, to measurement uncertainties.  Since Type~Ia SNe 
likely contribute material in differing proportions to the bulge, thick 
and thin disks, the similarity between the [Al/Mg] trends for these 
different populations is observational evidence that Type~Ia SNe do not 
produce significant quantities of Al and Mg, as expected.  The Al/Mg trend 
at low metallicity does show a positive slope, as predicted from the 
\citet{a71} dependence of Al/Mg yield on neutron excess; but above 
[Fe/H]$\sim$$-$0.5 the trend flattens. The Na/O trend for the three 
populations have similar slope, with no flattening, but small shifts exist 
between the three populations, and the thin disk points show larger 
scatter than expected, which may implicate an additional source of Na in 
the thin disk. Generally, the odd-even effect in the bulge seems to follow 
a universal trend of decreasing odd-even differences with increasing 
metallicity, as expected from nucleosynthesis predictions (e.g. Arnett 
1971).

\begin{figure}
\plotone{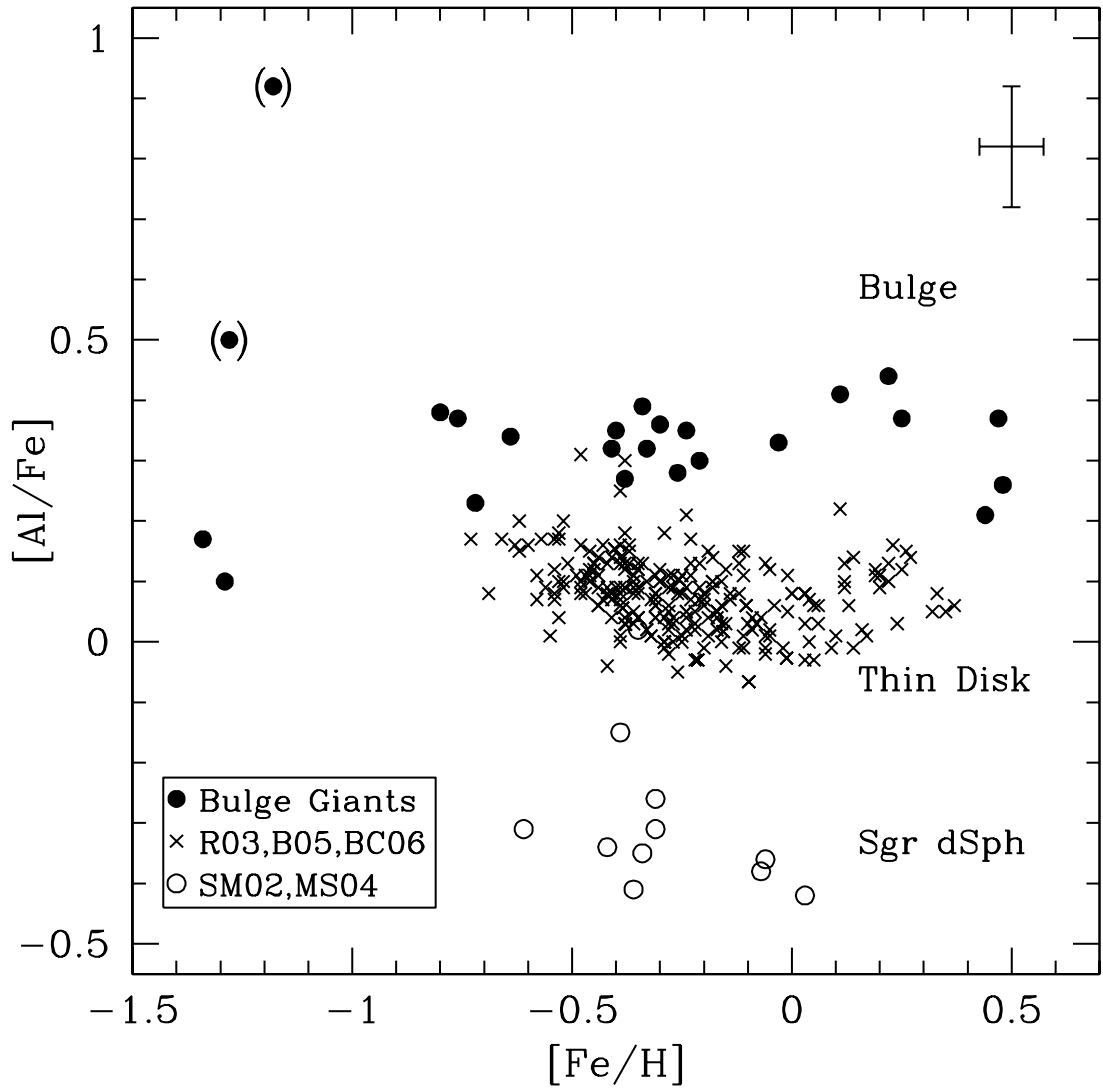}
\caption{A comparison of the distinctly different [Al/Fe] versus [Fe/H] 
trends in the Galactic bulge (filled circles), the Galactic thin disk 
(crosses), and the Sagittarius dwarf spheroidal galaxy (open circles). 
Our data points have been shifted by $-0.08$~dex.
Points in parentheses represent 
stars we have identified as being affected by envelope proton burning, 
with O and Al abundances that do not reflect their primordial 
compositions.  Clearly, the evolution of [Al/Fe] with [Fe/H] depends upon 
the environmental parameters for stellar systems.}
\label{fig-alfesgr}
\end{figure}

To demonstrate the sensitivity of Al to chemical evolution history we
present Figure \ref{fig-alfesgr}, which compares our Galactic bulge [Al/Fe]
ratios with those in the thin disk and the Sagittarius dwarf spheroidal
galaxy (Sgr dSph).  These three systems show remarkably distinct trends
of [Al/Fe] with [Fe/H], with a total range in [Al/Fe] of almost 1 dex.
Odd-Z elements should not be underestimated as a diagnostic of the chemical
enrichment history of stellar populations.

Assuming that Al is made entirely from Type~II SNe and Fe from both Type~Ia
and Type~II SNe; the 0.7~dex difference in the [Al/Fe] ratio between the
bulge and Sgr dSph giants indicates that the fraction of Fe from Type~II 
SNe in the Sgr dSph is one-fifth the value found in the bulge.  Therefore,
at least 80 percent of the Fe in the metal-rich stars in the Sgr dSph
is from Type~Ia SNe.

\section{I-264, IV-203, and the Abundance Variations in Globular Clusters}

The bulge stars I-264 and IV-203 show high [Na/Fe] and 
[Al/Fe] abundance ratios, but a low [O/Fe] ratio.  This pattern is also seen
between stars in individual globular clusters:  the so-called 
``Na-O anti-correlation'' \citep{gr04,can98,k94}.
The only other non-cluster system containing stars with this abundance pattern
is the Sagittarius dwarf spheroidal galaxy \citep{sm02,ms05}.

We plot the [Na/Fe] and [Al/Fe] abundances of these two stars against [O/Fe]
in Figure~\ref{fig-jff15-nafealfeofe}.  Also plotted are the same of M4 stars 
from \citet{iii99} and M5 stars from \citet{iii01}.
M4 and M5 has a mean [Fe/H] values of $-1.18$ and $-1.21$, which are 
similar to the [Fe/H] value
of these two stars ($-1.10$ for I-264 and $-1.25$ for IV-203).  Also plotted
are the two other metal-poor ([Fe/H] $< -1$) bulge stars II-119 ($-1.22$)
and IV-003 ($-1.30$).  The bulge
stars lie slightly off the locus of the M5 stars, but very close to the
M4 trend. 

Some clusters, like M13 \citep{j05,s96} show variations of [Mg/Fe]
as well (slight decreases in stars with high [Al/Fe] ratios).  These
variations were not seen in M4 and M5, nor in two the bulge stars.

The origin of the intra-cluster composition variations has been the subject 
of some debate.  One possibility is that these giants have dredged-up material
that has undergone nucleosynthetic processing in H-burning shells.
\citet{gr04} and \citet{can98} present discussions on both the
the nucleosynthesis processes involved.  
Another possibility is that the
variations were ``primordial'', that is, imprinted upon the stars by
some event during the birth or early life of the cluster.  The most
popular theory is that massive AGB stars ``salted'' neighboring
stars with highly processed material blown off during the AGB star's
death throes.  

The ``salting'' theory gained a boost from observations of less-evolved
cluster stars that showed that these stars show the same kind of 
abundance variations.  These stars cannot have internal temperatures
hot enough to create the reactions necessary to create the abundance 
pattern.

Both theories rely on some special property of globular clusters
to explain why cluster stars show the variations and field stars
do not.  The ``deep-mixing'' theory invoked rotation-induced meridional
circulation \citep{sm79}, while the AGB ``salting''
theory relies on the close proximity of the stars in the young cluster.
Independent of which theory is correct, the presence of this abundance
pattern tells us something important about the formation of the early bulge in
comparison to the rest of the metal-poor field.  Half (two of four stars) 
of our metal-poor
sample shows this pattern, yet none of the many dozens of metal-poor
Milky Way field stars studied to date do.  

One possibility is that the metal-poor stars in the bulge were originally
in globular cluster systems that were eventually disrupted and spread
out.  It should be noted that there is a metal-poor globular in Baade's
Window, NGC 6522, but our two bulge giants are far from the cluster
on the sky (over two half-light radii), have radial velocities inconsistent
with cluster membership (7.6 and 19.6 km s$^{-1}$ for I-264 and IV-203 while
\citet{rut97} found $v_{\rm helio} = -18.3$ for NGC~6522), and 
[Fe/H] values too high (Rutledge et al. found [Fe/H] $= -1.44$ for NGC~6522)
for these stars to be likely cluster members.

The other obvious possibility is that star formation conditions found in
the early bulge were similar to those found in globular clusters.  If the
``salting'' theory is correct, it would require the kind of relatively 
dense star formation not seen anywhere else but the bulge and in clusters.
Further study of this phenomenon in globular clusters and metal-poor bulge
stars may help constrain the earliest formation of the bulge.

\section{Summary}

We have performed a detailed chemical abundance analysis of 27 RGB stars 
towards the Galactic bulge in Baade's Window, based on Keck/HIRES echelle 
spectra.  In this paper we focus on light elements thought to be produced 
by massive stars:  the alpha elements O, Mg, Si, Ca and Ti, and also Al and 
Na.  We employed a differential analysis, line by line, relative to the 
high resolution, high S/N, spectral atlas of the red giant Arcturus by 
\citet{hw00}.  We used only unblended lines in the equivalent width  
abundance analysis; this list should be useful for future 
chemical abundance studies of red giants.  The advantage of 
differential analysis is that systematic errors, due to various 
inadequacies of the model atmospheres, cancel out when 
performed relative to similar stars; certainly the oscillator strengths of 
the atomic lines do not enter into the abundance determinations.  
In order to relate differential abundances 
relative to Arcturus to the solar differential scale we have determined 
the composition of Arcturus, relative to the Sun, using the \citet{kfb84}
solar atlas and the \citet{hw00} Arcturus atlas, which have 
superior wavelength coverage, S/N, and resolution than our science 
spectra.  Our abundance results for Arcturus show that its chemical 
abundance ratios are completely consistent with the composition of thick 
disk stars; the membership of Arcturus in the thick disk population is also
supported by its kinematic parameters. 

In addition the bulge spectra, we analyse 17 disk giants that have been 
observed at slightly lower resolution using other telescopes.
These local objects enable a literature comparison of the abundance results 
with large disk dwarf surveys.  The comparison confirms that the abundance 
results for our giants are in good general agreement with published dwarf 
surveys; but it uncovered possible small ($< 0.1$ dex) systematic zero-point 
differences for Al and Si.

The stellar model atmosphere parameters we use are derived from the analysis
method performed in Paper~I of this series.  The analysis relies on V$-$K 
photometry, differential Fe excitation temperatures and differential Fe 
ionization temperatures.  Plots showing the ionization equilibrium of Fe 
and Ti indicate a small, $\sim$0.05 dex difference between Fe~I and Fe~II, 
and no obvious difference between Ti~I and Ti~II; although the titanium 
results were affected by more scatter than those for iron.  The ionization 
plots show that any non-LTE effects are the same, to within $\sim$0.05 dex, 
across our entire sample of disk and bulge giant stars, which cover a 
range in [Fe/H] of 2 dex and T$_{\rm eff}$ by 1000K.

Our main result is that all five alpha elements, plus Na and Al, are 
enhanced in the bulge stars relative to the Galactic thin and thick 
disks; oxygen, however, may be enhanced only marginally relative to the thin 
disk. These elements are thought to be produced in mostly massive stars,
either during hydrostatic burning phases or during explosive nucleosynthesis
in Type II SNe events.  Consequently, our abundance results indicate that
massive stars contributed more to the chemical enrichment of the bulge than
to the disk.

The enhancement of these massive star products indicates that the ratio of 
Type~II/Type~Ia supernova was higher in the bulge than in the disk.  
\citet{t79} propsed what is now the established paradigm, 
that the declining trend of [O/Fe] with [Fe/H] observed from the halo to the 
solar neighborhood was due to the late addition of Fe, but 
not O, from Type~Ia SNe, due to the longer progenitor lifetimes of Type~Ia SNe.
These enhancements of O and other alpha elements relative to iron are often
associated with shorter formation timescales, and, thus, we conclude that
the bulge formed faster than the disk.
Chemical evolution models of the abundances of MR94 (whose results are similar
to those found here) by \citet{m99}, \citet{fws03}
and \citet{m06} find a $\sim 500$~Myr (rapid)
timescale for the chemical enrichment of the bulge as a central conclusion.  
While we believe that a short formation timescale for the bulge is the most
attractive explanation for our findings, we cannot ignore other possible
scenarios (IMF skewed to high mass, lower binary 
fraction in the bulge reducing the relative contribution Type~Ia SN 
ejecta, etc.).  For this reason the search for abundance patterns in the bulge 
that are characteristic of long-lived stars (e.g. low-mass AGB stars) will 
provide very useful constraints on the bulge formation timescale.

Overall, we find that Mg is strikingly enhanced relative the thick and thin 
disk populations, and that the trend of [Mg/Fe] with repsect to [Fe/H] to well 
above solar metallicity.  The strong enhancement of Mg with respect to the 
other alpha extralactic Mg enahncements; Mg has long been considered a proxy
for the measurement of alphas in galaxies, and this view may have to change.  
Indications of relatively low Ca abundances \citep{tmb03} in external galaxies 
may reflect a general dichotomy between Mg and the rest of the alphas.  Future 
studies should explore these findings.  Al and 
to some extent Na follow the same trend as Mg, but O and the other alphas are 
only marginally enhanced relative to the thick disk, and follow declining 
trends with [Fe/H].   This result was hinted at in \citet{rm00}
and in \citet{mr04}.  But we see these trends clearly in our dataset now.  

Our abundance trends indicate that the bulge is chemically distinct from the 
thick and thin disk stars in the Solar vicinity.  This does not rule out some 
kind of secular evolution \citep{pn90} from a primordial massive thick disk.   
Because they have similar chemical compositions below [Fe/H]$\sim -0.9$, it is 
possible that the early bulge gas could have come from a primordial thick disk.
However, the bulge and thin disk chemistry is so disjoint that no relationship 
seems possible between those populations.

When we intercompare indivdiual alpha element raios we find our second major 
result, that the explosive nucleosynthesis alphas (Si, Ca, and Ti) have very 
similar trends with [Fe/H], but O and Mg (produced during hydrostatic burning) 
have unique trends.  We note that Venn et al.'s (2004) review of dwarf galaxy 
abundances also reports differing trends for the explosive and hydrostatic 
alphas (in that case, Mg vs Ca and Ti). 

We find that [O/Fe] declines significantly more steeply than the other alphas 
for [Fe/H]$\geq -0.05$ dex.  We suggest that this is obsevational evidence of 
the reduced oxygen yields due to stellar winds and the Wolf-Rayet phenomenon 
that was predicted by Maeder (1980, 1991).  While MR04 proposed this idea based 
on analysis of 8 bulge stars from this sample, we believe that the oxygen trend 
in the present data requires that this idea be considered seriously.  Maeder's 
predictions suggest that the decline in oxygen yield should e accompanied by an
increase int he yield of carbon; that would provide a test of this idea.  
Another possible test comes from the prediction of \citet{ma00} who propose 
that Wolf-Rayet winds are significant source of flourine (although this is 
debated; see Placis et al. 2005).   Abundances of C and F in the 
Galactic bulge stars could provide a crucial test of the wind hypothesis.

While the bulge trend of [O/Fe] with [Fe/H] is more steep than that of the 
exposive alphas, the rend of [Mg/Fe] with metallicity is quite shallow, with 
[Mg/Fe]$=+0.3$ even for the the most metal-rich bulge stars.  We consider that 
the bulge Mg trend poses a significant challenge for understanding 
nucleosynthesis 
by Type II SNe and the chemical evolution of the Galactic bulge.  Given the 
importance of this question, we have been very careful to check this finding 
(even though it is well known that Mg is enhanced in E and S0 galaxies).  
However, the agreement of the locak disk giant [Mg/Fe] values with the dwarf 
studies and the internal consistency of our abundances from several Mg lines 
makes it very unlikely that our results are in error.  Furthermore, the bulge 
[Al/Mg] trend, which is in excellent agreement with that seen for the thick 
and think disks, removes any lingering doubt about our measured [Mg/Fe] ratios.

To explain the unusual [Mg/$<$SiCaTi$>$] trend in the bulge compared with the 
disk, either more Mg was produced by sources present in the bulge and not in 
the disk, or there are sources of Si, Ca, and Ti in the disk that are less 
common or absent in the bulge.  If excess Mg is produced in the bulge, there 
must be an accompanying source of Al that exactly matches the amount needed to 
maintain the [Al/Mg] trend with [Fe/H] seen in the thick and thin disks.   
If there is an additional source of Mg and Al, it is most unlikely to arise in 
Type~Ia SNe (see Iwamoto et al. 1999);
thus the unusual composition of the bulge must rest with the Type~II SNe.

Considering the dominance of the products of massive star SNe in the bulge 
relative to the disk, we believe that the bulge preserves the true metallicity 
dependent yield of hydrostatic (e.g. O, Mg, Al) versus explosive (Si Ca, Ti) 
elements.  We suggest that the [Mg/SiCaTi] trend in the bulge is due to Mg from 
Type~II SNe, combined with a metallicity dependent decline in the yields
of Si, Ca, and Ti.  Unfortunately, the predicted yields from WW95 are contrary 
to the required enhancement of Mg relative to Si, Ca, and Ti at high 
metallicity.  It is significant that WW95 did not include metal dependent winds 
in their models; these winds result in significant mass loss for metal-rich 
massive stars and deplete the star of mass that would contribute to the 
hydrostatic alphas.

If mass loss in metal-rich massive stars explains the Mg vs O disparity, then 
we may ask why differences between Mg, Si, Ca, and Ti are not so evident in 
the disk.   The solution is to resort 
to the idea that in the disk the Type~Ia SNe produce significant amounts of Si, 
Ca, and Ti; this is expected based on the predictions of \citet{iwa99}.

We do not favor the possibility that massive Type~II SNe progenitors eject Mg 
more efficiently at high metallicity, making them the source of metal-dependent 
Mg enhancements in the bulge.  This would 
require that Type~Ia SNe in the disk produce large amounts of Mg (and Al) 
contrary to theory.

An alternative scenario that might reproduce our unusual [Mg/SiCaTi] trend 
would posit a massive star IMF skewed to more massive stars as [Fe/H] 
increases.  A related alternative requires a sub-population of massive stars 
in the bulge that grows in importance with metallicity.  The difficulties 
presented by this scenario are the non-universal IMF, the unknown reason for 
the metal-dpeendent IMF, and the question of why the bulge abundance patterns 
are seen nowhere else.

Given the difficulties with the second scenario we favor the former idea: 
that the bulge [Mg/$<$SiCaTiI$>$] trend is due to a metallicity-dependent 
decline in the yield of Si, Ca and Ti from very massive Type~II SN 
progenitors, most likely driven by stellar winds.  The mechanism has the 
advantage that it presents fewer discrepancies with the current 
understanding of chemical evolution.

Another major finding is that the trend of [$<$SiCaTiI$>$/Fe] with [Fe/H] 
in the Galactic bulge has a remarkably small dispersion, $\sim$0.05 dex, 
in stark contrast to the scatter of this ratio in Galactic halo stars.  
The result indicates that the bulge evolved with a high degree of 
homogeneity, suggesting that either efficient mixing processes occurred 
during bulge evolution, or that large numbers of SN events completely 
sampled the nucleosynthesis yield function in all bulge locations; this is 
not surprising, given the current bulge infall time of $\sim$10$^6$ years. 
Our abundance results underscore the inhomogeneous evolution of the Galactic
halo as evidence by its abundance pattern, consistent with ideas of accretion 
and low density chemical evolution.

The bulge evolution with efficient mixing, indicated by the high degree of 
homogeneity in [$<$SiCaTiI$>$/Fe] ratios suggests the presence of energetic 
processes; combined with the enhancement of nucleosynthesis products from 
massive stars, this might reasonably be expected to favor the so-called 
ELS scenario \citep{els62}, in which the bulge formed early and 
rapidly through violent relaxation.

We also found that the [$<$SiCaTiI$>$/Fe] trend in the metal-poor bulge 
stars lies at a value equal to the maximum seen in the halo, exceeding the 
mean in the halo by $\sim$0.13 dex; well in excess of the estimated 
maximum systematic uncertainty of 0.04 dex. Indeed the metal-poor bulge 
overlaps with the top $\sim$20\% of halo [$<$SiCaTiI$>$/Fe] ratios. This 
indicates that the metal-poor bulge could not have formed out of Galactic 
halo gas with the present day halo composition.  However, it is possible 
for the metal-poor bulge to have formed from halo gas, as suggested by 
\citet{wg92}, if the halo composition at the onset of bulge 
formation was different than today, characterized by high 
[$<$SiCaTiI$>$/Fe] ratios.  For this to have occurred the mass in halo 
stars must be greater today than when the metal-poor bulge was formed, and 
the onset of bulge formation must have occurred very early in Galactic 
history, at roughly the same time as the halo.

Our plot comparing [Al/Fe] versus [Fe/H] in the Sgr dSph galaxy 
with the Galactic bulge and thin disk shows that [Al/Fe] depends very 
sensitively on environmental parameters.  Therefore, it should provide a 
useful probe for chemical evolution.  Further studies of the Al abundances
in many different populations and environments should be undertaken, both
to exploit this finding and to understand its underlying origin.

Compared to the original study of McWilliam \& Rich (1994) we have increased 
our resolution by roughly a factor of four, and our S/N by a factor of 3.  
Thanks to the Keck telescope and the HIRES spectrograph, abundance analysis of 
Galactic bulge giants can be done on data that is as good as those avaiable 
for the solar vicinity and globular cluster stars.  We have developed new 
approaches that place the iron abundance scale on a far more firm footing than 
before, and undertake a differential analysis relative to Arcturus that to a 
great extent, mitigates systematic errors associated with the imperfect nature 
of the stellar atmosphere modeling.  With these improvements, we have confirmed 
largely the results of MR94, finding that the bulge has strong enanchements of 
Mg, Al, and Na and lesser, but significant enhancements of the other alphas.  
However, we find important new results.  Oxygen and Mg have disjoint abundance 
trends, perhaps requiring that winds in the most metal-rich massive stars 
deplete the hydrostatic burning zones of mass.  The explosive alphas follow a 
different trend from Mg, Na, and Al.  Consideration of the explosive alphas 
shows that the bulge has a far more homogeneous composition than the halo, 
extending even to the most metal-poor bulge members.  The notion of a bulge 
that formed early, and rapidly, remains attractive.  This conclusion is 
consistent with observations of galaxy formation at high redshift.  However, 
it remains vital to study additional elements and larger samples of stars, 
because the fossil record locked in the bulge's composition has the potential 
to provide a detailed record of its history that is unmatched by anything that 
can be inferred from the observations of distant galaxies.

\acknowledgments
We are especially grateful to the staff of Keck
Observatory for their assistance, and S. Vogt and his team for building
HIRES.  We acknowledge support from grant AST-0098612 from the 
National Science Foundation.  RMR acknowleges partial support from grant 
AST-0098739 from the National Science Foundation.
JPF acknowledges support through grants from the W.M. Keck
Foundation and the Gordon and Betty Moore Foundation, to establish
a program of data-intensive science at to the Johns Hopkins University.
For the Arcturus abundance analysis we gratefully
acknowledge partial support from a NASA-SIM Key Project grant, entitled
{\it ``Anchoring the Population II Distance Scale: Accurate Ages for Globular 
Clusters and Field Halo Stars''}.    The authors acknowledge the cultural role 
that the summit of Mauna Kea has had within the indigenous Hawaiian community.
We are fortunate to have the opportunity to conduct observations from this
mountain.  This publication makes use of data products from the Two Micron 
All Sky Survey, which is a joint project of the University of Massachusetts
and the Infrared Processing and Analysis Center/California Institute of
Technology, funded by the National Aeronautics and Space Administration  
and the National Science Foundation.  This research has also made use of the
SIMBAD database, operated at CDS, Strasbourg, France.


\begin{deluxetable}{cccrr}
\tablenum{1}
\tablewidth{0pt}
\tablecaption{Arcturus-Sun Differential Line List}
\tablehead{
\colhead{Ion} & \colhead{Wavelength} & \colhead{E.P.} & \colhead{EW $\alpha$ Boo
} & \colhead{EW Sun} \\
 & \colhead{\AA} & \colhead{eV} & \colhead{m\AA} & \colhead{m\AA} \\ }
\startdata
[O I] &  6300.30 & 0.00	&  68.4	&   5.0 \\
Na I  &  6160.75 & 2.10	&  90.6	&  55.0 \\
Mg I  &  6318.72 & 5.10	&  68.3	&  39.4 \\
Mg I  &  7387.69 & 5.75	&  80.4	&  77.1 \\
Mg I  &  8473.69 & 5.93	&  21.6	&  13.0 \\
Mg I  &  8923.57 & 5.39	&  76.3	&  57.3 \\
Mg I  &  8997.15 & 5.93	&  26.4	&  20.6 \\
Al I  &  6698.67 & 3.14	&  60.9	&  21.0 \\
Al I  &  7835.31 & 4.02	&  64.7	&  41.1 \\
Al I  &  8772.86 & 4.02	&  97.0	&  78.2 \\
Si I  &  5488.98 & 5.61	&  23.5	&  22.9 \\
Si I  &  5701.10 & 4.93	&  49.5	&  38.5 \\
Si I  &  6125.02 & 5.61	&  31.5	&  32.4 \\
Si I  &  6142.48 & 5.61	&  33.1	&  33.7 \\
Si I  &  6155.13 & 5.61	&  75.0	&  89.3 \\
Si I  &  6555.46 & 5.98	&  34.4	&  45.2 \\
Si I  &  6583.71 & 5.95	&  16.3	&  17.0 \\
Si I  &  7226.21 & 5.61	&  39.4	&  38.5 \\
Si I  &  7250.63 & 5.61	&  52.4	&  64.5 \\
Si I  &  7680.27 & 5.86	&  65.6	&  87.2 \\
Si I  &  7932.35 & 5.96	&  69.3	&  99.7 \\
Si I  &  8443.97 & 5.87	&  27.4	&  31.5 \\
Si I  &  8728.01 & 6.18	&  56.8	&  93.5 \\
Si I  &  8892.72 & 5.98	&  51.5	&  67.2 \\
Ca I  &  5867.56 & 2.93	&  63.0	&  22.2 \\
Ca I  &  6156.02 & 2.52	&  42.0	&   8.5 \\
Ti I  &  5043.58 & 0.84	& 113.8	&  16.7 \\
Ti I  &  5062.10 & 2.16	&  86.9	&  15.9 \\
Ti I  &  5295.77 & 1.07	& 104.3	&  12.7 \\
Ti I  &  5453.64 & 1.44	&  73.3	&   5.1 \\
Ti I  &  5471.19 & 1.44	&  85.7	&   7.9 \\
Ti I  &  5648.56 & 2.49	&  65.7	&  10.3 \\
Ti I  &  5679.92 & 2.47	&  47.8	&   5.0 \\
Ti I  &  5739.47 & 2.25	&  63.4	&   8.0 \\
Ti I  &  5766.36 & 3.29	&  41.3	&   8.9 \\
Ti I  &  5978.54 & 1.87	& 108.0	&  23.2 \\
Ti I  &  6064.63 & 1.05	& 103.0	&   8.8 \\
Ti I  &  6303.76 & 1.44	&  88.2	&   8.1 \\
Ti I  &  6312.24 & 1.46	&  89.2	&   8.1 \\
Ti I  &  6554.22 & 1.44	& 115.2	&  16.5 \\
Ti I  &  6599.10 & 0.90	& 115.1	&   9.9 \\
Ti I  &  7138.91 & 1.44	&  91.4	&   7.1 \\
Ti I  &  7440.58 & 2.25	&  61.3	&   6.4 \\
Ti II &  5013.68 & 1.58	& 103.4	&  51.5 \\
Ti II &  5185.91 & 1.89	& 113.3	&  65.7 \\
Ti II &  5396.23 & 1.58	&  44.1	&  10.6 \\
Ti II &  5418.75 & 1.58	&  99.7	&  48.8 \\
Ti II &  5492.86 & 1.58	&  47.9	&  11.0 \\
Ti II &  6606.95 & 2.06	&  32.4	&   8.3 \\
Ti II &  8979.19 & 2.59	&  48.9	&  20.1 \\
\enddata
\end{deluxetable}

\begin{deluxetable}{rccrccc}
\tablenum{2}
\tablewidth{0pt}
\tablecaption{Adopted Arcturus Abundances}
\tablehead{
\colhead{Element} & \colhead{Diff. $\alpha$Boo $-$ Sun} & \colhead{$\sigma$} & \colhead{N$_{\rm lines}$} & \colhead{Solar\tablenotemark{a} log$\epsilon$} & \colhead{$\alpha$ Boo log$\epsilon$}  & \colhead{[M/Fe]\rlap{$_{\rm \alpha Boo}$}}
}
\startdata
O  & $-0.02$ & \nodata &  1 & 8.69 & 8.67 &  $+$0.48 \\
Na & $-0.41$ & \nodata &  1 & 6.30 & 5.89 &  $+$0.09 \\
Mg & $-0.11$ &    0.06 &  5 & 7.55 & 7.44 &  $+$0.39 \\
Al & $-0.12$ &    0.03 &  3 & 6.46 & 6.34 &  $+$0.38 \\
Si & $-0.15$ &    0.05 & 15 & 7.54 & 7.39 &  $+$0.35 \\
Ca & $-0.29$ &    0.01 &  2 & 6.34 & 6.05 &  $+$0.21 \\
Ti\tablenotemark{b} & $-0.24$ &    0.04 & 24 & 4.92 & 4.68 & $+$0.26 \\
\enddata
\tablenotetext{a}{From Lodders 2003.}
\tablenotetext{b}{Includes both Ti I lines (N = 17) and Ti II (N = 7) lines.}
\end{deluxetable}

\begin{deluxetable}{rrcr}
\tablenum{3}
\tablewidth{0pt}
\tablecaption{Line List}
\tablehead{
 \colhead{Ion} & \colhead{Wavelength} & \colhead{E. P.} & \colhead{E. W. $\alpha$ Boo} \\
 & \colhead{\AA} & \colhead{eV} & \colhead{m\AA} }
\startdata
{\rm [O I]} & 5577.34 & 1.97 &  12 \\
{\rm [O I]} & 6300.31 & 0.00 &  67 \\
{\rm [O I]} & 6363.79 & 0.02 &  30 \\
 Na I & 5682.63 & 2.10 & 119 \\
 Na I & 5688.20 & 2.10 & 150 \\
 Na I & 6154.23 & 2.10 &  73 \\
 Na I & 6160.75 & 2.10 &  95 \\
 Mg I & 5711.09 & 4.34 & 153 \\
 Mg I & 6318.72 & 5.10 &  70 \\
 Mg I & 6319.24 & 5.10 &  53 \\
 Mg I & 6319.49 & 5.10 &  27 \\
 Mg I & 6765.45 & 5.75 &  25 \\
 Mg I & 6799.00 & 5.75 &  34 \\
 Mg I & 6841.08 & 5.75 &  41 \\
 Mg I & 6894.92 & 5.75 &  43 \\
 Mg I & 6965.41 & 5.75 &  54 \\
 Mg I & 7387.69 & 5.75 &  83 \\
 Mg I & 7875.43 & 5.93 &  14 \\
 Al I & 5557.06 & 3.14 &  26 \\
 Al I & 6696.02 & 3.14 &  88 \\
 Al I & 6696.79 & 4.02 &  21 \\
 Al I & 6698.67 & 3.14 &  60 \\
 Al I & 7835.31 & 4.02 &  66 \\
 Al I & 7836.13 & 4.02 &  79 \\
 Si I & 5488.98 & 5.61 &  24 \\
 Si I & 5517.53 & 5.08 &  25 \\
 Si I & 5701.10 & 4.93 &  50 \\
 Si I & 6142.48 & 5.61 &  32 \\
 Si I & 6145.02 & 5.61 &  38 \\
 Si I & 6155.13 & 5.61 &  76 \\
 Si I & 6635.69 & 5.86 &  11 \\
 Si I & 7235.33 & 5.61 &  33 \\
 Si I & 7235.82 & 5.61 &  22 \\
 Si I & 7250.63 & 5.61 &  52 \\
 Si I & 7423.50 & 5.61 &  92 \\
 Si I & 7800.00 & 6.18 &  38 \\
 Ca I & 5512.98 & 2.93 & 122 \\
 Ca I & 5590.11 & 2.52 & 140 \\
 Ca I & 5867.56 & 2.93 &  60 \\
 Ca I & 6156.02 & 2.52 &  38 \\
 Ca I & 6161.30 & 2.52 & 111 \\
 Ca I & 6166.44 & 2.52 & 122 \\
 Ca I & 6169.04 & 2.52 & 146 \\
 Ca I & 6455.60 & 2.52 & 113 \\
 Ca I & 6471.66 & 2.52 & 150 \\
 Ca I & 6499.65 & 2.52 & 143 \\
 Ca I & 6798.48 & 2.71 &  27 \\
 Ca I & 7695.14 & 4.44 &   4 \\
 Ti I & 5453.64 & 1.44 &  74 \\
 Ti I & 5618.35 & 1.50 &  19 \\
 Ti I & 5648.56 & 2.49 &  66 \\
 Ti I & 5739.47 & 2.25 &  61 \\
 Ti I & 5766.36 & 3.29 &  41 \\
 Ti I & 5913.73 & 0.02 &  44 \\
 Ti I & 5918.53 & 1.07 & 111 \\
 Ti I & 5944.66 & 0.00 &  61 \\
 Ti I & 6092.79 & 1.89 &  54 \\
 Ti I & 6273.39 & 0.02 &  42 \\
 Ti I & 6706.29 & 1.50 &  11 \\
 Ti I & 6716.67 & 2.49 &  26 \\
 Ti I & 6746.33 & 1.89 &  19 \\
 Ti I & 7138.07 & 1.43 &  16 \\
 Ti I & 7271.51 & 1.44 &  59 \\
 Ti I & 7352.12 & 2.49 &  28 \\
 Ti I & 7391.51 & 1.50 &  20 \\
 Ti I & 7738.96 & 1.46 &  10 \\
Ti II & 5418.75 & 1.58 &  99 \\
Ti II & 5492.86 & 1.58 &  48 \\
Ti II & 6606.95 & 2.06 &  32 \\
Ti II & 7214.72 & 2.59 &  54 \\
\enddata
\end{deluxetable}

\LongTables
\begin{deluxetable}{rrcccccccc}
\tablenum{5}
\tablewidth{0pt}
\tablecaption{Final Parameters}
\tablehead{
\colhead{Name} & \colhead{Grid} & \colhead{T$_{\rm eff}$} & \colhead{$\sigma$T} & \colhead{$\log{g}$} & \colhead{$\sigma\log{g}$} & \colhead{[m/H]} & \colhead{$\sigma$[m/H]} & \colhead{v$_{\rm t}$} &\colhead{$\sigma$v$_{\rm t}$}  \\
 & & \colhead{K} & \colhead{K} & & & & & \colhead{km s$^{-1}$} & \colhead{km s$^{-1}$} }
\startdata
\multicolumn{10}{c}{BAADE's WINDOW BULGE STARS} \\
  I-012&  Kurucz & 4257 & 49 & 1.55 & 0.10 &$-0.37$& 0.08 & 1.54 & 0.03 \\
       &  ODFNEW & 4248 & 42 & 1.54 & 0.10 &$-0.39$& 0.08 & 1.64 & 0.04 \\
       & AODFNEW & 4246 & 42 & 1.54 & 0.10 &$-0.39$& 0.08 & 1.64 & 0.03 \\
  \\
  I-025&  Kurucz & 4340 & 49 & 2.02 & 0.10 &$+0.51$& 0.09 & 1.62 & 0.05 \\
       &  ODFNEW & 4323 & 38 & 2.01 & 0.10 &$+0.47$& 0.09 & 1.69 & 0.05 \\
       & AODFNEW & 4303 & 41 & 2.00 & 0.10 &$+0.49$& 0.10 & 1.75 & 0.04 \\
  \\
  I-039&  Kurucz & 4386 & 49 & 2.13 & 0.10 &$+0.50$& 0.09 & 1.47 & 0.06 \\
       &  ODFNEW & 4364 & 60 & 2.11 & 0.10 &$+0.45$& 0.09 & 1.55 & 0.03 \\
       & AODFNEW & 4351 & 70 & 2.10 & 0.10 &$+0.48$& 0.09 & 1.60 & 0.05 \\
  \\
  I-141&  Kurucz & 4335 & 49 & 1.68 & 0.10 &$-0.27$& 0.09 & 1.33 & 0.05 \\
       &  ODFNEW & 4331 & 41 & 1.68 & 0.10 &$-0.30$& 0.10 & 1.43 & 0.06 \\
       & AODFNEW & 4330 & 49 & 1.68 & 0.10 &$-0.30$& 0.10 & 1.44 & 0.05 \\
  \\
  I-151&  Kurucz & 4405 & 49 & 1.70 & 0.09 &$-0.77$& 0.08 & 1.22 & 0.06 \\
       &  ODFNEW & 4406 & 38 & 1.70 & 0.09 &$-0.77$& 0.08 & 1.35 & 0.06 \\
       & AODFNEW & 4407 & 41 & 1.70 & 0.09 &$-0.80$& 0.09 & 1.34 & 0.04 \\
  \\
  I-152&  Kurucz & 4646 & 49 & 2.20 & 0.09 &$-0.01$& 0.10 & 1.22 & 0.06 \\
       &  ODFNEW & 4632 & 38 & 2.20 & 0.09 &$-0.02$& 0.10 & 1.30 & 0.05 \\
       & AODFNEW & 4635 & 41 & 2.20 & 0.09 &$-0.03$& 0.10 & 1.31 & 0.05 \\
  \\
  I-156&  Kurucz & 4312 & 49 & 1.74 & 0.10 &$-0.71$& 0.07 & 1.16 & 0.05 \\
       &  ODFNEW & 4314 & 38 & 1.75 & 0.10 &$-0.71$& 0.07 & 1.29 & 0.05 \\
       & AODFNEW & 4315 & 49 & 1.75 & 0.10 &$-0.72$& 0.08 & 1.29 & 0.05 \\
  \\
  I-158&  Kurucz & 4349 & 49 & 1.93 & 0.10 &$-0.20$& 0.10 & 1.27 & 0.06 \\
       &  ODFNEW & 4338 & 38 & 1.93 & 0.10 &$-0.22$& 0.10 & 1.37 & 0.05 \\
       & AODFNEW & 4335 & 44 & 1.92 & 0.10 &$-0.22$& 0.10 & 1.39 & 0.05 \\
  \\
  I-194&  Kurucz & 4176 & 49 & 1.65 & 0.10 &$-0.25$& 0.10 & 1.34 & 0.05 \\
       &  ODFNEW & 4167 & 43 & 1.64 & 0.10 &$-0.28$& 0.10 & 1.44 & 0.05 \\
       & AODFNEW & 4157 & 49 & 1.63 & 0.10 &$-0.25$& 0.10 & 1.46 & 0.03 \\
  \\
  I-202&  Kurucz & 4184 & 49 & 1.53 & 0.10 &$+0.16$& 0.11 & 1.12 & 0.04 \\
       &  ODFNEW & 4164 & 47 & 1.51 & 0.10 &$+0.10$& 0.11 & 1.20 & 0.04 \\
       & AODFNEW & 4148 & 92 & 1.50 & 0.10 &$+0.12$& 0.11 & 1.23 & 0.04 \\
  \\
  I-264&  Kurucz & 4097 & 49 & 0.87 & 0.10 &$-1.15$& 0.08 & 1.67 & 0.10 \\
       &  ODFNEW & 4095 & 38 & 0.87 & 0.10 &$-1.14$& 0.08 & 1.85 & 0.05 \\
       & AODFNEW & 4100 & 41 & 0.88 & 0.10 &$-1.16$& 0.09 & 1.82 & 0.05 \\
  \\
  I-322&  Kurucz & 4106 & 49 & 0.89 & 0.11 &$-0.25$& 0.09 & 1.63 & 0.04 \\
       &  ODFNEW & 4100 & 38 & 0.88 & 0.11 &$-0.28$& 0.08 & 1.73 & 0.04 \\
       & AODFNEW & 4091 & 41 & 0.87 & 0.11 &$-0.27$& 0.09 & 1.74 & 0.04 \\
  \\
 II-033&  Kurucz & 4277 & 49 & 1.41 & 0.10 &$-0.75$& 0.07 & 1.41 & 0.03 \\
       &  ODFNEW & 4285 & 38 & 1.42 & 0.10 &$-0.75$& 0.07 & 1.55 & 0.03 \\
       & AODFNEW & 4290 & 41 & 1.42 & 0.10 &$-0.76$& 0.07 & 1.53 & 0.03 \\
  \\
 II-119&  Kurucz & 4554 & 57 & 1.66 & 0.10 &$-1.22$& 0.10 & 1.24 & 0.13 \\
       &  ODFNEW & 4560 & 51 & 1.67 & 0.10 &$-1.21$& 0.10 & 1.43 & 0.12 \\
       & AODFNEW & 4565 & 74 & 1.67 & 0.10 &$-1.26$& 0.10 & 1.37 & 0.10 \\
  \\
 II-154&  Kurucz & 4650 & 49 & 2.03 & 0.09 &$-0.61$& 0.08 & 1.00 & 0.05 \\
       &  ODFNEW & 4660 & 38 & 2.03 & 0.09 &$-0.60$& 0.08 & 1.11 & 0.05 \\
       & AODFNEW & 4661 & 45 & 2.03 & 0.09 &$-0.64$& 0.08 & 1.10 & 0.05 \\
  \\
 II-172&  Kurucz & 4480 & 49 & 2.14 & 0.09 &$-0.30$& 0.11 & 1.03 & 0.05 \\
       &  ODFNEW & 4491 & 38 & 2.15 & 0.09 &$-0.31$& 0.12 & 1.17 & 0.03 \\
       & AODFNEW & 4478 & 41 & 2.14 & 0.09 &$-0.32$& 0.12 & 1.19 & 0.04 \\
  \\
III-152&  Kurucz & 4157 & 49 & 1.58 & 0.10 &$-0.41$& 0.08 & 1.21 & 0.04 \\
       &  ODFNEW & 4152 & 38 & 1.58 & 0.10 &$-0.42$& 0.08 & 1.33 & 0.03 \\
       & AODFNEW & 4159 & 41 & 1.59 & 0.10 &$-0.41$& 0.08 & 1.36 & 0.04 \\
  \\
III-220&  Kurucz & 4550 & 49 & 1.99 & 0.09 &$-0.31$& 0.08 & 1.27 & 0.01 \\
       &  ODFNEW & 4543 & 38 & 1.99 & 0.09 &$-0.32$& 0.08 & 1.35 & 0.05 \\
       & AODFNEW & 4556 & 41 & 2.00 & 0.09 &$-0.34$& 0.08 & 1.35 & 0.04 \\
  \\
  \\
 IV-003&  Kurucz & 4433 & 49 & 1.41 & 0.09 &$-1.29$& 0.07 & 1.35 & 0.11 \\
       &  ODFNEW & 4435 & 38 & 1.42 & 0.09 &$-1.28$& 0.07 & 1.59 & 0.11 \\
       & AODFNEW & 4438 & 41 & 1.42 & 0.09 &$-1.33$& 0.07 & 1.50 & 0.11 \\
  \\
 IV-047&  Kurucz & 4556 & 49 & 2.28 & 0.09 &$-0.40$& 0.11 & 1.54 & 0.06 \\
       &  ODFNEW & 4559 & 38 & 2.28 & 0.09 &$-0.40$& 0.10 & 1.65 & 0.07 \\
       & AODFNEW & 4569 & 41 & 2.29 & 0.09 &$-0.40$& 0.11 & 1.66 & 0.06 \\
  \\
 IV-072&  Kurucz & 4272 & 49 & 1.78 & 0.09 &$+0.26$& 0.10 & 1.31 & 0.05 \\
       &  ODFNEW & 4256 & 38 & 1.77 & 0.09 &$+0.20$& 0.10 & 1.40 & 0.04 \\
       & AODFNEW & 4229 & 41 & 1.75 & 0.09 &$+0.22$& 0.10 & 1.43 & 0.05 \\
  \\
 IV-167&  Kurucz & 4301 & 49 & 2.07 & 0.10 &$+0.46$& 0.10 & 1.38 & 0.05 \\
       &  ODFNEW & 4279 & 38 & 2.05 & 0.10 &$+0.40$& 0.10 & 1.47 & 0.03 \\
       & AODFNEW & 4266 & 41 & 2.04 & 0.10 &$+0.44$& 0.11 & 1.51 & 0.05 \\
  \\
 IV-203&  Kurucz & 3902 & 59 & 0.51 & 0.12 &$-1.29$& 0.07 & 1.88 & 0.09 \\
       &  ODFNEW & 3902 & 66 & 0.51 & 0.12 &$-1.26$& 0.07 & 1.78 & 0.07 \\
       & AODFNEW & 3895 & 67 & 0.50 & 0.12 &$-1.28$& 0.04 & 2.00 & 0.04 \\
  \\
 IV-325&  Kurucz & 4289 & 49 & 2.05 & 0.10 &$+0.28$& 0.09 & 1.69 & 0.05 \\
       &  ODFNEW & 4269 & 38 & 2.05 & 0.10 &$+0.23$& 0.09 & 1.78 & 0.05 \\
       & AODFNEW & 4264 & 41 & 2.04 & 0.10 &$+0.26$& 0.09 & 1.82 & 0.04 \\
  \\
 IV-329&  Kurucz & 4197 & 49 & 1.29 & 0.10 &$-0.94$& 0.06 & 1.47 & 0.04 \\
       &  ODFNEW & 4197 & 38 & 1.29 & 0.10 &$-0.94$& 0.06 & 1.62 & 0.04 \\
       & AODFNEW & 4195 & 41 & 1.29 & 0.10 &$-0.95$& 0.06 & 1.59 & 0.02 \\
  \\
\multicolumn{10}{c}{BAADE's WINDOW NON-BULGE STARS} \\
 II-122&  Kurucz & 3912 & 59 & 0.14 & 0.12 &$-0.79$& 0.09 & 1.53 & 0.05 \\
       &  ODFNEW & 3902 & 66 & 0.13 & 0.12 &$-0.79$& 0.09 & 1.62 & 0.05 \\
       & AODFNEW & 3896 & 53 & 0.13 & 0.12 &$-0.79$& 0.09 & 1.61 & 0.05 \\
  \\
 IV-025&  Kurucz & 4614 & 66 & 2.64 & 0.10 &$+0.21$& 0.11 & 1.12 & 0.07 \\
       &  ODFNEW & 4598 & 74 & 2.63 & 0.10 &$+0.18$& 0.11 & 1.22 & 0.04 \\
       & AODFNEW & 4604 & 71 & 2.64 & 0.10 &$+0.19$& 0.11 & 1.28 & 0.06 \\
  \\
\multicolumn{10}{c}{LOCAL DISK STARS} \\
HR1184 &  Kurucz & 4753 & 49 & 2.27 & 0.09 &$-0.38$& 0.09 & 1.39 & 0.10 \\
       &  ODFNEW & 4739 & 41 & 2.26 & 0.09 &$-0.39$& 0.10 & 1.50 & 0.06 \\
       & AODFNEW & 4742 & 53 & 2.27 & 0.09 &$-0.42$& 0.10 & 1.50 & 0.09 \\
  \\
HR1346 &  Kurucz & 4823 & 50 & 2.43 & 0.08 &$+0.15$& 0.09 & 1.57 & 0.04 \\
       &  ODFNEW & 4837 & 38 & 2.44 & 0.08 &$+0.17$& 0.09 & 1.64 & 0.04 \\
       & AODFNEW & 4840 & 41 & 2.44 & 0.08 &$+0.14$& 0.09 & 1.66 & 0.04 \\
  \\
HR1348 &  Kurucz & 4409 & 49 & 1.96 & 0.08 &$-0.36$& 0.09 & 1.56 & 0.04 \\
       &  ODFNEW & 4396 & 52 & 1.95 & 0.08 &$-0.38$& 0.09 & 1.66 & 0.04 \\
       & AODFNEW & 4394 & 51 & 1.95 & 0.08 &$-0.38$& 0.09 & 1.67 & 0.04 \\
  \\
HR1409 &  Kurucz & 4838 & 49 & 2.52 & 0.08 &$+0.20$& 0.09 & 1.63 & 0.04 \\
       &  ODFNEW & 4846 & 46 & 2.52 & 0.08 &$+0.21$& 0.09 & 1.70 & 0.04 \\
       & AODFNEW & 4857 & 44 & 2.53 & 0.08 &$+0.19$& 0.09 & 1.73 & 0.04 \\
  \\
HR1411 &  Kurucz & 4961 & 63 & 2.69 & 0.08 &$+0.17$& 0.08 & 1.48 & 0.05 \\
       &  ODFNEW & 4983 & 62 & 2.70 & 0.08 &$+0.19$& 0.08 & 1.57 & 0.05 \\
       & AODFNEW & 4986 & 42 & 2.70 & 0.08 &$+0.16$& 0.08 & 1.59 & 0.05 \\
  \\
HR1585 &  Kurucz & 4333 & 49 & 1.63 & 0.09 &$-0.35$& 0.08 & 1.63 & 0.04 \\
       &  ODFNEW & 4330 & 44 & 1.63 & 0.09 &$-0.35$& 0.08 & 1.75 & 0.06 \\
       & AODFNEW & 4329 & 47 & 1.63 & 0.09 &$-0.36$& 0.09 & 1.77 & 0.04 \\
  \\
HR2035 &  Kurucz & 4624 & 49 & 2.32 & 0.09 &$-0.64$& 0.07 & 1.15 & 0.05 \\
       &  ODFNEW & 4614 & 42 & 2.31 & 0.09 &$-0.65$& 0.08 & 1.29 & 0.05 \\
       & AODFNEW & 4623 & 41 & 2.32 & 0.09 &$-0.66$& 0.08 & 1.28 & 0.05 \\
  \\
HR2113 &  Kurucz & 4239 & 49 & 1.29 & 0.09 &$-0.66$& 0.07 & 1.57 & 0.03 \\
       &  ODFNEW & 4226 & 68 & 1.28 & 0.09 &$-0.67$& 0.07 & 1.70 & 0.03 \\
       & AODFNEW & 4224 & 82 & 1.28 & 0.09 &$-0.67$& 0.07 & 1.70 & 0.03 \\
  \\
HR2443 &  Kurucz & 4429 & 49 & 1.50 & 0.08 &$-0.21$& 0.07 & 1.61 & 0.03 \\
       &  ODFNEW & 4422 & 43 & 1.49 & 0.08 &$-0.22$& 0.07 & 1.63 & 0.03 \\
       & AODFNEW & 4419 & 45 & 1.49 & 0.08 &$-0.23$& 0.07 & 1.70 & 0.03 \\
  \\
  \\
HR3418 &  Kurucz & 4470 & 70 & 1.91 & 0.08 &$+0.16$& 0.08 & 1.77 & 0.04 \\
       &  ODFNEW & 4459 & 70 & 1.90 & 0.08 &$+0.15$& 0.08 & 1.81 & 0.04 \\
       & AODFNEW & 4452 & 64 & 1.89 & 0.08 &$+0.15$& 0.08 & 1.85 & 0.03 \\
  \\
HR3733 &  Kurucz & 5001 & 50 & 2.89 & 0.08 &$-0.12$& 0.08 & 1.50 & 0.01 \\
       &  ODFNEW & 5019 & 57 & 2.89 & 0.08 &$-0.10$& 0.08 & 1.63 & 0.05 \\
       & AODFNEW & 5028 & 41 & 2.90 & 0.08 &$-0.13$& 0.08 & 1.65 & 0.05 \\
  \\
HR3905 &  Kurucz & 4531 & 49 & 2.34 & 0.08 &$+0.22$& 0.09 & 1.50 & 0.01 \\
       &  ODFNEW & 4514 & 38 & 2.33 & 0.08 &$+0.32$& 0.09 & 1.48 & 0.05 \\
       & AODFNEW & 4517 & 41 & 2.33 & 0.08 &$+0.34$& 0.09 & 1.53 & 0.04 \\
  \\
HR4104 &  Kurucz & 4046 & 49 & 1.07 & 0.10 &$-0.36$& 0.10 & 1.87 & 0.05 \\
       &  ODFNEW & 4026 & 38 & 1.05 & 0.10 &$-0.39$& 0.10 & 1.98 & 0.05 \\
       & AODFNEW & 4025 & 41 & 1.05 & 0.10 &$-0.37$& 0.10 & 2.00 & 0.05 \\
  \\
HR4382 &  Kurucz & 4475 & 72 & 1.72 & 0.10 &$-0.54$& 0.10 & 1.55 & 0.05 \\
       &  ODFNEW & 4458 & 64 & 1.71 & 0.10 &$-0.59$& 0.10 & 1.66 & 0.05 \\
       & AODFNEW & 4455 & 64 & 1.71 & 0.10 &$-0.57$& 0.10 & 1.67 & 0.05 \\
  \\
HR4450 &  Kurucz & 4975 & 51 & 2.53 & 0.08 &$+0.13$& 0.10 & 1.39 & 0.07 \\
       &  ODFNEW & 4963 & 83 & 2.53 & 0.08 &$+0.14$& 0.10 & 1.46 & 0.05 \\
       & AODFNEW & 4970 & 93 & 2.53 & 0.08 &$+0.10$& 0.10 & 1.47 & 0.04 \\
  \\
HR4608 &  Kurucz & 4853 & 49 & 2.46 & 0.08 &$-0.41$& 0.09 & 1.44 & 0.06 \\
       &  ODFNEW & 4864 & 38 & 2.47 & 0.08 &$-0.39$& 0.09 & 1.57 & 0.06 \\
       & AODFNEW & 4869 & 41 & 2.47 & 0.08 &$-0.43$& 0.09 & 1.55 & 0.06 \\
  \\
HR5340 &  Kurucz & 4283 & 49 & 1.55 & 0.10 &$-0.50$& 0.07 & 1.61 & 0.03 \\
       &  ODFNEW & 4292 & 38 & 1.56 & 0.10 &$-0.58$& 0.07 & 1.62 & 0.02 \\
       & AODFNEW & 4285 & 42 & 1.55 & 0.10 &$-0.54$& 0.07 & 1.62 & 0.02 \\
\enddata
\end{deluxetable}

\clearpage
\LongTables
\begin{landscape}
\begin{deluxetable}{rrcccccccccccccccccccc}
\tablenum{6}
\tablewidth{0pt}
\tablecaption{Derived Abundances}
\tablehead{
\colhead{Name} & \colhead{Grid} & \colhead{[FeI/H]} & \colhead{$\sigma$} & \colhead{[FeII/H]} & \colhead{$\sigma$} & \colhead{[O/Fe]} & \colhead{$\sigma$} & \colhead{[Na/Fe]} & \colhead{$\sigma$} & \colhead{[Mg/Fe]} & \colhead{$\sigma$} & \colhead{[Al/Fe]} & \colhead{$\sigma$} & \colhead{[Si/Fe]} & \colhead{$\sigma$} & \colhead{[Ca/Fe]} & \colhead{$\sigma$} & \colhead{[TiI/Fe]} & \colhead{$\sigma$} & \colhead{[TiII/Fe]} & \colhead{$\sigma$} }
\startdata
\multicolumn{21}{c}{BAADE's WINDOW BULGE STARS} \\
I-012   &   ODFNEW &$-0.38$& 0.07 &$-0.36$& 0.12 &$+0.36$& 0.15 &$+0.10$& 0.15 &$+0.39$& 0.08 &$+0.34$& 0.09 &$+0.36$& 0.09 &$+0.17$& 0.11 &$+0.28$& 0.10 &$+0.14$& 0.14\\ 
        &  AODFNEW &$-0.38$& 0.08 &$-0.35$& 0.12 &$+0.47$& 0.15 &$+0.15$& 0.13 &$+0.44$& 0.08 &$+0.35$& 0.09 &$+0.45$& 0.09 &$+0.20$& 0.10 &$+0.32$& 0.10 &$+0.26$& 0.14\\ 
        &    Final &$-0.38$& 0.08 &$-0.35$& 0.12 &$+0.48$& 0.15 &$+0.16$& 0.14 &$+0.45$& 0.08 &$+0.35$& 0.09 &$+0.46$& 0.09 &$+0.20$& 0.11 &$+0.32$& 0.10 &$+0.27$& 0.14\\ 
I-025   &   ODFNEW &$+0.47$& 0.09 &$+0.47$& 0.14 &$-0.09$& 0.17 &$+0.16$& 0.16 &$+0.24$& 0.12 &$+0.32$& 0.12 &$+0.10$& 0.10 &$+0.07$& 0.11 &$-0.12$& 0.11 &$-0.05$& 0.14\\ 
        &  AODFNEW &$+0.48$& 0.09 &$+0.49$& 0.15 &$-0.01$& 0.18 &$+0.24$& 0.15 &$+0.29$& 0.12 &$+0.35$& 0.12 &$+0.20$& 0.10 &$+0.10$& 0.11 &$-0.07$& 0.11 &$+0.07$& 0.15\\ 
        &    Final &$+0.48$& 0.09 &$+0.48$& 0.15 &$-0.04$& 0.17 &$+0.21$& 0.15 &$+0.27$& 0.12 &$+0.34$& 0.12 &$+0.17$& 0.10 &$+0.09$& 0.11 &$-0.09$& 0.11 &$+0.03$& 0.15\\ 
I-039   &   ODFNEW &$+0.46$& 0.07 &$+0.39$& 0.13 &$+0.19$& 0.16 &$+0.20$& 0.07 &$+0.24$& 0.11 &$+0.42$& 0.14 &$+0.17$& 0.11 &$-0.01$& 0.12 &$+0.04$& 0.11 &$+0.30$& 0.22\\ 
        &  AODFNEW &$+0.48$& 0.08 &$+0.42$& 0.15 &$+0.26$& 0.17 &$+0.27$& 0.08 &$+0.29$& 0.13 &$+0.47$& 0.12 &$+0.28$& 0.12 &$+0.03$& 0.11 &$+0.09$& 0.12 &$+0.41$& 0.23\\ 
        &    Final &$+0.47$& 0.08 &$+0.41$& 0.14 &$+0.24$& 0.17 &$+0.25$& 0.08 &$+0.27$& 0.12 &$+0.45$& 0.13 &$+0.25$& 0.11 &$+0.02$& 0.11 &$+0.07$& 0.11 &$+0.38$& 0.23\\ 
I-141   &   ODFNEW &$-0.29$& 0.06 &$-0.31$& 0.11 &$+0.33$& 0.13 &$+0.16$& 0.08 &$+0.37$& 0.08 &$+0.42$& 0.10 &$+0.34$& 0.07 &$+0.25$& 0.09 &$+0.30$& 0.09 &$+0.22$& 0.14\\ 
        &  AODFNEW &$-0.30$& 0.06 &$-0.31$& 0.12 &$+0.45$& 0.14 &$+0.22$& 0.10 &$+0.42$& 0.08 &$+0.44$& 0.10 &$+0.44$& 0.07 &$+0.30$& 0.11 &$+0.34$& 0.10 &$+0.34$& 0.16\\ 
        &    Final &$-0.30$& 0.06 &$-0.31$& 0.11 &$+0.46$& 0.14 &$+0.22$& 0.09 &$+0.42$& 0.08 &$+0.44$& 0.10 &$+0.45$& 0.07 &$+0.30$& 0.10 &$+0.34$& 0.10 &$+0.35$& 0.15\\ 
I-151   &   ODFNEW &$-0.76$& 0.07 &$-0.79$& 0.09 &$+0.61$& 0.12 &$+0.09$& 0.07 &$+0.44$& 0.13 &$+0.42$& 0.07 &$+0.39$& 0.09 &$+0.30$& 0.12 &$+0.34$& 0.10 &$+0.29$& 0.14\\ 
        &  AODFNEW &$-0.79$& 0.07 &$-0.82$& 0.10 &$+0.75$& 0.13 &$+0.15$& 0.07 &$+0.50$& 0.12 &$+0.45$& 0.07 &$+0.49$& 0.09 &$+0.36$& 0.12 &$+0.38$& 0.10 &$+0.44$& 0.14\\ 
        &    Final &$-0.80$& 0.07 &$-0.83$& 0.10 &$+0.79$& 0.12 &$+0.17$& 0.07 &$+0.52$& 0.12 &$+0.46$& 0.07 &$+0.52$& 0.09 &$+0.38$& 0.12 &$+0.39$& 0.10 &$+0.48$& 0.14\\ 
I-152   &   ODFNEW &$-0.01$& 0.09 &$-0.04$& 0.11 &$+0.41$& 0.15 &$+0.14$& 0.15 &$+0.24$& 0.12 &$+0.37$& 0.16 &$+0.16$& 0.13 &$+0.30$& 0.10 &$+0.29$& 0.15 &$+0.26$& 0.13\\ 
        &  AODFNEW &$-0.04$& 0.09 &$-0.06$& 0.12 &$+0.53$& 0.16 &$+0.22$& 0.14 &$+0.30$& 0.13 &$+0.42$& 0.16 &$+0.28$& 0.13 &$+0.36$& 0.10 &$+0.35$& 0.15 &$+0.41$& 0.13\\ 
        &    Final &$-0.03$& 0.09 &$-0.05$& 0.11 &$+0.49$& 0.15 &$+0.20$& 0.15 &$+0.28$& 0.12 &$+0.41$& 0.16 &$+0.24$& 0.13 &$+0.34$& 0.10 &$+0.33$& 0.15 &$+0.37$& 0.13\\ 
I-156   &   ODFNEW &$-0.70$& 0.08 &$-0.72$& 0.12 &$+0.41$& 0.14 &$+0.14$& 0.11 &$+0.40$& 0.08 &$+0.27$& 0.08 &$+0.35$& 0.08 &$+0.20$& 0.09 &$+0.24$& 0.10 &$+0.25$& 0.12\\ 
        &  AODFNEW &$-0.72$& 0.08 &$-0.72$& 0.14 &$+0.52$& 0.16 &$+0.20$& 0.12 &$+0.46$& 0.08 &$+0.30$& 0.09 &$+0.46$& 0.09 &$+0.26$& 0.10 &$+0.28$& 0.09 &$+0.38$& 0.14\\ 
        &    Final &$-0.72$& 0.08 &$-0.72$& 0.13 &$+0.54$& 0.15 &$+0.21$& 0.11 &$+0.47$& 0.08 &$+0.31$& 0.08 &$+0.48$& 0.08 &$+0.27$& 0.10 &$+0.29$& 0.10 &$+0.40$& 0.13\\ 
I-158   &   ODFNEW &$-0.23$& 0.08 &$-0.30$& 0.14 &$+0.48$& 0.16 &$-0.07$& 0.08 &$+0.26$& 0.11 &$+0.36$& 0.09 &$+0.27$& 0.12 &$+0.13$& 0.14 &$+0.10$& 0.12 &$-0.04$& 0.17\\ 
        &  AODFNEW &$-0.21$& 0.09 &$-0.27$& 0.18 &$+0.56$& 0.20 &$-0.01$& 0.09 &$+0.32$& 0.10 &$+0.39$& 0.09 &$+0.36$& 0.11 &$+0.17$& 0.12 &$+0.13$& 0.12 &$+0.07$& 0.21\\ 
        &    Final &$-0.21$& 0.08 &$-0.28$& 0.16 &$+0.54$& 0.18 &$-0.02$& 0.08 &$+0.31$& 0.11 &$+0.38$& 0.09 &$+0.34$& 0.11 &$+0.16$& 0.13 &$+0.12$& 0.12 &$+0.04$& 0.19\\ 
I-194   &   ODFNEW &$-0.31$& 0.07 &$-0.45$& 0.13 &$+0.34$& 0.15 &$+0.11$& 0.08 &$+0.44$& 0.09 &$+0.44$& 0.08 &$+0.22$& 0.09 &$+0.18$& 0.11 &$+0.26$& 0.10 &$+0.24$& 0.13\\ 
        &  AODFNEW &$-0.25$& 0.07 &$-0.29$& 0.13 &$+0.36$& 0.15 &$+0.10$& 0.08 &$+0.48$& 0.08 &$+0.43$& 0.07 &$+0.33$& 0.09 &$+0.18$& 0.10 &$+0.25$& 0.09 &$+0.30$& 0.13\\ 
        &    Final &$-0.24$& 0.07 &$-0.25$& 0.13 &$+0.36$& 0.15 &$+0.10$& 0.08 &$+0.49$& 0.08 &$+0.43$& 0.08 &$+0.35$& 0.09 &$+0.18$& 0.11 &$+0.25$& 0.10 &$+0.31$& 0.13\\ 
I-202   &   ODFNEW &$+0.11$& 0.07 &$+0.02$& 0.14 &$+0.04$& 0.16 &$+0.28$& 0.10 &$+0.13$& 0.09 &$+0.47$& 0.09 &$+0.14$& 0.10 &$+0.20$& 0.11 &$+0.13$& 0.09 &$+0.18$& 0.14\\ 
        &  AODFNEW &$+0.11$& 0.08 &$+0.03$& 0.18 &$+0.11$& 0.20 &$+0.38$& 0.10 &$+0.19$& 0.10 &$+0.52$& 0.09 &$+0.23$& 0.10 &$+0.26$& 0.10 &$+0.20$& 0.10 &$+0.30$& 0.19\\ 
        &    Final &$+0.11$& 0.08 &$+0.02$& 0.16 &$+0.07$& 0.18 &$+0.32$& 0.10 &$+0.15$& 0.10 &$+0.49$& 0.09 &$+0.17$& 0.10 &$+0.22$& 0.11 &$+0.16$& 0.10 &$+0.23$& 0.17\\ 
I-264   &   ODFNEW &$-1.14$& 0.05 &$-1.14$& 0.11 &$-0.06$& 0.13 &$+0.51$& 0.06 &$+0.45$& 0.07 &$+0.96$& 0.06 &$+0.42$& 0.08 &$+0.39$& 0.09 &$+0.35$& 0.10 &$+0.25$& 0.12\\ 
        &  AODFNEW &$-1.17$& 0.06 &$-1.17$& 0.12 &$+0.05$& 0.14 &$+0.56$& 0.06 &$+0.49$& 0.07 &$+0.99$& 0.06 &$+0.52$& 0.08 &$+0.44$& 0.09 &$+0.40$& 0.09 &$+0.40$& 0.12\\ 
        &    Final &$-1.18$& 0.06 &$-1.18$& 0.11 &$+0.08$& 0.14 &$+0.57$& 0.06 &$+0.50$& 0.07 &$+1.00$& 0.06 &$+0.55$& 0.08 &$+0.45$& 0.09 &$+0.41$& 0.10 &$+0.44$& 0.12\\ 
I-322   &   ODFNEW &$-0.26$& 0.05 &$-0.37$& 0.12 &$+0.15$& 0.13 &$+0.16$& 0.07 &$+0.19$& 0.09 &$+0.35$& 0.09 &$+0.16$& 0.07 &$+0.11$& 0.09 &$-0.02$& 0.07 &$+0.07$& 0.12\\ 
        &  AODFNEW &$-0.26$& 0.06 &$-0.35$& 0.12 &$+0.22$& 0.14 &$+0.21$& 0.07 &$+0.23$& 0.08 &$+0.37$& 0.07 &$+0.25$& 0.07 &$+0.14$& 0.07 &$+0.02$& 0.07 &$+0.18$& 0.13\\ 
        &    Final &$-0.26$& 0.06 &$-0.36$& 0.12 &$+0.19$& 0.14 &$+0.19$& 0.07 &$+0.21$& 0.08 &$+0.36$& 0.08 &$+0.21$& 0.07 &$+0.13$& 0.08 &$+0.00$& 0.07 &$+0.13$& 0.12\\ 
II-033  &   ODFNEW &$-0.74$& 0.06 &$-0.78$& 0.10 &$+0.44$& 0.12 &$+0.01$& 0.10 &$+0.39$& 0.09 &$+0.42$& 0.07 &$+0.36$& 0.08 &$+0.32$& 0.06 &$+0.31$& 0.09 &$+0.22$& 0.12\\ 
        &  AODFNEW &$-0.76$& 0.06 &$-0.79$& 0.11 &$+0.57$& 0.13 &$+0.07$& 0.12 &$+0.44$& 0.08 &$+0.45$& 0.07 &$+0.46$& 0.08 &$+0.38$& 0.07 &$+0.36$& 0.09 &$+0.36$& 0.13\\ 
        &    Final &$-0.76$& 0.06 &$-0.79$& 0.11 &$+0.58$& 0.12 &$+0.08$& 0.11 &$+0.45$& 0.08 &$+0.45$& 0.07 &$+0.47$& 0.08 &$+0.39$& 0.07 &$+0.37$& 0.09 &$+0.38$& 0.12\\ 
II-119  &   ODFNEW &$-1.21$& 0.10 &$-1.26$& 0.10 &$+0.64$& 0.14 &$-0.02$& 0.12 &$+0.44$& 0.11 &$+0.09$& 0.17 &$+0.30$& 0.12 &$+0.35$& 0.15 &$+0.47$& 0.11 &$+0.35$& 0.10\\ 
        &  AODFNEW &$-1.27$& 0.11 &$-1.34$& 0.11 &$+0.79$& 0.14 &$+0.05$& 0.13 &$+0.51$& 0.12 &$+0.16$& 0.18 &$+0.40$& 0.13 &$+0.43$& 0.16 &$+0.54$& 0.11 &$+0.52$& 0.11\\ 
        &    Final &$-1.29$& 0.11 &$-1.37$& 0.11 &$+0.84$& 0.14 &$+0.07$& 0.12 &$+0.53$& 0.11 &$+0.18$& 0.17 &$+0.43$& 0.12 &$+0.46$& 0.15 &$+0.56$& 0.11 &$+0.58$& 0.11\\ 
II-154  &   ODFNEW &$-0.59$& 0.08 &$-0.64$& 0.10 &$+0.28$& 0.13 &$+0.01$& 0.14 &$+0.31$& 0.12 &$+0.37$& 0.12 &$+0.35$& 0.09 &$+0.34$& 0.12 &$+0.17$& 0.11 &$+0.21$& 0.11\\ 
        &  AODFNEW &$-0.64$& 0.08 &$-0.68$& 0.11 &$+0.40$& 0.14 &$+0.09$& 0.15 &$+0.37$& 0.13 &$+0.42$& 0.12 &$+0.47$& 0.09 &$+0.41$& 0.12 &$+0.22$& 0.11 &$+0.35$& 0.12\\ 
        &    Final &$-0.64$& 0.08 &$-0.68$& 0.11 &$+0.39$& 0.14 &$+0.08$& 0.15 &$+0.36$& 0.12 &$+0.42$& 0.12 &$+0.46$& 0.09 &$+0.40$& 0.12 &$+0.22$& 0.11 &$+0.34$& 0.11\\ 
II-172  &   ODFNEW &$-0.30$& 0.07 &$-0.27$& 0.11 &$+0.30$& 0.14 &$+0.04$& 0.10 &$+0.31$& 0.11 &$+0.37$& 0.08 &$+0.21$& 0.09 &$+0.22$& 0.14 &$+0.23$& 0.08 &$-0.02$& 0.14\\ 
        &  AODFNEW &$-0.33$& 0.08 &$-0.27$& 0.12 &$+0.40$& 0.15 &$+0.10$& 0.08 &$+0.38$& 0.10 &$+0.40$& 0.08 &$+0.33$& 0.09 &$+0.28$& 0.13 &$+0.27$& 0.09 &$+0.10$& 0.15\\ 
        &    Final &$-0.33$& 0.08 &$-0.27$& 0.11 &$+0.39$& 0.15 &$+0.10$& 0.09 &$+0.38$& 0.11 &$+0.40$& 0.08 &$+0.32$& 0.09 &$+0.28$& 0.14 &$+0.27$& 0.08 &$+0.09$& 0.15\\ 
III-152 &   ODFNEW &$-0.42$& 0.06 &$-0.42$& 0.11 &$+0.30$& 0.14 &$+0.16$& 0.10 &$+0.40$& 0.08 &$+0.36$& 0.09 &$+0.28$& 0.09 &$+0.20$& 0.11 &$+0.25$& 0.08 &$+0.17$& 0.22\\ 
        &  AODFNEW &$-0.41$& 0.06 &$-0.41$& 0.12 &$+0.41$& 0.14 &$+0.23$& 0.09 &$+0.46$& 0.09 &$+0.39$& 0.10 &$+0.37$& 0.09 &$+0.25$& 0.10 &$+0.31$& 0.09 &$+0.30$& 0.23\\ 
        &    Final &$-0.41$& 0.06 &$-0.41$& 0.11 &$+0.43$& 0.14 &$+0.24$& 0.10 &$+0.47$& 0.08 &$+0.40$& 0.10 &$+0.39$& 0.09 &$+0.26$& 0.11 &$+0.32$& 0.08 &$+0.32$& 0.23\\ 
III-220 &   ODFNEW &$-0.31$& 0.08 &$-0.33$& 0.11 &$+0.37$& 0.13 &$-0.07$& 0.08 &$+0.30$& 0.13 &$+0.43$& 0.10 &$+0.24$& 0.08 &$+0.29$& 0.13 &$+0.38$& 0.09 &$+0.30$& 0.23\\ 
        &  AODFNEW &$-0.34$& 0.08 &$-0.37$& 0.12 &$+0.51$& 0.14 &$-0.01$& 0.08 &$+0.36$& 0.11 &$+0.47$& 0.10 &$+0.35$& 0.08 &$+0.35$& 0.14 &$+0.44$& 0.08 &$+0.44$& 0.24\\ 
        &    Final &$-0.34$& 0.08 &$-0.36$& 0.11 &$+0.49$& 0.14 &$-0.02$& 0.08 &$+0.35$& 0.12 &$+0.47$& 0.10 &$+0.34$& 0.08 &$+0.34$& 0.14 &$+0.43$& 0.08 &$+0.42$& 0.23\\ 
IV-003  &   ODFNEW &$-1.28$& 0.05 &$-1.30$& 0.07 &$+0.54$& 0.10 &$+0.00$& 0.05 &$+0.41$& 0.10 &$+0.19$& 0.08 &$+0.37$& 0.10 &$+0.37$& 0.08 &$+0.32$& 0.09 &$+0.21$& 0.12\\ 
        &  AODFNEW &$-1.33$& 0.06 &$-1.36$& 0.07 &$+0.69$& 0.10 &$+0.07$& 0.06 &$+0.47$& 0.09 &$+0.24$& 0.08 &$+0.48$& 0.09 &$+0.45$& 0.10 &$+0.37$& 0.08 &$+0.36$& 0.12\\ 
        &    Final &$-1.34$& 0.06 &$-1.37$& 0.07 &$+0.72$& 0.10 &$+0.08$& 0.06 &$+0.48$& 0.10 &$+0.25$& 0.08 &$+0.50$& 0.10 &$+0.47$& 0.09 &$+0.38$& 0.08 &$+0.39$& 0.12\\ 
IV-047  &   ODFNEW &$-0.39$& 0.11 &$-0.38$& 0.13 &$+0.38$& 0.16 &$+0.06$& 0.25 &$+0.42$& 0.15 &$+0.39$& 0.11 &$+0.27$& 0.12 &$+0.22$& 0.13 &$+0.37$& 0.14 &$+0.34$& 0.15\\ 
        &  AODFNEW &$-0.40$& 0.11 &$-0.39$& 0.14 &$+0.51$& 0.17 &$+0.14$& 0.28 &$+0.48$& 0.14 &$+0.42$& 0.11 &$+0.37$& 0.12 &$+0.28$& 0.12 &$+0.42$& 0.14 &$+0.48$& 0.16\\ 
        &    Final &$-0.40$& 0.11 &$-0.39$& 0.14 &$+0.54$& 0.17 &$+0.16$& 0.27 &$+0.49$& 0.15 &$+0.43$& 0.11 &$+0.39$& 0.12 &$+0.29$& 0.12 &$+0.43$& 0.14 &$+0.51$& 0.15\\ 
IV-072  &   ODFNEW &$+0.20$& 0.06 &$+0.17$& 0.11 &$+0.01$& 0.13 &$+0.22$& 0.06 &$+0.33$& 0.07 &$+0.45$& 0.09 &$+0.12$& 0.09 &$+0.16$& 0.09 &$+0.20$& 0.10 &$+0.16$& 0.14\\ 
        &  AODFNEW &$+0.22$& 0.06 &$+0.21$& 0.11 &$+0.07$& 0.13 &$+0.29$& 0.07 &$+0.39$& 0.08 &$+0.52$& 0.10 &$+0.22$& 0.10 &$+0.17$& 0.09 &$+0.24$& 0.09 &$+0.26$& 0.14\\ 
        &    Final &$+0.22$& 0.06 &$+0.21$& 0.11 &$+0.07$& 0.13 &$+0.29$& 0.07 &$+0.39$& 0.08 &$+0.52$& 0.10 &$+0.22$& 0.10 &$+0.17$& 0.09 &$+0.24$& 0.10 &$+0.26$& 0.14\\ 
IV-167  &   ODFNEW &$+0.42$& 0.07 &$+0.41$& 0.13 &~...~& 0.13 &$-0.23$& 0.07 &$+0.27$& 0.12 &$+0.25$& 0.10 &$+0.06$& 0.10 &$+0.06$& 0.11 &$-0.05$& 0.10 &$+0.06$& 0.17\\ 
        &  AODFNEW &$+0.44$& 0.08 &$+0.44$& 0.14 &~...~& 0.14 &$-0.12$& 0.08 &$+0.34$& 0.13 &$+0.30$& 0.12 &$+0.16$& 0.10 &$+0.10$& 0.10 &$+0.02$& 0.11 &$+0.18$& 0.17\\ 
        &    Final &$+0.44$& 0.08 &$+0.43$& 0.14 &~...~& 0.14 &$-0.14$& 0.08 &$+0.33$& 0.12 &$+0.29$& 0.11 &$+0.14$& 0.10 &$+0.09$& 0.11 &$+0.01$& 0.11 &$+0.16$& 0.17\\ 
IV-203  &   ODFNEW &$-1.26$& 0.06 &$-1.11$& 0.15 &$+0.13$& 0.17 &$+0.60$& 0.09 &$+0.41$& 0.09 &$+0.56$& 0.07 &$+0.47$& 0.10 &$+0.31$& 0.10 &$+0.32$& 0.09 &$+0.14$& 0.15\\ 
        &  AODFNEW &$-1.28$& 0.07 &$-1.14$& 0.16 &$+0.23$& 0.17 &$+0.65$& 0.08 &$+0.46$& 0.09 &$+0.58$& 0.07 &$+0.56$& 0.10 &$+0.35$& 0.09 &$+0.37$& 0.09 &$+0.27$& 0.16\\ 
        &    Final &$-1.28$& 0.07 &$-1.15$& 0.15 &$+0.25$& 0.17 &$+0.66$& 0.08 &$+0.47$& 0.09 &$+0.58$& 0.07 &$+0.58$& 0.10 &$+0.36$& 0.10 &$+0.38$& 0.09 &$+0.29$& 0.15\\ 
IV-325  &   ODFNEW &$+0.23$& 0.07 &$+0.25$& 0.13 &$-0.16$& 0.20 &~...~& 0.07 &$+0.29$& 0.12 &$+0.42$& 0.13 &$+0.25$& 0.10 &$-0.14$& 0.13 &$-0.02$& 0.11 &$+0.02$& 0.21\\ 
        &  AODFNEW &$+0.26$& 0.07 &$+0.27$& 0.13 &$-0.07$& 0.19 &~...~& 0.07 &$+0.34$& 0.12 &$+0.46$& 0.14 &$+0.35$& 0.10 &$-0.10$& 0.12 &$+0.04$& 0.11 &$+0.13$& 0.21\\ 
        &    Final &$+0.25$& 0.07 &$+0.27$& 0.13 &$-0.09$& 0.20 &~...~& 0.07 &$+0.33$& 0.12 &$+0.45$& 0.14 &$+0.33$& 0.10 &$-0.11$& 0.12 &$+0.03$& 0.11 &$+0.11$& 0.21\\ 
IV-329  &   ODFNEW &$-0.93$& 0.05 &$-0.93$& 0.09 &$+0.39$& 0.11 &$+0.13$& 0.10 &$+0.40$& 0.12 &$+0.30$& 0.07 &$+0.45$& 0.07 &$+0.24$& 0.08 &$+0.28$& 0.09 &$+0.24$& 0.13\\ 
        &  AODFNEW &$-0.95$& 0.05 &$-0.93$& 0.10 &$+0.50$& 0.12 &$+0.17$& 0.10 &$+0.45$& 0.12 &$+0.32$& 0.07 &$+0.54$& 0.07 &$+0.28$& 0.09 &$+0.32$& 0.09 &$+0.37$& 0.13\\ 
        &    Final &$-0.95$& 0.05 &$-0.93$& 0.10 &$+0.52$& 0.11 &$+0.18$& 0.10 &$+0.46$& 0.12 &$+0.32$& 0.07 &$+0.55$& 0.07 &$+0.29$& 0.08 &$+0.33$& 0.09 &$+0.39$& 0.13\\ 

\multicolumn{21}{c}{BAADE's WINDOW NON-BULGE STARS} \\
II-122  &   ODFNEW &$-0.79$& 0.07 &$-0.79$& 0.15 &$+0.38$& 0.16 &$+0.47$& 0.11 &$+0.38$& 0.11 &$+0.63$& 0.08 &$+0.12$& 0.10 &$+0.22$& 0.12 &$+0.41$& 0.11 &$+0.27$& 0.17\\ 
        &  AODFNEW &$-0.79$& 0.07 &$-0.76$& 0.14 &$+0.46$& 0.16 &$+0.52$& 0.11 &$+0.46$& 0.11 &$+0.65$& 0.08 &$+0.27$& 0.10 &$+0.27$& 0.12 &$+0.46$& 0.12 &$+0.39$& 0.18\\ 
        &    Final &$-0.79$& 0.07 &$-0.75$& 0.15 &$+0.48$& 0.16 &$+0.53$& 0.11 &$+0.48$& 0.11 &$+0.65$& 0.08 &$+0.30$& 0.10 &$+0.28$& 0.12 &$+0.47$& 0.11 &$+0.41$& 0.17\\ 
IV-025  &   ODFNEW &$+0.20$& 0.10 &$+0.14$& 0.16 &$+0.16$& 0.20 &$-0.30$& 0.18 &$-0.03$& 0.13 &$+0.35$& 0.12 &$-0.03$& 0.11 &$-0.06$& 0.14 &$+0.11$& 0.12 &$+0.14$& 0.20\\ 
        &  AODFNEW &$+0.18$& 0.10 &$+0.13$& 0.16 &$+0.28$& 0.20 &$-0.20$& 0.20 &$+0.06$& 0.14 &$+0.41$& 0.11 &$+0.11$& 0.10 &$+0.02$& 0.13 &$+0.17$& 0.12 &$+0.29$& 0.19\\ 
        &    Final &$+0.20$& 0.10 &$+0.14$& 0.16 &$+0.19$& 0.20 &$-0.28$& 0.19 &$-0.01$& 0.14 &$+0.36$& 0.11 &$+0.01$& 0.11 &$-0.04$& 0.14 &$+0.12$& 0.12 &$+0.18$& 0.20\\ 

\multicolumn{22}{c}{LOCAL DISK STARS} \\
HR1184  &   ODFNEW &$-0.38$& 0.06 &$-0.47$& 0.08 &$+0.26$& 0.11 &$-0.10$& 0.11 &$+0.04$& 0.09 &$+0.22$& 0.06 &$+0.10$& 0.10 &$+0.06$& 0.09 &$+0.18$& 0.09 &$+0.25$& 0.08\\ 
        &  AODFNEW &$-0.42$& 0.06 &$-0.50$& 0.09 &$+0.40$& 0.12 &$-0.04$& 0.12 &$+0.11$& 0.09 &$+0.26$& 0.07 &$+0.21$& 0.10 &$+0.11$& 0.09 &$+0.23$& 0.09 &$+0.41$& 0.09\\ 
        &    Final &$-0.39$& 0.06 &$-0.47$& 0.08 &$+0.28$& 0.11 &$-0.09$& 0.11 &$+0.05$& 0.09 &$+0.22$& 0.07 &$+0.11$& 0.10 &$+0.07$& 0.09 &$+0.19$& 0.09 &$+0.27$& 0.08\\ 
HR1346  &   ODFNEW &$+0.18$& 0.05 &$+0.11$& 0.08 &$-0.08$& 0.11 &$+0.01$& 0.11 &$-0.07$& 0.07 &$+0.17$& 0.08 &$-0.03$& 0.09 &$-0.07$& 0.12 &$-0.07$& 0.10 &$-0.10$& 0.16\\ 
        &  AODFNEW &$+0.14$& 0.05 &$+0.07$& 0.09 &$+0.06$& 0.11 &$+0.08$& 0.11 &$-0.01$& 0.07 &$+0.21$& 0.07 &$+0.06$& 0.09 &$-0.02$& 0.11 &$-0.02$& 0.10 &$+0.05$& 0.16\\ 
        &    Final &$+0.17$& 0.05 &$+0.10$& 0.08 &$-0.04$& 0.11 &$+0.03$& 0.11 &$-0.06$& 0.07 &$+0.18$& 0.08 &$-0.01$& 0.09 &$-0.06$& 0.11 &$-0.06$& 0.10 &$-0.06$& 0.16\\ 
HR1348  &   ODFNEW &$-0.36$& 0.06 &$-0.43$& 0.11 &$+0.22$& 0.13 &$+0.00$& 0.08 &$+0.07$& 0.10 &$+0.18$& 0.06 &$+0.12$& 0.08 &$+0.00$& 0.08 &$+0.05$& 0.09 &$+0.05$& 0.17\\ 
        &  AODFNEW &$-0.38$& 0.06 &$-0.44$& 0.11 &$+0.34$& 0.13 &$+0.04$& 0.08 &$+0.13$& 0.08 &$+0.21$& 0.07 &$+0.23$& 0.08 &$+0.04$& 0.08 &$+0.09$& 0.09 &$+0.18$& 0.18\\ 
        &    Final &$-0.36$& 0.06 &$-0.43$& 0.11 &$+0.24$& 0.13 &$+0.01$& 0.08 &$+0.08$& 0.09 &$+0.19$& 0.07 &$+0.14$& 0.08 &$+0.01$& 0.08 &$+0.06$& 0.09 &$+0.08$& 0.17\\ 
HR1409  &   ODFNEW &$+0.22$& 0.07 &$+0.24$& 0.10 &$-0.09$& 0.13 &$+0.00$& 0.11 &$-0.03$& 0.08 &$+0.15$& 0.09 &$+0.06$& 0.10 &$-0.04$& 0.10 &$-0.03$& 0.08 &$-0.23$& 0.12\\ 
        &  AODFNEW &$+0.19$& 0.07 &$+0.21$& 0.10 &$+0.05$& 0.13 &$+0.07$& 0.10 &$+0.02$& 0.08 &$+0.19$& 0.09 &$+0.15$& 0.10 &$+0.02$& 0.10 &$+0.02$& 0.08 &$-0.08$& 0.12\\ 
        &    Final &$+0.21$& 0.07 &$+0.23$& 0.10 &$-0.05$& 0.13 &$+0.02$& 0.11 &$-0.02$& 0.08 &$+0.16$& 0.09 &$+0.08$& 0.10 &$-0.03$& 0.10 &$-0.02$& 0.08 &$-0.19$& 0.12\\ 
HR1411  &   ODFNEW &$+0.20$& 0.10 &$+0.19$& 0.12 &$-0.13$& 0.15 &$+0.14$& 0.13 &$-0.15$& 0.10 &$+0.06$& 0.10 &$+0.04$& 0.11 &$+0.05$& 0.12 &$-0.01$& 0.12 &$-0.07$& 0.18\\ 
        &  AODFNEW &$+0.16$& 0.09 &$+0.15$& 0.12 &$+0.04$& 0.14 &$+0.22$& 0.12 &$-0.08$& 0.10 &$+0.11$& 0.10 &$+0.15$& 0.10 &$+0.12$& 0.12 &$+0.04$& 0.11 &$+0.08$& 0.17\\ 
        &    Final &$+0.19$& 0.10 &$+0.18$& 0.12 &$-0.09$& 0.15 &$+0.16$& 0.12 &$-0.13$& 0.10 &$+0.07$& 0.10 &$+0.07$& 0.11 &$+0.07$& 0.12 &$+0.00$& 0.11 &$-0.03$& 0.17\\ 
HR1585  &   ODFNEW &$-0.35$& 0.10 &$-0.40$& 0.13 &$+0.17$& 0.15 &$+0.00$& 0.11 &$+0.17$& 0.10 &$+0.13$& 0.10 &$+0.15$& 0.17 &$+0.03$& 0.14 &$+0.04$& 0.12 &$+0.04$& 0.14\\ 
        &  AODFNEW &$-0.36$& 0.10 &$-0.40$& 0.13 &$+0.27$& 0.15 &$+0.04$& 0.12 &$+0.24$& 0.10 &$+0.15$& 0.10 &$+0.27$& 0.17 &$+0.07$& 0.13 &$+0.09$& 0.12 &$+0.15$& 0.15\\ 
        &    Final &$-0.36$& 0.10 &$-0.40$& 0.13 &$+0.22$& 0.15 &$+0.02$& 0.11 &$+0.21$& 0.10 &$+0.14$& 0.10 &$+0.21$& 0.17 &$+0.05$& 0.14 &$+0.07$& 0.12 &$+0.10$& 0.15\\ 
HR2035  &   ODFNEW &$-0.63$& 0.05 &$-0.70$& 0.08 &$+0.50$& 0.11 &$+0.06$& 0.09 &$+0.24$& 0.07 &$+0.33$& 0.06 &$+0.20$& 0.09 &$+0.26$& 0.08 &$+0.27$& 0.07 &$+0.30$& 0.08\\ 
        &  AODFNEW &$-0.66$& 0.05 &$-0.74$& 0.08 &$+0.64$& 0.11 &$+0.12$& 0.10 &$+0.29$& 0.07 &$+0.37$& 0.07 &$+0.29$& 0.09 &$+0.33$& 0.07 &$+0.32$& 0.07 &$+0.46$& 0.09\\ 
        &    Final &$-0.65$& 0.05 &$-0.73$& 0.08 &$+0.60$& 0.11 &$+0.10$& 0.10 &$+0.27$& 0.07 &$+0.36$& 0.07 &$+0.26$& 0.09 &$+0.31$& 0.08 &$+0.30$& 0.07 &$+0.41$& 0.08\\ 
HR2113  &   ODFNEW &$-0.66$& 0.05 &$-0.77$& 0.11 &$+0.43$& 0.13 &$+0.14$& 0.06 &$+0.30$& 0.07 &$+0.44$& 0.06 &$+0.15$& 0.06 &$+0.16$& 0.09 &$+0.28$& 0.08 &$+0.37$& 0.12\\ 
        &  AODFNEW &$-0.67$& 0.05 &$-0.77$& 0.13 &$+0.53$& 0.15 &$+0.17$& 0.06 &$+0.36$& 0.07 &$+0.46$& 0.07 &$+0.26$& 0.06 &$+0.20$& 0.09 &$+0.31$& 0.08 &$+0.49$& 0.14\\ 
        &    Final &$-0.67$& 0.05 &$-0.77$& 0.12 &$+0.52$& 0.14 &$+0.17$& 0.06 &$+0.35$& 0.07 &$+0.46$& 0.07 &$+0.25$& 0.06 &$+0.20$& 0.09 &$+0.31$& 0.08 &$+0.48$& 0.13\\ 
HR2443  &   ODFNEW &$-0.21$& 0.05 &$-0.30$& 0.09 &$+0.10$& 0.11 &$+0.00$& 0.07 &$+0.04$& 0.07 &$+0.17$& 0.11 &$+0.00$& 0.07 &$+0.06$& 0.07 &$+0.03$& 0.09 &$+0.01$& 0.17\\ 
        &  AODFNEW &$-0.23$& 0.05 &$-0.32$& 0.09 &$+0.25$& 0.11 &$+0.03$& 0.06 &$+0.09$& 0.06 &$+0.19$& 0.10 &$+0.09$& 0.07 &$+0.10$& 0.06 &$+0.07$& 0.09 &$+0.14$& 0.16\\ 
        &    Final &$-0.21$& 0.05 &$-0.30$& 0.09 &$+0.12$& 0.11 &$+0.00$& 0.07 &$+0.05$& 0.07 &$+0.17$& 0.11 &$+0.01$& 0.07 &$+0.06$& 0.07 &$+0.03$& 0.09 &$+0.02$& 0.17\\ 
HR3418  &   ODFNEW &$+0.15$& 0.05 &$+0.21$& 0.11 &$-0.10$& 0.13 &$+0.18$& 0.12 &$-0.07$& 0.10 &$+0.07$& 0.08 &$+0.09$& 0.08 &$-0.13$& 0.11 &$-0.16$& 0.09 &$-0.20$& 0.20\\ 
        &  AODFNEW &$+0.15$& 0.05 &$+0.21$& 0.11 &$+0.04$& 0.13 &$+0.22$& 0.11 &$-0.01$& 0.10 &$+0.10$& 0.09 &$+0.20$& 0.08 &$-0.09$& 0.11 &$-0.12$& 0.09 &$-0.08$& 0.20\\ 
        &    Final &$+0.15$& 0.05 &$+0.21$& 0.11 &$-0.07$& 0.13 &$+0.19$& 0.11 &$-0.06$& 0.10 &$+0.08$& 0.08 &$+0.12$& 0.08 &$-0.12$& 0.11 &$-0.15$& 0.09 &$-0.17$& 0.20\\ 
HR3733  &   ODFNEW &$-0.08$& 0.06 &$-0.06$& 0.07 &$-0.02$& 0.10 &$-0.12$& 0.06 &$-0.03$& 0.09 &$-0.04$& 0.10 &$+0.11$& 0.07 &$-0.02$& 0.07 &$+0.08$& 0.08 &$-0.04$& 0.09\\ 
        &  AODFNEW &$-0.13$& 0.05 &$-0.12$& 0.07 &$+0.13$& 0.10 &$-0.05$& 0.06 &$+0.03$& 0.08 &$+0.01$& 0.09 &$+0.20$& 0.07 &$+0.04$& 0.07 &$+0.13$& 0.07 &$+0.12$& 0.09\\ 
        &    Final &$-0.09$& 0.06 &$-0.07$& 0.07 &$+0.02$& 0.10 &$-0.10$& 0.06 &$-0.01$& 0.08 &$-0.03$& 0.10 &$+0.13$& 0.07 &$-0.01$& 0.07 &$+0.09$& 0.08 &$+0.00$& 0.09\\ 
HR3905  &   ODFNEW &$+0.32$& 0.05 &$+0.33$& 0.10 &$+0.11$& 0.12 &$+0.43$& 0.12 &$+0.01$& 0.08 &$+0.34$& 0.06 &$+0.25$& 0.12 &$+0.06$& 0.08 &$+0.05$& 0.07 &$-0.04$& 0.10\\ 
        &  AODFNEW &$+0.33$& 0.05 &$+0.34$& 0.10 &$+0.23$& 0.13 &$+0.54$& 0.11 &$+0.07$& 0.08 &$+0.38$& 0.06 &$+0.36$& 0.11 &$+0.10$& 0.08 &$+0.10$& 0.07 &$+0.07$& 0.10\\ 
        &    Final &$+0.32$& 0.05 &$+0.33$& 0.10 &$+0.11$& 0.12 &$+0.43$& 0.11 &$+0.01$& 0.08 &$+0.34$& 0.06 &$+0.25$& 0.11 &$+0.06$& 0.08 &$+0.05$& 0.07 &$-0.04$& 0.10\\ 
HR4104  &   ODFNEW &$-0.38$& 0.05 &$-0.39$& 0.11 &$+0.08$& 0.14 &$+0.09$& 0.10 &$+0.10$& 0.08 &$+0.24$& 0.10 &$+0.18$& 0.10 &$+0.03$& 0.12 &$+0.01$& 0.11 &$+0.02$& 0.16\\ 
        &  AODFNEW &$-0.37$& 0.06 &$-0.37$& 0.12 &$+0.19$& 0.14 &$+0.12$& 0.09 &$+0.14$& 0.07 &$+0.26$& 0.10 &$+0.27$& 0.09 &$+0.06$& 0.10 &$+0.06$& 0.10 &$+0.14$& 0.16\\ 
        &    Final &$-0.38$& 0.06 &$-0.38$& 0.11 &$+0.11$& 0.14 &$+0.10$& 0.10 &$+0.11$& 0.08 &$+0.25$& 0.10 &$+0.20$& 0.10 &$+0.04$& 0.11 &$+0.02$& 0.11 &$+0.05$& 0.16\\ 
HR4382  &   ODFNEW &$-0.55$& 0.05 &$-0.59$& 0.09 &$+0.23$& 0.12 &$+0.11$& 0.08 &$+0.09$& 0.08 &$+0.15$& 0.09 &$+0.18$& 0.11 &$+0.10$& 0.08 &$+0.04$& 0.08 &$+0.11$& 0.09\\ 
        &  AODFNEW &$-0.58$& 0.05 &$-0.62$& 0.10 &$+0.33$& 0.12 &$+0.17$& 0.09 &$+0.15$& 0.08 &$+0.18$& 0.08 &$+0.29$& 0.11 &$+0.15$& 0.08 &$+0.08$& 0.07 &$+0.25$& 0.10\\ 
        &    Final &$-0.56$& 0.05 &$-0.60$& 0.10 &$+0.26$& 0.12 &$+0.13$& 0.08 &$+0.11$& 0.08 &$+0.16$& 0.08 &$+0.21$& 0.11 &$+0.11$& 0.08 &$+0.05$& 0.08 &$+0.15$& 0.10\\ 
HR4450  &   ODFNEW &$+0.15$& 0.08 &$+0.00$& 0.09 &$-0.03$& 0.14 &$-0.03$& 0.11 &$-0.30$& 0.10 &$-0.04$& 0.09 &$-0.26$& 0.10 &$-0.10$& 0.12 &$+0.02$& 0.09 &$+0.03$& 0.15\\ 
        &  AODFNEW &$+0.10$& 0.08 &$-0.06$& 0.10 &$+0.17$& 0.13 &$+0.04$& 0.10 &$-0.25$& 0.10 &$+0.01$& 0.09 &$-0.18$& 0.10 &$-0.03$& 0.12 &$+0.07$& 0.09 &$+0.19$& 0.15\\ 
        &    Final &$+0.15$& 0.08 &$+0.00$& 0.09 &$-0.03$& 0.14 &$-0.03$& 0.11 &$-0.30$& 0.10 &$-0.04$& 0.09 &$-0.26$& 0.10 &$-0.10$& 0.12 &$+0.02$& 0.09 &$+0.03$& 0.15\\ 
HR4608  &   ODFNEW &$-0.38$& 0.05 &$-0.40$& 0.07 &$+0.34$& 0.10 &$+0.03$& 0.05 &$+0.05$& 0.08 &$+0.14$& 0.06 &$+0.09$& 0.07 &$+0.04$& 0.07 &$+0.10$& 0.07 &$+0.06$& 0.10\\ 
        &  AODFNEW &$-0.43$& 0.05 &$-0.43$& 0.08 &$+0.48$& 0.10 &$+0.09$& 0.05 &$+0.13$& 0.07 &$+0.19$& 0.06 &$+0.22$& 0.07 &$+0.10$& 0.07 &$+0.14$& 0.07 &$+0.21$& 0.10\\ 
        &    Final &$-0.39$& 0.05 &$-0.40$& 0.08 &$+0.36$& 0.10 &$+0.04$& 0.05 &$+0.06$& 0.08 &$+0.15$& 0.06 &$+0.11$& 0.07 &$+0.05$& 0.07 &$+0.11$& 0.07 &$+0.08$& 0.10\\ 
HR5340  &   ODFNEW &$-0.56$& 0.04 &$-0.55$& 0.10 &$+0.40$& 0.12 &$+0.10$& 0.05 &$+0.42$& 0.07 &$+0.37$& 0.08 &$+0.34$& 0.09 &$+0.20$& 0.06 &$+0.32$& 0.07 &$+0.16$& 0.11\\ 
        &  AODFNEW &$-0.57$& 0.05 &$-0.55$& 0.11 &$+0.52$& 0.12 &$+0.15$& 0.05 &$+0.47$& 0.07 &$+0.40$& 0.09 &$+0.43$& 0.09 &$+0.25$& 0.06 &$+0.36$& 0.08 &$+0.29$& 0.11\\ 
        &    Final &$-0.57$& 0.04 &$-0.55$& 0.11 &$+0.54$& 0.12 &$+0.16$& 0.05 &$+0.48$& 0.07 &$+0.41$& 0.08 &$+0.45$& 0.09 &$+0.26$& 0.06 &$+0.37$& 0.08 &$+0.32$& 0.11\\ 
\enddata
\end{deluxetable}
\clearpage
\end{landscape}

\clearpage

%
%

\clearpage

\clearpage

\clearpage

\clearpage

\begin{figure}[h]
\plotone{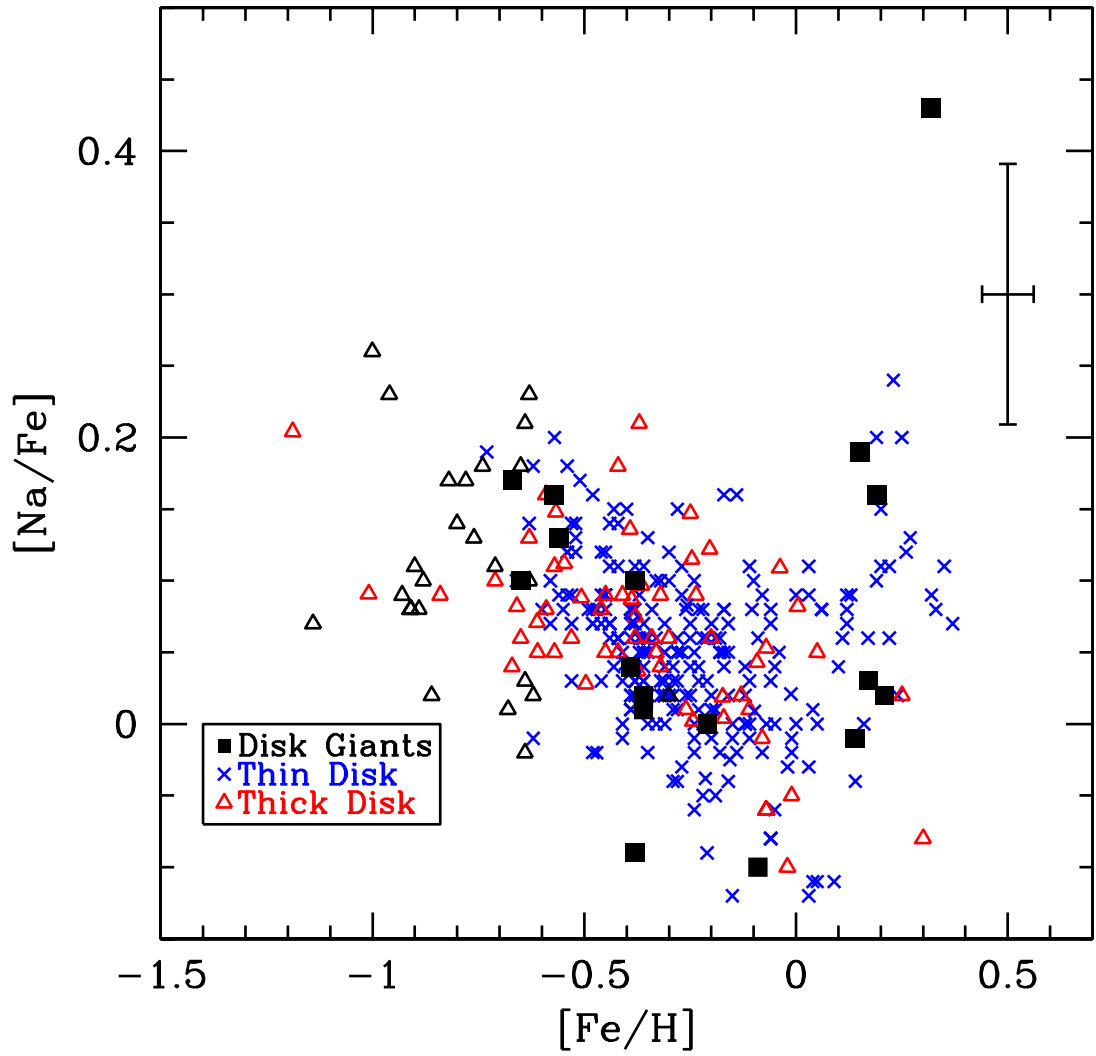}
\caption{Similar to Figure~\ref{fig-ofe1dg}, but for [Na/Fe].  Again, $\mu$~Leo
shows enhanced [Na/Fe], but the rest of the sample shows good agreement
with the literature sample.} 
\end{figure}
\clearpage

\clearpage

\clearpage

\clearpage

\clearpage

\clearpage

\clearpage

\clearpage

\begin{figure}
\plotone{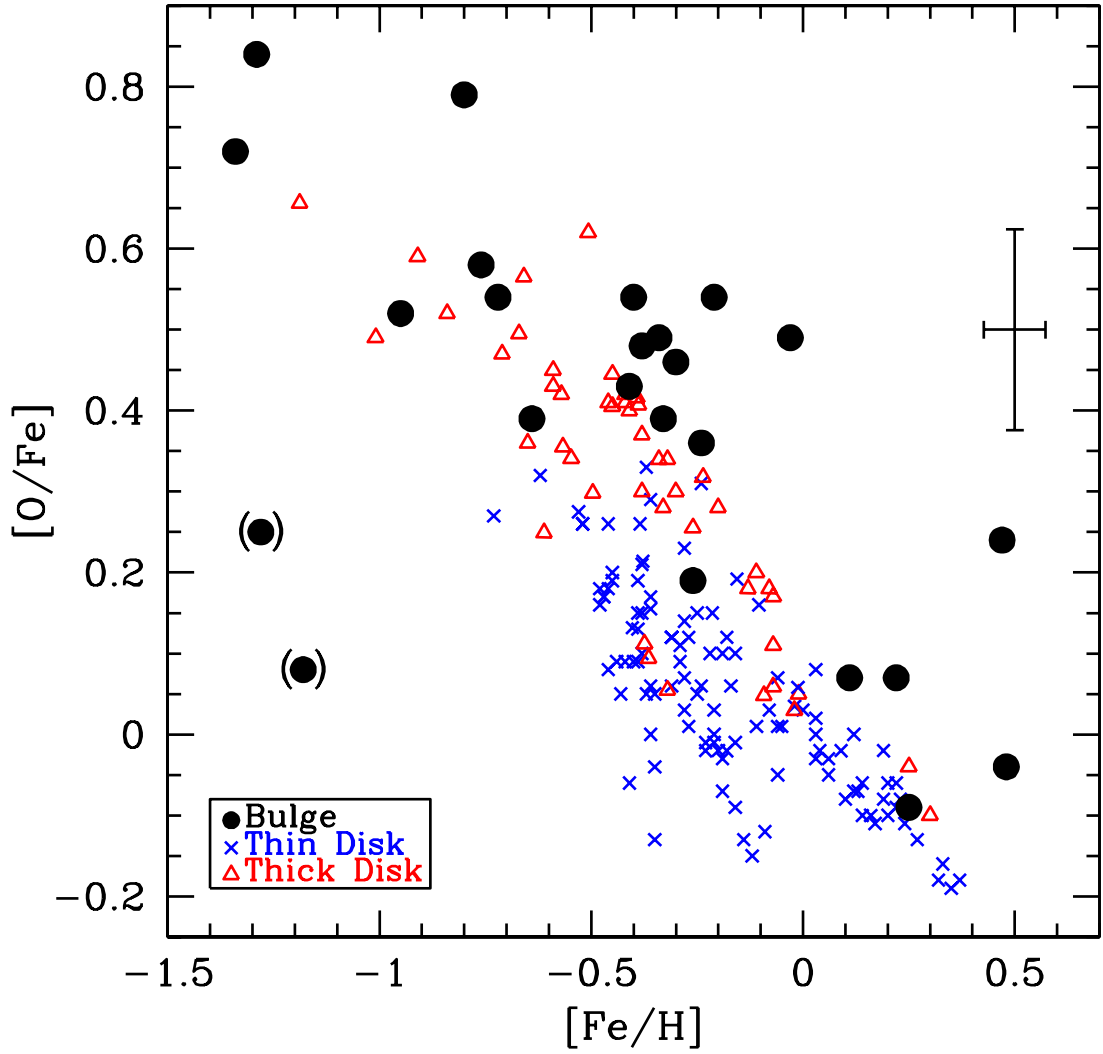}
\caption{The distribution of [O/Fe] vs. [Fe/H] for our bulge 
sample and literature values.  The points are the same as in 
Figure~\ref{fig-ofe1dg}--\ref{fig-tife1dg}. 
For the bulge stars, the [O/Fe] ratio
drops with increasing [Fe/H], but the mean value of the distribution
stays higher than the disk at the highest metallicity.  The two metal-poor
bulge stars with low [O/Fe] values are I-264 and IV-203 (both marked by 
paratheses in this and following plots). The nature of
these stars are discussed in Section~9.}\label{fig-jff5-ofebulgedisk}
\end{figure}
\clearpage

\begin{figure} 
\plotone{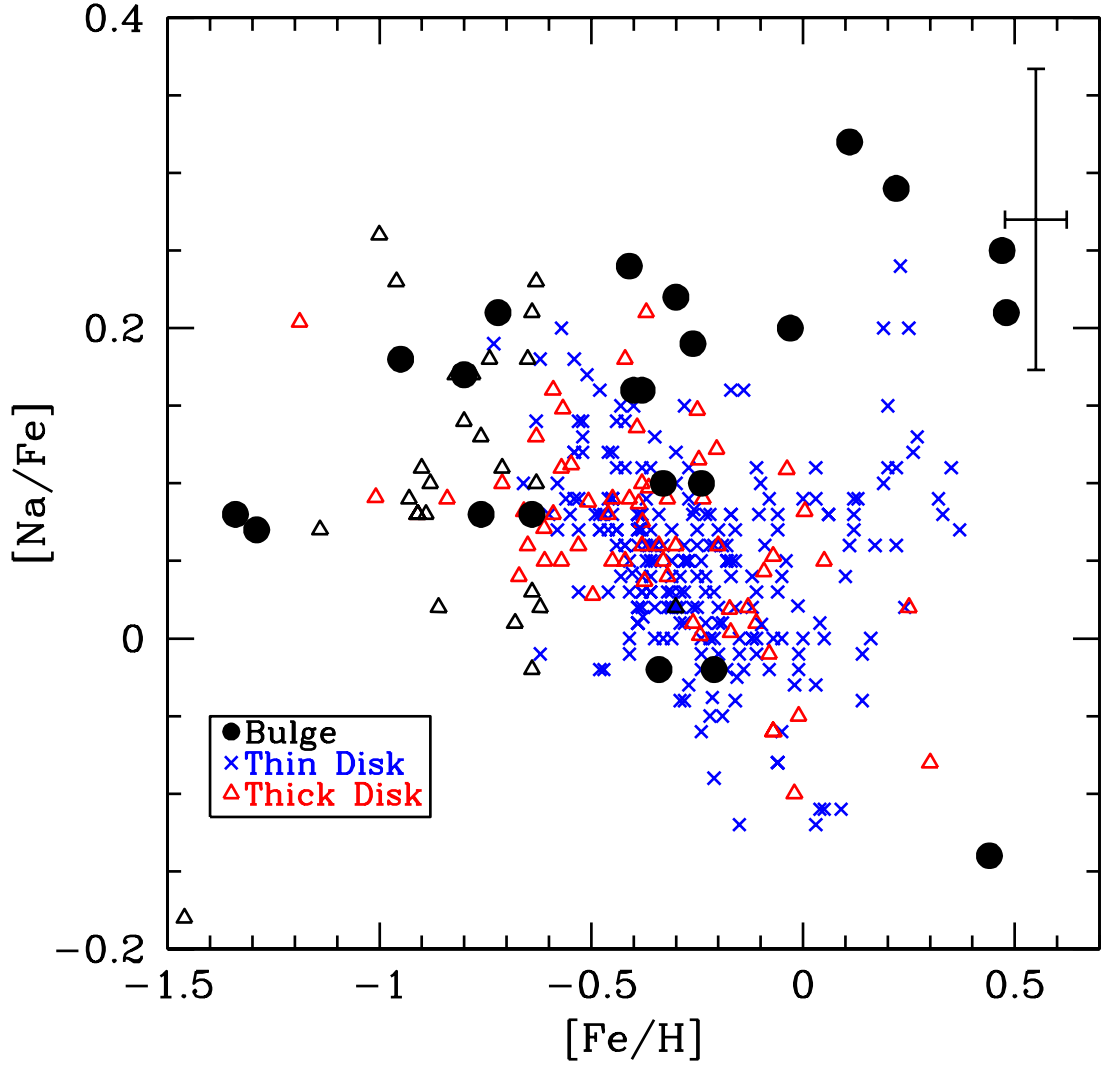} 
\caption{Same as Figure~\ref{fig-jff5-ofebulgedisk}, but for [Na/Fe].  
The black open triangles represent thick disk stars
from Fulbright (2000) using the populations identifications of 
Venn et al. (2004).
The bulge stars show slightly-rising [Na/Fe] values with increasing 
[Fe/H], with [Na/Fe] near $+0.25$ for the most metal-rich stars.  
Two metal-poor bulge stars with high [Na/Fe] are excluded by the scale
of this plot:  I-264 has [Fe/H] $= -1.18$ and [Na/Fe] $= +0.57$ and 
IV-203 has [Fe/H] $= -1.28$ and [Na/Fe] $= +0.66$.}
\label{fig-jff6-nafebulgedisk} 
\end{figure} 
\clearpage

\begin{figure}
\plotone{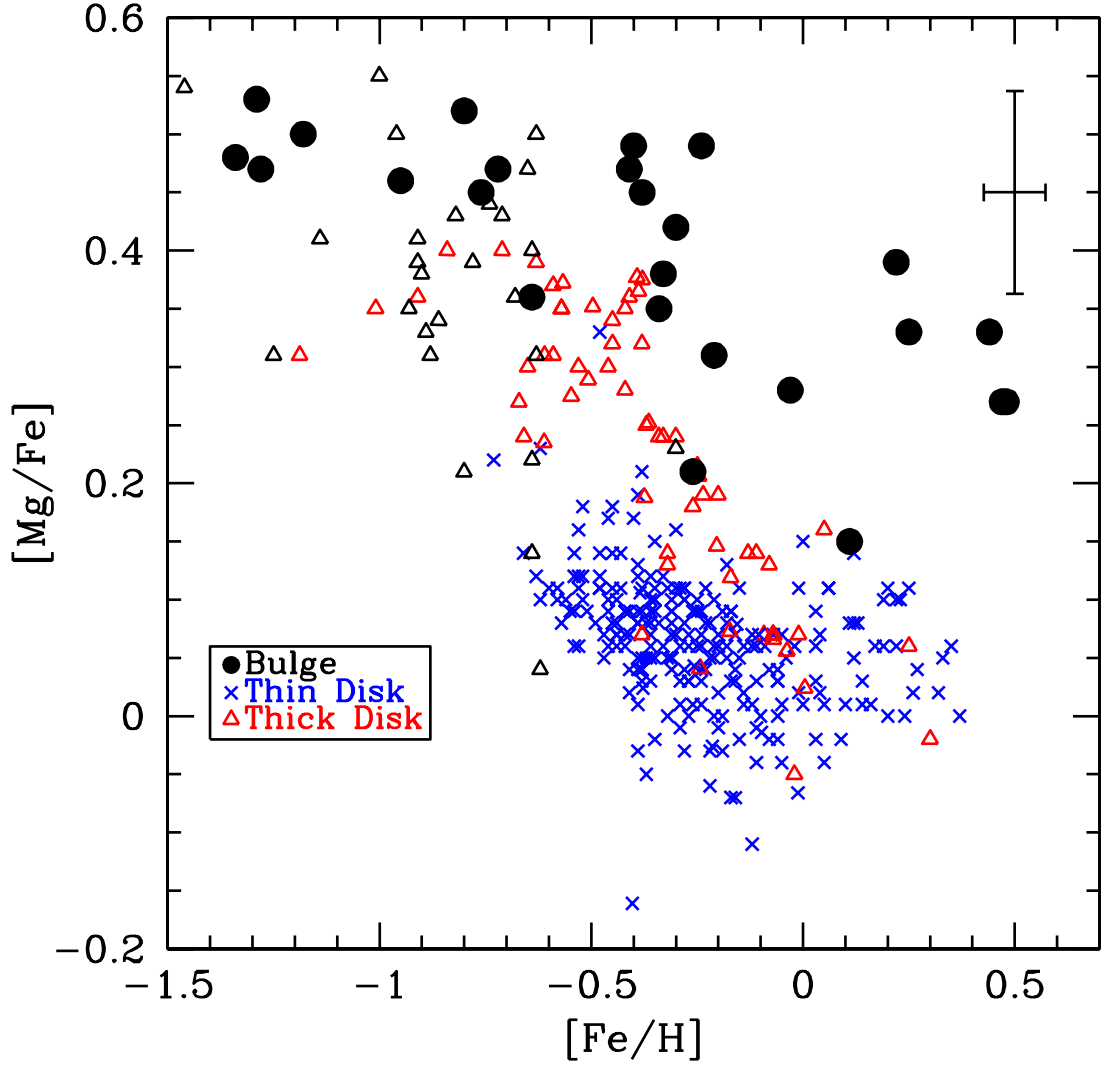} 
\caption{Same as Figure~\ref{fig-jff5-ofebulgedisk}, but for 
[Mg/Fe].  The bulge stars show enhanced [Mg/Fe] values at all metallicities,
which was previously seen by MR94.}\label{fig-jff7-mgfebulgedisk} 
\end{figure} 
\clearpage

\clearpage

\begin{figure}
\plotone{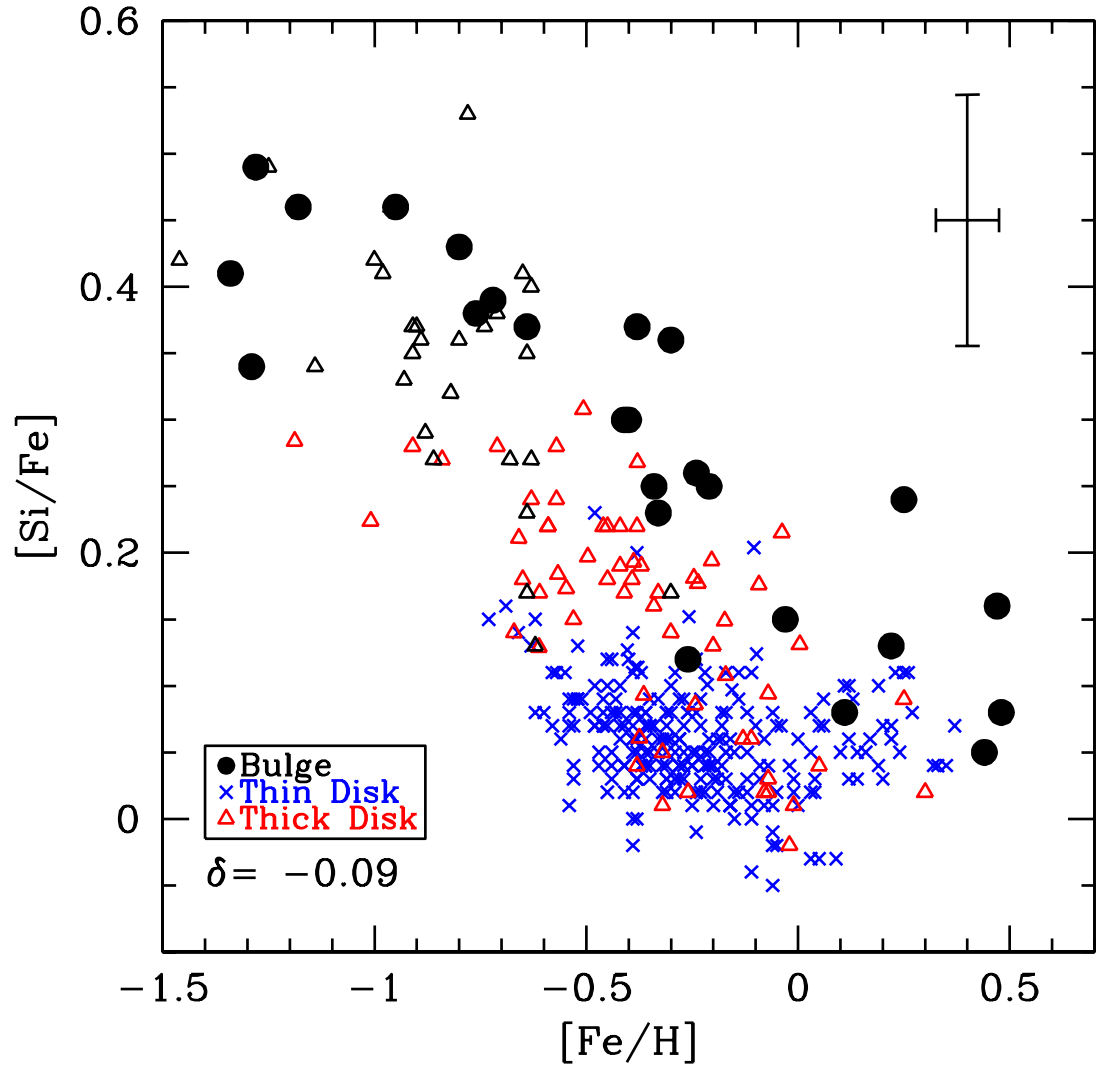}
\caption{Same as Figure~\ref{fig-jff5-ofebulgedisk}, but for 
[Si/Fe].  Like Figure~\ref{fig-sife1dg}, our data points have been shifted
by $-0.09$~dex.}\label{fig-jff9-sifebulgedisk}
\end{figure}
\clearpage

\begin{figure}
\plotone{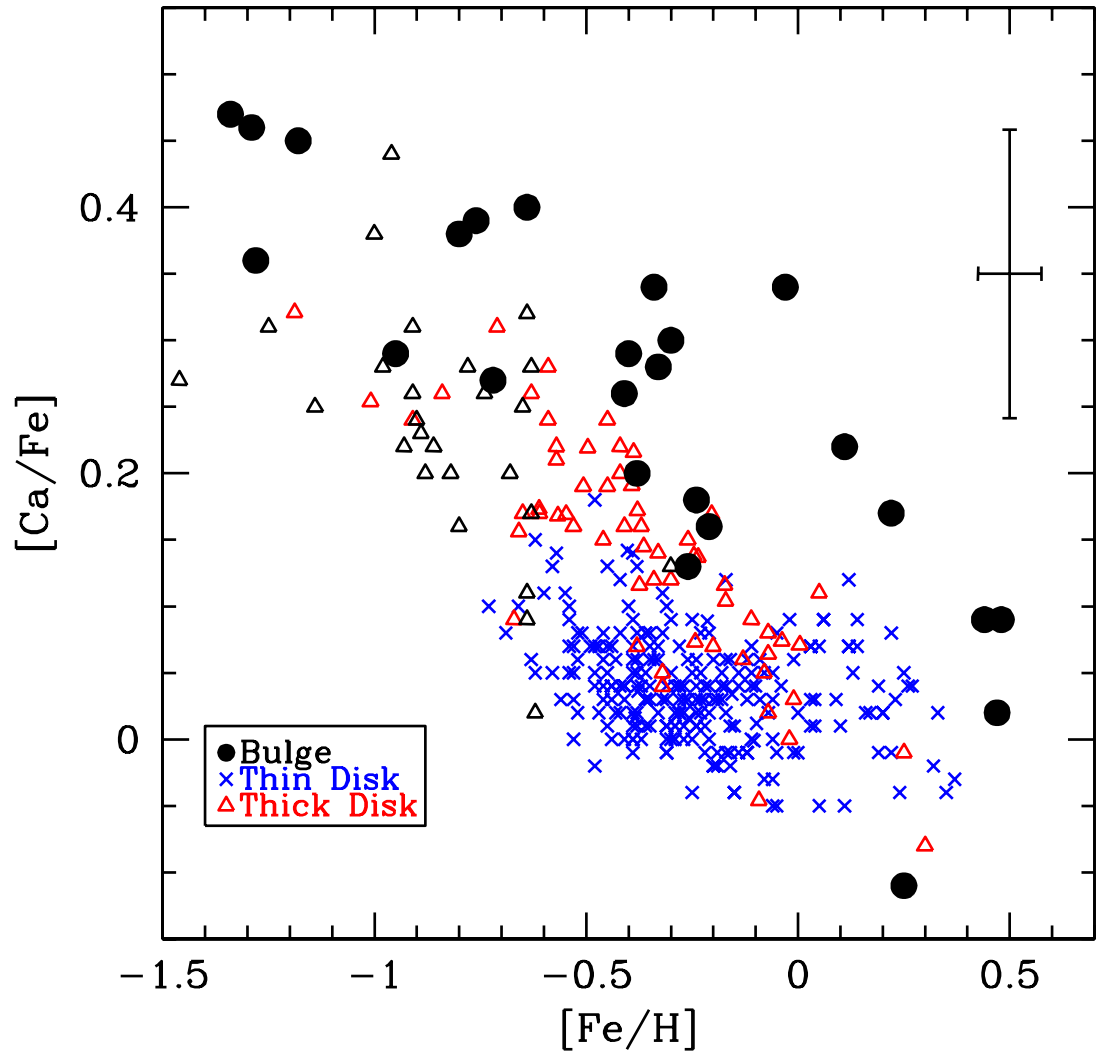}
\caption{Same as Figure~\ref{fig-jff5-ofebulgedisk}, but for 
[Ca/Fe].}\label{fig-jff10-cafebulgedisk}
\end{figure}
\clearpage

\begin{figure}
\plotone{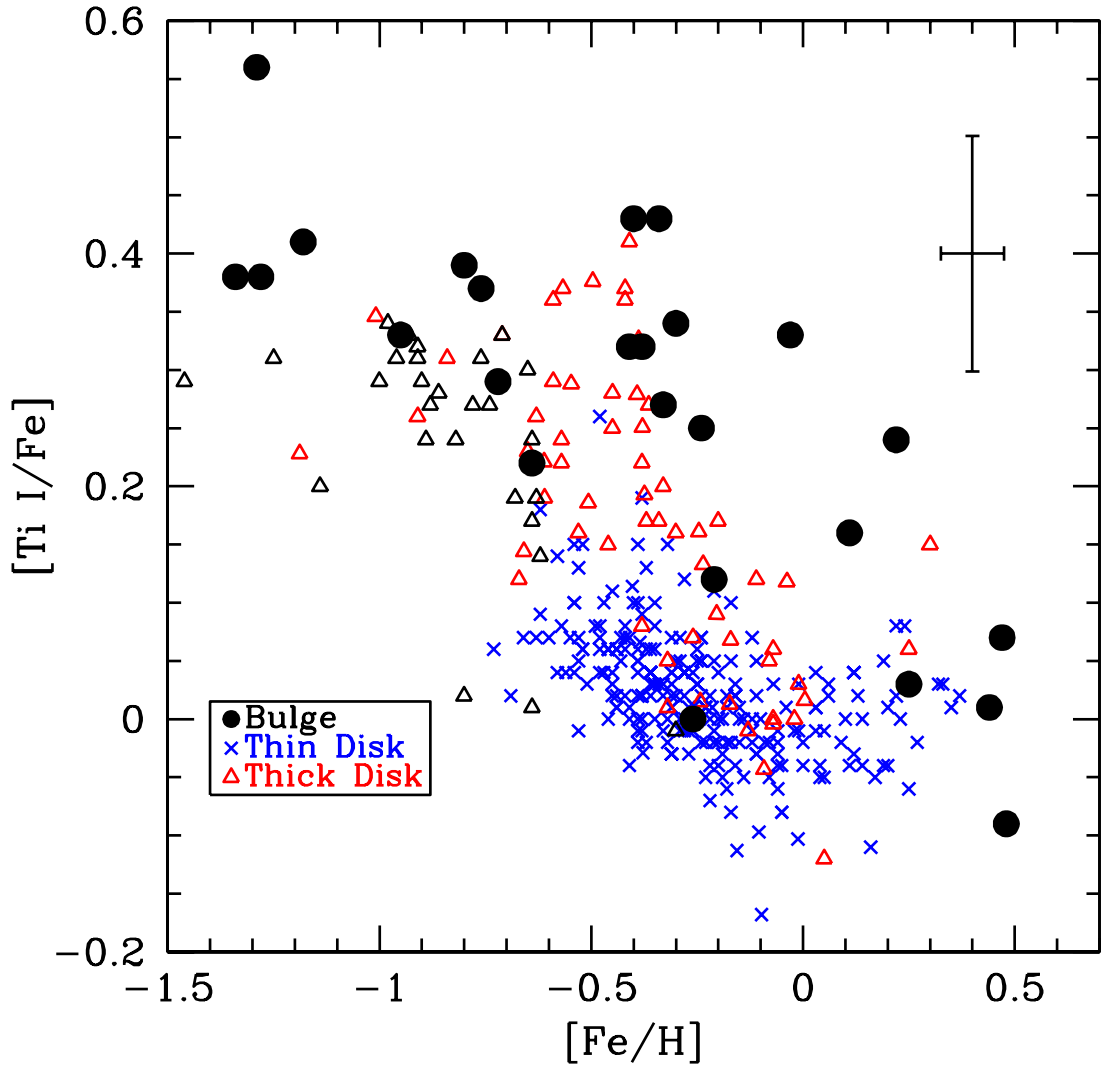}
\caption{Same as Figure~\ref{fig-jff5-ofebulgedisk}, but for [Ti/Fe].  
Only the Ti~I results were used for this 
plot.}\label{fig-jff11-tifebulgedisk}
\end{figure}
\clearpage

\clearpage

\clearpage

\clearpage

\clearpage

\clearpage

\clearpage

\clearpage

\begin{figure}
\plotone{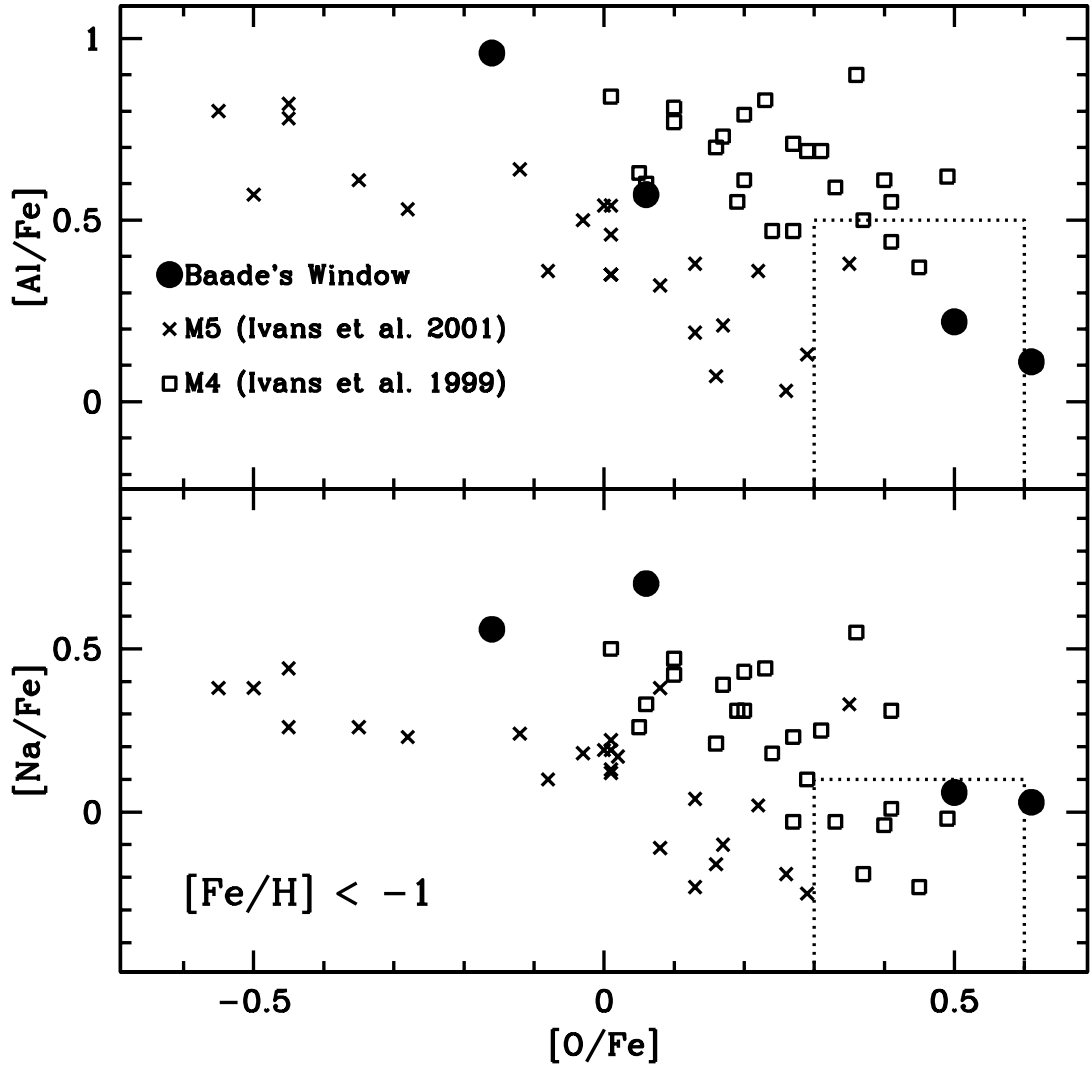}
\caption{The panels show the anti-correlation of Na and Al to O in the 
globular clusters M4 and M5 plus the four metal-poor [Fe/H] $< -1$ bulge 
giants in our survey.  Both globular clusters have [Fe/H] values similar 
to the bulge stars ([Fe/H] $\sim -1.2$).  The bulge stars lie very close 
to the locus defined by M4.  Similar-metallicity field stars mostly lie 
in the regions denoted in the lower right corners by the dotted lines 
(see, for example, Figures~\ref{fig-jff5-ofebulgedisk}, 
\ref{fig-jff6-nafebulgedisk} 
and~\ref{fig-alfenosgr}.}\label{fig-jff15-nafealfeofe}
\end{figure}
\clearpage


\begin{thebibliography}{}

\bibitem[Adelberger et al.(2003)]{adel03} Adelberger, K. L., Steidel, C. C., Shapley, A., E. \& Pettini, M. 2003, \apj, 584, 45
\bibitem[Allende Prieto et al.(2001)]{ap01} Allende Prieto, C., Lambert, 
D. L. \& Asplund, M., 2001, \apj, 556, L63 
\bibitem[Alonso et al.(1999)]{a99} Alonso, A., Arribas, S. \& Martinez-Roger, C. 1999, A\&AS 
140, 261 
\bibitem[Arnett(1971)]{a71} Arnett, W. D. 1971, \apj, 166, 153 
\bibitem[Baade(1963)]{b63} Baade, W. 1963, in Evolution of Stars and Galaxies, ed. W. Baade \& C. P. Gaposhkin (Cambridge: Harvard Univ. Press), 279 
\bibitem[Ballero et al.(2006)]{m06} Ballero, S.K., Matteucci, F., Origlia, L., \& Rich, R.M. 2006, \aap, submitted.
\bibitem[Barbuy et al.(2006) and references therein]{bar06} Barbuy, B. et al. 2006, 449, 349
\bibitem[Bensby et al.(2003)]{b03} Bensby T., Feltzing S. \& Lundstr\"om 
I. 2003, \aap, 410, 527 
\bibitem[Bensby et al.(2005)]{b05} Bensby T., 
Feltzing S., Lundstr\"om I. \& Ilyin I. 2005, \aap, 433, 185 
\bibitem[Bihain et al.(2004)]{birbm04} Bihain, G., Israelian, G., Rebolo, 
R., Bonifacio, P., \& Molaro, P. 2004, \aap, 423, 777 
\bibitem[Blanco \& Blanco(1984)]{bb84} Blanco, V. M. \& Blanco, B. M. 1984, \pasp, 96, 603
\bibitem[Brewer \& Carney (2006)]{bc06} Brewer, M. M., \& Carney, B.W. 2006, 
\aj, 131, 431 
\bibitem[Brown, Wallerstein \& Zucker (1997)]{bwz97} Brown, J.A., 
Wallerstein, G. \& Zucker, D. 1997, \aj, 114, 180 
\bibitem[Cannon et al.(1998)]{can98} Cannon, R. D., Croke, B. F. W., 
Bell, R. A., Hesser, J. E. \& Stathakis, R. A. 1998, MNRAS, 298, 601 
\bibitem[e.g., Carretta et al.(2001)]{car01} Carretta, E., Cohen, J. G., Gratton, R. G. \& Behr, B. B. 2001, \aj, 122, 1469
\bibitem[Castro et al.(1996)]{c96} Castro, S., Rich, R. M., McWilliam, A., 
Ho, L. C., Spinrad, H. \& Filippenko, A. V. 1996, \aj, 111, 2439 
\bibitem[Cayrel et al.(2004)]{cayrel04} Cayrel, R., Depagne, E., Spite, M., 
Hill, V., Spite, F., et al. 2004, \aap, 416, 1117 
\bibitem[Combes(2001)]{c01} Combes, F. 2001, in Galaxy Disks and Disk Galaxies, ASP Conf Ser Vol 320, J. Fumes and M. Corsini (eds) San Francisco: ASP p. 213
\bibitem[Cunha \& Smith(2006)]{cs06} Cunha, K. \& Smith, V. V. 2006, \apj, In Press
\bibitem[Dwek et al.(1995)]{dw95} Dwek, E. et al. 1995, \apj, 445, 716
\bibitem[Eggen et al.(1962)]{els62} Eggen, O. J., Lynden-Bell, D. \& Sandage, A. R. 1962, \apj, 136, 748
\bibitem[Elmegreen(1999)]{elm99} Elmegreen, B. G. 1999, \apj, 515, 323
\bibitem[Ferraras et al.(2003)]{fws03} Ferreras, I., Wyse, R. F. G. \& 
Silk, J. 2003, \mnras, 345, 1381 
\bibitem[Figer et al.(2004)]{figer04} Figer, D. F., Rich, R. M., Kim, S. S., Morris, M. \& Serabyn, E. 2004, \apj, 601, 319
\bibitem[Fulbright(2000)]{f00} Fulbright, J. P. 2000, \aj, 120, 1841 (F00) 
\bibitem[Fulbright(2002)]{f02} Fulbright, J. P. 2002, \aj, 123, 404 
\bibitem[Fulbright \& Johnson(2003)]{fj03} Fulbright, J. P. \& Johnson, J. A. 
2003, \apj, 595, 1154 
\bibitem[Fulbright et al.(2006a; hereafter Paper~I)]{f06} Fulbright, J. P., 
McWilliam, A. \& Rich, R. M. 2006, \apj, 636, 821 
\bibitem[Fulbright et al.(2006b)]{fulb06c} Fulbright, J. P., Rich, R. M. \& McWilliam, A. 2006, in The Metal Rich Universe meeting proceedings, in prep.
\bibitem[Gratton et al.(2004)]{gr04} Gratton, R. F., Sneden, C. \& Carretta, E.
2003, \araa, 42, 385 
\bibitem[Gratton \& Sneden(1990)]{gs90} Gratton, R. F. \& Sneden, C. 1990, 
\aap, 234, 366 
\bibitem[Griffin(1964)]{gr64} Griffin, R. F. 1964, Obs, 84, 154 
\bibitem[Hinkle et al.(2000)]{hw00} Hinkle, K., Wallace, L., Valenti, J., \& 
Harmer, D., eds. 2000, Visible and Near Infrared Atlas of the Arcturus 
Spectrum 3727--9300~\AA{} (San Francisco: ASP) 
\bibitem[Immeli et al.(2004)]{imm04} Immeli, A, Samland, M., Gerhard, O., Westera, P. 2004, ApJ 611, 20
\bibitem[Ivans et al.(1999)]{iii99} Ivans, I. I., Sneden, C., Kraft, R. P., 
Suntzeff, N. B., Smith, V. V., Langer, G. E. \& Fulbright, J. P. 1999, \aj, 
118, 1273 
\bibitem[Ivans et al.(2001)]{iii01} Ivans, I. I., Kraft, R. P., Sneden, 
C., Smith, G. H., Rich, R. M. \& Shetrone, M. 2001, \aj, 122, 1438 
\bibitem[Iwamoto et al.(1999)]{iwa99} Iwamoto, K., Brachwitz, F., Nomoto, 
K., Kishimoto, N., Umeda, H., Hix, W.R. \& Thielemann, F.-K. 1999, \apjs, 
125, 439 
\bibitem[Johnson et. al.(2005)]{j05} Johnson, C. I., Kraft, R. 
P., Pilachowski, C. A., Sneden, C., Ivans, I. I. \& Benman, G. 2005, PASP, 
117, 1308 
\bibitem[Kormendy \& Kennicutt(1994)]{kk04} Kormendy, J. \& Kennicutt, R. C., Jr. 2004, \araa, 42, 603
\bibitem[Kraft(1994)]{k94} Kraft, R. P. 1994, PASP, 106, 553 
\bibitem[Kuijken \& Rich(2002)]{kr02} Kuijken, K. \& Rich, R. M. 2002, 124, 2054
\bibitem[Kurucz et al. (1984)]{kfb84} Kurucz, R.L., Furenlid, I., \& 
Brault, J. 1984, in National Solar Observatory Atlas, vol. 1, Solar Flux 
Atlas from 296 to 1300nm (Sunspot: NSO) 
\bibitem[Kurucz(1993)]{k93} Kurucz, R.L. 1993, IAU Commission 43, Ed. 
E.F.Milone, p.93 
\bibitem[Lambert \& Ries (1981)]{lr81} Lambert, D.L., \& Ries, L.M. 1981, 
\apj, 248, 228 
\bibitem[Luck \& Bond (1985)]{lb85} Luck, R.E., Bond, H.E. 1985, \apj, 292, 
559 
\bibitem[Lodders(2003)]{lodd03} Lodders, K. 2003, \apj, 591, 1220 
\bibitem[Mao \& Paczynski(2002)]{mao} Mao, S. \& Paczynski, B. 2002, MNRAS, 337, 895
\bibitem[Matteucci \& Brocato(1990)]{m90} Matteucci, F. \& Brocato, E. 1990, \apj, 365, 539
\bibitem[Matteucci et al.(1999)]{m99} Matteucci, F., Romano, D. \& Molaro, P.  
\bibitem[Maeder (1980)]{maeder80} Maeder, A. 1980, \aap, 90, 311 
\bibitem[Maeder (1991)]{maeder91} Maeder, A. 1991, \aap, 242, 93 
\bibitem[Maeder (1992)]{maeder92} Maeder, A. 1992, \aap, 264, 105 
\bibitem[McWilliam (1990)]{mcw90} McWilliam, A. 1990, ApJS, 74, 1075 
\bibitem[McWilliam \& Rich(1994; hereafter MR94)]{mr94} McWilliam, A., \& 
Rich, R.M. 1994, ApJS, 91, 749 (MR94) 
\bibitem[McWilliam et al.(1995)]{mcw95a} McWilliam, A., Preston, G.W., Sneden, 
C., \& Shectman, S. 1995a, AJ, 109, 2736 
\bibitem[McWilliam et al.(1995)]{mcw95b} McWilliam, A., Preston, G.W., Sneden, 
C., \& Searle, L. 1995b, AJ, 109, 2757 
\bibitem[McWilliam(1997)]{mcw97} McWilliam, A. 1997, ARA\&A, 35, 503 
\bibitem[McWilliam \& Rich (2004; hereafter MR04)]{mr04} McWilliam, A., \& 
Rich, R.M. 2004, in Origin and Evolution of the Elements, ed. A. McWilliam \& 
M. Rauch, Carnegie Observatories: Pasadena, 
(http://www.ociw.edu/ociw/symposia/series/symposium4/proceedings.html) 
\bibitem[McWilliam \& Smecker-Hane (2005)]{ms05} McWilliam, A., \& 
Smecker-Hane, T.A. 2005, in Cosmic Abundances as Records of Stellar 
Evolution and Nucleosynthesis in honor of David L. Lambert, ASP Conference 
Series, Vol. 336, eds. T.G.Barnes \& F.N.Bash.  San Francisco: 
Astronomical Society of the Pacific, 2005, p.221 
\bibitem[e.g., Mele\'ndez et al.(2003)]{mel03} Mene\'ndez, J. et al. 2003, \aap, 411, 417
\bibitem[Meynet \& Arnould (2000)]{ma00} Meynet, G. \& Arnould, M. 2000, \aap, 
355, 176 
\bibitem[Meynet \& Maeder (2002)]{mm02} Meynet, G. \& Maeder, A. 2002, 
\aap, 390, 561 
\bibitem[Mitchell \& Mohler(1965)]{mm65} Mitchell, W. E., Jr. \& Mohler, O. C. 
1965, \apj, 141, 1126 
\bibitem[Nissen \& Schuster(1997)]{ns97} Nissen, P.E., \& Schuster, W.J. 
1997, \aap, 326, 751 
\bibitem[Origlia et al.(2005)]{orig05} Origlia, L., Valenti, E., Rich, M. R. \& Ferraro, F. R. 2005, \mnras, 363, 897
\bibitem[Ortolani et al.(1995)]{ort95} Ortolani, S. et al. 1995, Nature, 377, 701
\bibitem[Palacios et al.(2005)]{pal05} Palacios A., Arnould, M. \& Meynet, 
G. 2005, \aap, 443, 243 
\bibitem[Pettini et al.(2002)]{pett02} Pettini, M., Rix, S.A., Steidel, C.C., Adelburger, K.L., Hunt, M.P., \& Shapley, A.E. 2002, ApJ, 569, 742
\bibitem[Pfenniger \& Norman (1990)]{pn90} Pfenniger, D., \& Norman, C.A. 
1990, ApJ, 363, 391 
\bibitem[Prochaska et al.(2000)]{pro00} Prochaska, J.X., Naumov, S.O., Carney, 
B.W., McWilliam, A., \& Wolfet, A.M. 2000, \apj, 120, 2513 
\bibitem[Puzia et al.(2002)]{puzia2002} Puzia, T. H. 2002, \aap, 395, 45
\bibitem[Rami\'rez et al.(2000)]{ram00} Rami\'rez, S. V., Stephens, A. W., Frogel, J. A., \& DePoy, D. L. 2000, \aj, 120, 833
\bibitem[Reddy et al.(2003)]{r03} Reddy, B. E., Tomkin, J., Lambert, D. L. \& 
Allende Prieto, C. 2003, \mnras, 340, 304 
\bibitem[Reddy et al.(2006)]{rd06} Reddy, B. E., Lambert, D. L. \& Allende 
Prieto, C. 2006, \mnras, 367, 1329 
\bibitem[Renda et al.(2004)]{ren04} Renda, A., Fenner, Y., Gibson, B. K., 
Karakas, A. I., Lattanzio, J. C.; Campbell, S., Chieffi, A., Cunha, K. \& 
Smith, V. V. 2004, \mnras, 354, 575 
\bibitem[Rich(1988)]{r88} Rich, R. M. 1988, \aj, 95, 828
\bibitem[Rich \& McWilliam(2000; hereafter RM00)]{rm00} Rich, R.M., \& 
McWilliam, A. 2000, SPIE, 4005, 150 (RM00) 
\bibitem[Rich \& Origlia(2005)]{ro05} Rich, R. M. \& Origlia, L. 2005, \apj, 634, 1293
\bibitem[Rich et al.(2006)]{r06} Rich, R.M., Fulbright, J., McWilliam, A., \& Origlia, L.  2006, in Stellar Populations Cozumel meeting ASP conf series, in press
\bibitem[Ruland et al.(1980)]{rul80} Ruland, F., Holweger, H., 
Griffin, R., \& Biehl, D.  1980, \aaps, 42, 391 
\bibitem[Rutledge et al.(1997)]{rut97} Rutledge, G. A., Hesser, J. E. \& 
Stetson, P. B. 1997, \pasp, 109, 907 
\bibitem[Shetrone(1996)]{s96} Shetrone, M. D. 1996, \aj, 112, 1511 
\bibitem[Shetrone et al. (2003)]{s03} Shetrone, M. D., Venn, K.A., Tolstoy, 
E., Primas, F., Hill, V., \& Kaufer, A. 2003, \aj, 125, 684
\bibitem[Smith \& Ruck(2000)]{sr00} Smith, G. \& Ruck, M. J. 2000, \aap, 356, 
570
\bibitem[Sneden(1973)]{moog} Sneden, C. 1973, \apj, 184, 839
\bibitem[Soto Vicencio et al.(2006)]{soto06} Soto Vicencio, Rich, R. M. \& Kuijken, K. 2006, in preparation
\bibitem[Steenbock(1975)]{s85} Steenbock, W. 1985, in Cool Stars with Excesses 
of Heavy Elements, eds. M.Jascheck \& P.C. Keenan (Dordrecht: Reidel), 231
\bibitem[Steidel et al.(1996)]{st96} Steidel, C. C., Giavalisco, M., Pettini, M., Dickinson, M., \& Adelberger, K.L. 1996, \apj, 462, 17
\bibitem[Smecker-Hane \& McWilliam (2002)]{sm02} Smecker-Hane, T. \& McWilliam, A. 2002, astro-ph/0205411
\bibitem[Sweigart \& Mengel (1979)]{sm79} Sweigart, A.V., \& Mengel, J.G. 
1979, \apj, 229, 624
\bibitem[Sumi et al.(2003)]{sumi03} Sumi, T. Eyer, L., Wozniak, P.R.  2003, MNRAS, 340, 1346
\bibitem[Thielemann et al.(1996)]{tnh96} Thielemann, F.-K., Nomoto, K. \& Hashimoto, M. 1996, \apj, 460, 408
\bibitem[Thomas et al.(2003)]{tmb03} Thomas, D., Marastron, C. and Bender, R. 2003, \mnras, 343, 279.

\bibitem[Timmes et al.(1995)]{t95} Timmes, F. X., Woosley, S. E. \& Weaver, T. A. 1995, \apjs, 98, 617
\bibitem[Tinsley(1979)]{t79} Tinsley, B. M 1979, \apj, 229, 1046
\bibitem[Venn et al.(2004)]{venn04} Venn, K.A., Irwin, M., Shetrone, M.D., 
Tout, C.A., Hill, V., \& Tolstoy, E. 2004, \aj, 128, 1177
\bibitem[Vogt et al.(1994)]{vogt} Vogt, S. S., et al. 1994, Proc. SPIE, 2198, 362
\bibitem[Weaver et al.(1989)]{w89} Wheeler, J. C., Sneden, C. \& Truran, J. W., Jr. 1989, \araa, 27, 279
\bibitem[Whitford(1978)]{w78} Whitford, A. E. 1978, \apj, 226, 777
\bibitem[Woosley \& Weaver(1995)]{ww95} Woosley, S.E., \& Weaver, T.A. 1995, 
ApJS, 101, 181 (WW95)
\bibitem[Worthey et al.(1992)]{wfg92} Worthey, G., Faber, S. M. \& Gonzalez, J. J. 1992, \apj, 398, 69
\bibitem[Wyse \& Gilmore (1992)]{wg92} Wyse, R.F.G., \& Gilmore, G. 1992, AJ, 
104, 144
\bibitem[Zhang \& Liu(2005)]{zl05} Zhang, Y. \& Liu, X.-W. 2005, \apj, 631, 61
\bibitem[Zhao et al.(1994)]{zhao94} Zhao, H., Spergel, D. N. \& Rich, R. M. 1994, \aj, 108, 2154
\bibitem[Zoccali et al.(2003)]{z03} Zoccali, M., et al. 2003, \aap, 399, 931
\end{thebibliography}
\end{document}